\documentclass{jfm}
\usepackage{graphicx}
\usepackage{epstopdf, epsfig}
\usepackage{amsmath}
\usepackage{xfrac}
\usepackage{url}

\usepackage{fancyvrb}
\fvset{fontsize=\footnotesize,xleftmargin=2em}
\usepackage[linesnumbered,ruled]{algorithm2e}
\SetKwComment{Comment}{$\triangleright$\ }{}
\newcommand{\Christoffel}[3][]{\genfrac{\{}{\}}{0pt}{}{#2}{#3}_{#1}}

\usepackage{subfigure}
\usepackage{tikz}
\usepackage{pgfplots}
\usepackage{lipsum}
\usepackage{caption}
\usepackage{placeins}
\usepackage{MnSymbol}
\usetikzlibrary{calc,matrix,chains,positioning,decorations.pathreplacing,arrows}

\shorttitle{Construction of Reduced Order Models for Fluid Flows}
\shortauthor{Hugo F. S. Lui and William R. Wolf}

\title{Construction of Reduced Order Models for Fluid Flows Using Deep Feedforward Neural Networks}

\author{Hugo F. S. Lui\aff{1},
  William R. Wolf\aff{1}\corresp{\email{wolf@fem.unicamp.br}}}

\affiliation{\aff{1}School of Mechanical Engineering, Universidade Estadual de Campinas, Campinas, SP, 13083-860, Brazil}

\begin{document}

\maketitle

\begin{abstract}
We present a numerical methodology for construction of reduced order models, ROMs, of fluid flows through the combination of flow modal decomposition and regression analysis. Spectral proper orthogonal decomposition, SPOD, is applied to reduce the dimensionality of the model and, at the same time, filter the POD temporal modes. The regression step is performed by a deep feedforward neural network, DNN, and the current framework is implemented in a context similar to the sparse identification of non-linear dynamics algorithm, SINDy. A discussion on the optimization of the DNN hyperparameters is provided for obtaining the best ROMs and an assessment of these models is presented for a canonical nonlinear oscillator and the compressible flow past a cylinder. Then, the method is tested on the reconstruction of a turbulent flow computed by a large eddy simulation of a plunging airfoil under dynamic stall.
The reduced order model is able to capture the dynamics of the leading edge stall vortex and the subsequent trailing edge vortex. For the cases analyzed, the numerical framework allows the prediction of the flowfield beyond the training window using larger time increments than those employed by the full order model. We also demonstrate the robustness of the current ROMs constructed via deep feedforward neural networks through a comparison with sparse regression. The DNN approach is able to learn transient features of the flow and presents more accurate and stable long-term predictions compared to sparse regression.
\end{abstract}

\begin{keywords}
\end{keywords}

\section{Introduction}
\label{sec:intro}

Current supercomputers allow the application of high fidelity numerical simulations of turbulent flows over large-scale industrial configurations. 
The results from these simulations certainly improve the understanding of complex physical phenomena such as mixing enhancement, drag reduction, heat transfer, noise generation, to name a few. Such simulations may lead to discretizations with billions of degrees of freedom in order to resolve the energetically relevant spatial and temporal scales. In these cases, the fluid flow data is generally obtained for long periods using small time steps to compute meaningful converged statistics of the flow with sufficient accuracy.

The analysis of unsteady flows by time-resolved simulations and experiments require the acquisition and treatment of large datasets. In recent years, data-driven algorithms have been developed and applied to perform statistical post-processing of such datasets of unsteady fluid flows allowing the investigation of complex physical mechanisms in turbulent flows and improving their analyses. For example, one can cite techniques of flow modal decomposition such as proper orthogonal decomposition (POD) and its variations \citep{Lumley1967,sieber2016,Jean2017,towne2018}, dynamic mode decomposition (DMD) and variations \citep{Schmid2010,bruntonDMD,HODMD}, Lagrangian coherent structures (LCS) \citep{Haller2015,GreenLCS,TairaLCS} and resolvent analysis \citep{Mckeon2014,Mckeon2016}, among others. Recently, \citet{Taira2017} published a review of such techniques in the context of fluid flows. 

The previous techniques of modal analysis can also be employed for the construction of reduced order models (ROMs) which are appealing since they can be used in preliminary stages of design due to their inherent reduction in the computational costs compared to large-scale simulations. Such techniques should be also useful in the context of optimization analyzes and studies of flow control \citep{BruntonControl}. However, in order to be employed for these applications, ROMs should be able to reproduce the main physical aspects of the full scale models. Several ROM techniques have been discussed in the literature such as Galerkin projection \citep{Rowley2004}, least-squares Petrov-Galerkin projection \citep{Carlberg2011}, eigensystem realization analysis \citep{ERA1985} and sparse regression of nonlinear dynamics \citep{Brunton3932}. 
The previous techniques can be used to reduce the original sets of partial differential equations to sets of ordinary differential equations. Sparse regression has also been applied for discovering sets of partial differential equations through spatio-temporal data collection \citep{Rudy2017}. These techniques have been mostly applied for canonical problems and their application to more complex turbulent flows is still a challenging task. 

In some cases, ROMs constructed using some of the above techniques may exhibit unstable behavior \citep{Carlberg2011}. For example, ROMs developed based on Galerkin projection employ POD to rewrite the Navier-Stokes equations as sets of dynamical systems for the evolution of the POD temporal modes. Due to the basis truncation typical of such methods, numerical instabilities can be created from the unbalance in the turbulent kinetic energy budget in the ROM. This issue may be addressed using turbulence models \citep{Cazemier1998, Osth2014, Protas2015}, however, this approach can destroy consistency between the original partial differential equations and the ODE system of the ROM. Other alternatives have been also proposed to deal with this problem via minimal rotation of the projected subspace \citep{Balajewicz2016}. This approach is able to account for the contribution of the truncated modes while keeping the consistency between the full and reduced order models. Recently, 
\citet{San2018} employed machine learning via neural networks to compute optimal coefficients for an eddy viscosity model which is able to stabilize a POD-Galerkin ROM.

Machine learning is a field that finds application in several areas from data classification to pattern recognition and non-linear function approximation. Several groups have applied machine learning for investigations concerning fluid flows. \cite{Ling:2015} evaluated different machine learning techniques to classify regions of high uncertainty in RANS calculations. \cite{Ling:2016} presented a novel deep neural network (DNN) architecture to improve RANS turbulence models. In a similar context, \cite{Wang:2017} developed a machine learning approach based on random forests for predicting discrepancies in Reynolds stresses obtained by RANS calculations. \cite{Strofer:2019} employed convolutional neural networks (CNNs), a machine learning technique well suited for image pattern recognition, to identify features in fluid flows.

Machine learning is also a natural candidate for the development of ROMs of large-scale dynamical systems, typical of numerical simulations of unsteady flows.	
A discussion on the application of deep learning in the context just described was  presented by \citet{Kutz2017}.
Recently, several authors have proposed algorithms for the prediction of high-dimensional complex dynamical systems using neural networks and their variants \citep{Rudy2018, San2018, Karthik2018, Wan2018, Vlachas2018}. \citet{Rudy2018} presented  a methodology that is able to learn the dynamics of a particular system and estimate the noise from measurements using feedforward neural networks (FNN). \citet{Karthik2018} performed a comparison between FNNs and sparse regression for modeling of dynamical systems and demonstrated the benefits of the former in terms of adaptability. They computed the Frobenius norm of the Jacobian of the neural network as the regularization term of the cost function to improve accuracy and robustness of the framework. \citet{Wan2018} developed a reduced order model methodology capable of modeling extreme events occurring in dynamical systems. These authors employed a long short-term memory (LSTM) approach to regularize a recurrent neural network (RNN) which obtains the complementary dynamics of a non-linear Galerkin projection of the dynamical system. \citet{Vlachas2018} combined a LSTM neural network with a mean stochastic model to propose a data-driven algorithm which has desirable short and long-term prediction capabilities.

In this work, we present a numerical methodology which combines flow modal decomposition via POD and regression analysis using machine learning and FNNs. The framework is implemented in a context similar to that of the sparse identification of non-linear dynamics (SINDy) algorithm \citep{Brunton3932}. In order to improve the regression step in the current approach, a spectral proper orthogonal decomposition (SPOD) \citep{sieber2016} is employed to extract a low-dimensional representation of the full order model. The SPOD technique is able to filter the temporal modes while preserving the information of the full order model by a redistribution of the energy of the system to higher POD modes. The current method is applied for the construction of ROMs of unsteady compressible flows. First, we test the method for a simple two degree of freedom non-linear dynamical system. Then, we evaluate the capability of the method to reproduce the compressible flow past a cylinder including its noise generation. In this case, we show that the current DNN approach is also able to reproduce transient stages of the flow. Finally, the method is tested for a turbulent flow involving dynamic stall of a plunging airfoil. We present a discussion on the optimization of hyperparameters for obtaining the best models for the deep neural networks. For both the cylinder and plunging airfoil cases, a comparison between DNN and sparse regression techniques is presented in terms of their predictive capability. The approach presented in this work allows to predict the flowfield beyond the training window and with larger time increments than those used by the full order model, demonstrating the robustness of the current ROMs constructed via deep feedforward neural networks. The current numerical tool for construction of reduced order models via DNNs can be downloaded from \url{<http://cces.unicamp.br/software/>}.

\section{Theoretical and Numerical Formulation}
\label{sec:formu}

\subsection{Flow modal decomposition}

The equations governing an unsteady three-dimensional compressible flow contain partial derivatives with respect to both the spatial coordinates $\boldsymbol{x} = \mathopen{\begin{bmatrix}x&y&z\end{bmatrix}}^T$ and time $t$.
Using the method of lines one can first approximate the spatial derivatives producing a system of nonlinear ordinary differential equations (ODEs). In the most general notation, for each mesh point, these ODEs would be expressed in the form
\begin{equation} \label{eq:1}
\frac{d\boldsymbol{q}}{dt} = \boldsymbol{F}(\boldsymbol{q},t) \mbox{ ,}
\end{equation}
where $\boldsymbol{F}$ is a nonlinear operator, $\boldsymbol{q} = \mathopen{\begin{bmatrix} \rho & \rho u & \rho v & \rho w & p \end{bmatrix}}^T$ is the vector of flow variables, $\rho$ is the density, $u$, $v$ and $w$ are respectively the $x$, $y$ and $z$-components of the velocity vector, and $p$ is the pressure.

\begin{figure}
	\centering
	{\includegraphics[trim = 10mm 10mm 5mm 0mm, clip,width=0.99\linewidth]{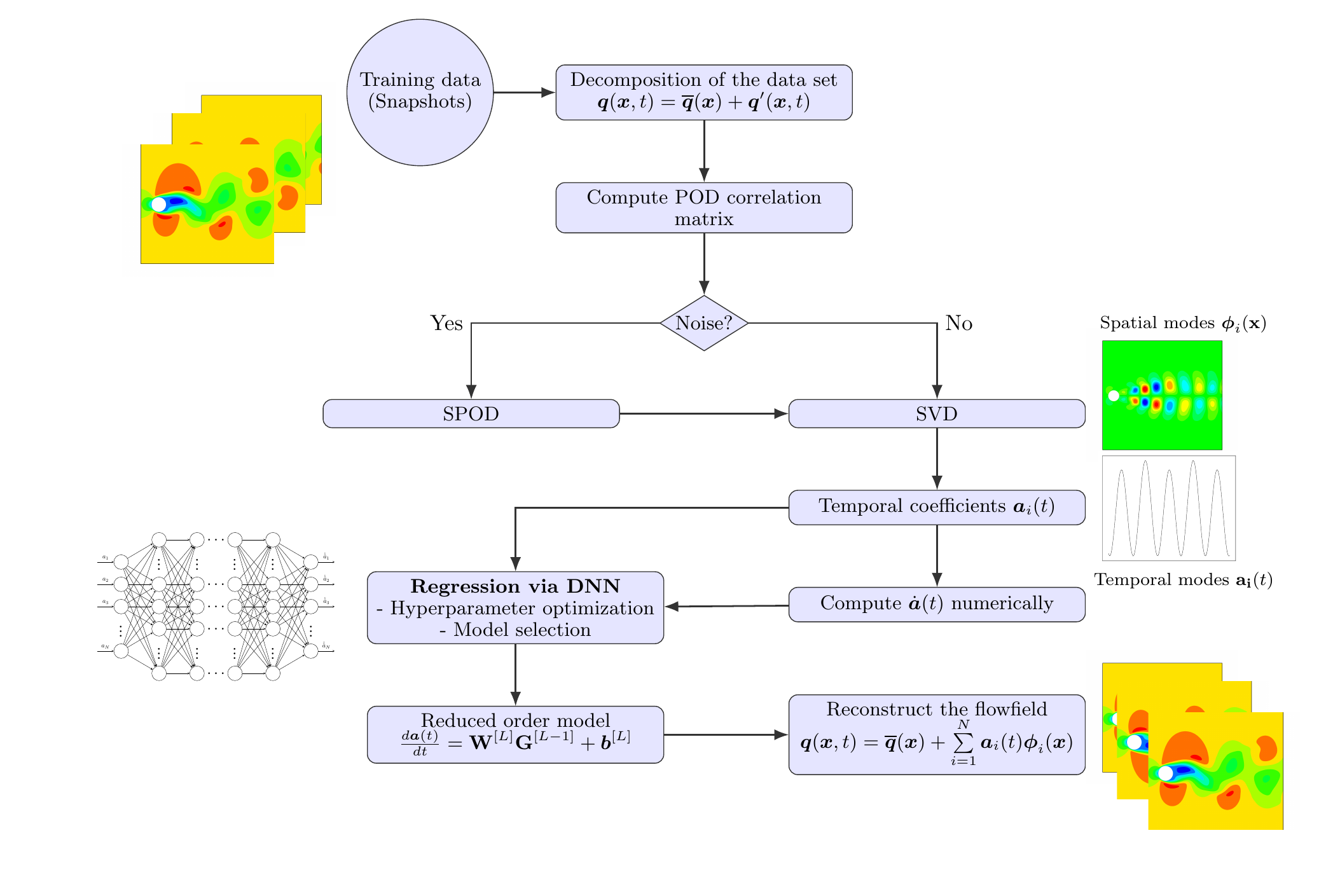}}
	\caption{Schematic of the current method used for construction of reduced order models via DNNs and flow modal decomposition.}
	\label{fig:chart}
\end{figure}

Here, we consider that a dynamical system given by Eq.~\ref{eq:1} is solved at each mesh point. To determine the nonlinear operator $\boldsymbol{F}$ from data, we follow the ideas of the sparse identification of nonlinear dynamics algorithm (SINDy) developed by \citet{Brunton3932} with some modifications. A schematic of the current method can be seen in figure \ref{fig:chart}. First, we collect snapshots of the variables which will be our training data. The data set is then arranged into a matrix $\mathsfbi{Q}$ as
\begin{equation} \label{eq:2}
\mathsfbi{Q} =
\begin{bmatrix}
\boldsymbol{q}(\boldsymbol{x_1},t_1) & \boldsymbol{q}(\boldsymbol{x_2},t_1) & ... & \boldsymbol{q}(\boldsymbol{x_{N_p}},t_1)  \\
\boldsymbol{q}(\boldsymbol{x_1},t_2) & \boldsymbol{q}(\boldsymbol{x_2},t_2) & ... & \boldsymbol{q}(\boldsymbol{x_{N_p}},t_2)  \\
\vdots & \vdots & \ddots & \vdots  \\
\boldsymbol{q}(\boldsymbol{x_1},t_{N_T}) & \boldsymbol{q}(\boldsymbol{x_2},t_{N_T}) & ... & \boldsymbol{q}(\boldsymbol{x_{N_p}},t_{N_T})  \\
\end{bmatrix} \mbox{ ,}
\end{equation}
where $N_p$ is the number of grid points in the computational domain and $N_T$ is the number of snapshots.

Because of the high dimensionality of the input data $\mathsfbi{Q}$, we first reduce the dimension of the dynamical system using the snapshot POD method \citep{Sirovich1986}. Proper orthogonal decomposition has been widely applied to obtain optimal bases that represent the most energetic content of the system dynamics with as few bases functions as possible \citep{Lumley1967}. 
The snapshot POD approach starts with a decomposition of the vector quantities $\boldsymbol{q}(\boldsymbol{x},t)$ into the mean flow, $\overline{\boldsymbol{q}}(\boldsymbol{x})$, and fluctuations, $\boldsymbol{q}'(\boldsymbol{x},t)$. The latter can be further expanded into a combination of spatial modes $\boldsymbol{\phi}_{i}(\boldsymbol{x})$ and their temporal coefficients $\boldsymbol{a}_{i} (t)$ for a defined number of $N$ modes as
\begin{equation} \label{eq:3}
\boldsymbol{q}(\boldsymbol{x},t) = \overline{\boldsymbol{q}}(\boldsymbol{x}) + \boldsymbol{q}'(\boldsymbol{x},t) = \overline{\boldsymbol{q}}(\boldsymbol{x}) + \sum\limits_{i=1}^{N} \boldsymbol{a}_i (t)\boldsymbol{\phi}_i(\boldsymbol{x}) \mbox{ .}
\end{equation}

To calculate the POD correlation matrix of the data set $\mathsfbi{Q}$ some specific norm must be used. For an incompressible flow, a kinetic energy norm provides an optimal result, however, for a compressible flow, other norms can be employed \citep{Rowley2004}. Hence, we define a vector $\boldsymbol{\eta} = \mathopen{\begin{bmatrix} \eta_1 & \eta_2 & \eta_3 & \eta_4 & \eta_5 \end{bmatrix}}^T$ that determines which norm should be used to compute the correlation between two snapshots. For example, a pressure-based norm uses $\boldsymbol{\eta} = \mathopen{\begin{bmatrix} 0 & 0 & 0 & 0 & 1 \end{bmatrix}}^T$ and a kinetic energy norm uses $\boldsymbol{\eta} = \mathopen{\begin{bmatrix} 0 & 1 & 1 & 1 & 0 \end{bmatrix}}^T$. The correlation between two snapshots is computed using the $L^2$ inner product. Therefore, considering $\boldsymbol{q'}(\boldsymbol{x},t_i) = \boldsymbol{q'}_{i}$ and $\boldsymbol{q'}(\boldsymbol{x},t_j) = \boldsymbol{q'}_{j}$, the elements of the correlation matrix $\mathsfbi{C}$ are given by
\begin{equation} \label{eq:4}
C_{ij} = \int_{\Omega} [\eta_1\rho_{i}'\rho_{j}' + \eta_2(\rho u)_{i}'(\rho u)_{j}' + \eta_3(\rho v)_{i}'(\rho v)_{j}' + \eta_4(\rho w)_{i}'(\rho w)_{j}' + \eta_5 p_{i}'p_{j}'] d\Omega \mbox{ ,}
\end{equation}
where $\Omega$ is the fluid region of interest for the reconstruction and the matrix $\mathsfbi{C}$ is of size $N \times N$.
For the problems studied in this work, we employ norms based on kinetic energy and  pressure. Despite the changes in the POD modes computed for the different norms, we observed that the ROMs obtained by either kinetic energy or pressure norms were similar. In both cases, stable and accurate models could be reconstructed by the DNNs and further comments are provided in the results section.

In turbulent flows, the POD temporal modes may contain contributions from several frequencies, including high-frequency noise. For such cases, in order to provide a better identification of the individual modes and to smooth out the temporal coefficients, we employ the spectral proper orthogonal decomposition (SPOD) described by \citet{sieber2016}. The SPOD technique is able to filter the temporal modes while preserving the information of the full order model by a redistribution of the energy of the system to higher POD modes. The technique consists of applying a filter function to the POD correlation matrix, which results in a matrix $\tilde{\mathsfbi{C}}$ with elements given as
\begin{equation} \label{eq:5}
\tilde{C}_{ij} = \sum\limits_{k=-N_f/2}^{N_f/2} g_{k} C_{i+k,j+k} \mbox{ .}
\end{equation}
Here, $g_k$ is the filter function and $N_f$ is the size of the filter window. We consider a periodic temporal series and assume that the correlation matrix is also periodic (see \citet{Jean2017} for details). Hence, $N_f = f_{snap}/ N_T$, where $f_{snap}$ is the number of snapshots used in the SPOD filter. Following this notation, if we apply 50\% of filter to the correlation matrix, it means that we are filtering 50\% of the total number of snapshots. Several filter functions can be applied to the POD correlation matrix and their effects are reported by \citet{Jean2017}. The filter function employed in this work is the box filter represented by $g_k = 1/(2N_f + 1)$.

The temporal coefficients $\boldsymbol{a}_i = \mathopen{\begin{bmatrix} a_i(t_1) & ... & a_i(t_{N_T}) \end{bmatrix}}^T$ and the POD eigenvalues $\lambda_i$ are determined from the filtered correlation matrix $\tilde{\mathsfbi{C}}$ as
\begin{equation} \label{eq:6}
\tilde{\mathsfbi{C}} \boldsymbol{a}_i = \lambda_i \boldsymbol{a}_i \mbox{ .}
\end{equation}
Singular value decomposition (SVD) can be employed to compute the eigenvalues $\lambda_i$ and eigenvectors $\boldsymbol{a}_{i}$ of $\tilde{\mathsfbi{C}}$ since the matrix is real symmetric positive-definite. The spatial modes are obtained from the projection of the fluctuation quantities onto the temporal coefficients
\begin{equation} \label{eq:7}
\boldsymbol{\phi}_i(\boldsymbol{x}) = \frac{1}{\lambda_i} \sum\limits_{j=1}^{N} \boldsymbol{a}_{i} (t_j) \boldsymbol{q}'(\boldsymbol{x},t_j) \mbox{ .}
\end{equation}
Finally, the temporal coefficients and spatial modes can be stored in matrices $\mathsfbi{A}$ and $\boldsymbol{\Phi}$ as
\begin{equation} \label{eq:8}
\mathsfbi{A} = 
\begin{bmatrix}
\boldsymbol{a}^{T}(t_1) \\
\boldsymbol{a}^{T}(t_2) \\
\vdots  \\
\boldsymbol{a}^{T}(t_{N_T}) \\
\end{bmatrix}
=
\begin{bmatrix}
a_1(t_1) & a_2(t_1) & ... & a_N(t_1)  \\
a_1(t_2) & a_2(t_2) & ... & a_N(t_2)  \\
\vdots & \vdots & \ddots & \vdots  \\
a_1(t_{N_T}) & a_2(t_{N_T}) & ... & a_N(t_{N_T})  \\
\end{bmatrix} \mbox{ ,}
\end{equation}
and
\begin{equation} \label{eq:9}
\boldsymbol{\Phi} =
\begin{bmatrix}
\phi_1(\boldsymbol{x_1}) & \phi_1(\boldsymbol{x_2}) & ... & \phi_1(\boldsymbol{x}_{N_p})  \\
\phi_2(\boldsymbol{x_1}) & \phi_2(\boldsymbol{x_2}) & ... & \phi_2(\boldsymbol{x}_{N_p})  \\
\vdots & \vdots & \ddots & \vdots  \\
\phi_N(\boldsymbol{x_1}) & \phi_N(\boldsymbol{x_2}) & ... & \phi_N(\boldsymbol{x}_{N_p})  \\
\end{bmatrix} \mbox{ ,}
\end{equation}
where the temporal coefficients are the columns of $\mathsfbi{A}$ and the spatial modes are the rows of $\boldsymbol{\Phi}$. Hence, the matrix of fluctuation quantities can be written as
\begin{equation} \label{eq:10}
\mathsfbi{Q}\,' = \mathsfbi{A} \, \boldsymbol{\Phi} \mbox{ .}
\end{equation}

Taking the time derivative of Eq. \ref{eq:3}, we arrive at the following set of equations
\begin{equation} \label{eq:11}
\frac{d\boldsymbol{q} (t)}{dt} = \sum\limits_{i=1}^{N} \boldsymbol{\phi}_i(\boldsymbol{x})\frac{d{\boldsymbol{a}_{i}} (t)}{dt} \mbox{ .}
\end{equation}
The last term of Eq. \ref{eq:11} represents the temporal evolution of coefficients $\boldsymbol{a}_i(t)$ associated with the $N$ modes retained in the SPOD modal basis. We can express this system of coupled ODEs as
\begin{equation} \label{eq:12}
\frac{d\boldsymbol{a}(t)}{dt} = \boldsymbol{\dot{a}}(t) = \boldsymbol{F}(\boldsymbol{a}(t)) \mbox{ .}
\end{equation}

Next, we compute the derivative $\boldsymbol{\dot{a}}(t)$ numerically using the data $\boldsymbol{a}(t)$ for each temporal mode. Different numerical schemes are tested for the computation of the temporal derivatives and a discussion on the application of explicit schemes and compact schemes will be provided in the results section. For now, lets consider that the numerical scheme employed for the temporal derivatives is a 10th-order accurate compact scheme \citep{Lele1992} which provides high spectral resolution being non-dissipative and low-dispersive. The derivative $\boldsymbol{\dot{a}}(t)$ is then obtained as
\begin{equation} \label{eq:13}
\delta_1 \boldsymbol{\dot{a}}_{i-2} + \delta_2 \boldsymbol{\dot{a}}_{i-1} + \boldsymbol{\dot{a}}_{i} + \delta_2 \boldsymbol{\dot{a}}_{i+1} + \delta_1 \boldsymbol{\dot{a}}_{i+2} = \delta_3 \frac{\boldsymbol{a}_{i+3} - \boldsymbol{a}_{i-3}}{6h} + \delta_4 \frac{\boldsymbol{a}_{i+2} - \boldsymbol{a}_{i-2}}{4h} + \delta_5 \frac{\boldsymbol{a}_{i+1} - \boldsymbol{a}_{i-1}}{2h} \mbox{ .}
\end{equation}
In the equation above, $h$ is the time step, and the coefficients of the numerical scheme are set as $\delta_1 = 1/20$, $\delta_2 = 1/2$, $\delta_3 = 1/100$,  $\delta_4 = 101/150$ and $\delta_5 = 17/12$ for a 10th-order discretization. 
The system of Eqs. \ref{eq:13} written for each temporal mode can be solved as a pentadiagonal linear system for the unknown derivatives $\boldsymbol{\dot{a}}(t)$. The boundary data points can be computed from the interior points following \citet{Britton}.

The derivatives $\boldsymbol{\dot{a}}(t)$ are then arranged into a matrix $\mathsfbi{\dot{A}}$
\begin{equation} \label{eq:17}
\mathsfbi{\dot{A}} = 
\begin{bmatrix}
\boldsymbol{\dot{a}}^{T}(t_1) \\
\boldsymbol{\dot{a}}^{T}(t_2) \\
\vdots  \\
\boldsymbol{\dot{a}}^{T}(t_{N_T})\\
\end{bmatrix}
=
\begin{bmatrix}
\dot{a}_1(t_1) & \dot{a}_2(t_1) & ... & \dot{a}_N(t_1)  \\
\dot{a_1}(t_2) & \dot{a_2}(t_2) & ... & \dot{a}_N(t_2)  \\
\vdots & \vdots & \ddots & \vdots  \\
\dot{a}_1(t_{N_T}) & \dot{a}_2(t_{N_T}) & ... & \dot{a}_N(t_{N_T})  \\ 
\end{bmatrix} \mbox{ .}
\end{equation}

\subsection{Regression via DNN}

Once the matrix of temporal derivatives is computed, we can set up a regression problem to find the weights $\mathsfbi{W}$ and biases $\boldsymbol{b}$ that determine the function $F(\boldsymbol{a}(t))$ presented in Eq. \ref{eq:12}
\begin{equation} \label{eq:18}
\mathsfbi{\dot{A}} = \mathsfbi{W}\boldsymbol{\Theta}(\mathsfbi{A}) + \boldsymbol{b} \mbox{ ,}
\end{equation}
where $\boldsymbol{\Theta}(\mathsfbi{A})$ is the matrix of features.
In the SINDy algorithm, \citet{Brunton3932} suggest that $\boldsymbol{\Theta}(\mathsfbi{A})$ may consist of constant, polynomial, exponential and trigonometric functions. However, in many cases, it is complicated to know what set of features should be extracted from the data. 
Hence, we use machine learning to circumvent the problem of finding the functions which represent the dynamics of the problem. Therefore, the methodology can discover not only the weights $\mathsfbi{W}$ and biases $\boldsymbol{b}$ but also the features $\boldsymbol{\Theta}(\mathsfbi{A})$. The ``learned'' features often result in a better performance when compared to those obtained using ``engineered'' features. A learning algorithm can find a proper set of features in minutes or hours, depending on the task complexity. On the other hand, manually engineered features would require a great amount of human time and effort for complex tasks \citep{Goodfellow-et-al}.

Deep learning methods are feature learning algorithms that can find a proper set of features using multiple layers, from higher layer features defined in terms of lower layer features. Automatically learning features at multiple processing layers allow the learning of complex functions through mapping the input to the output directly from a given data \citep{Bengio}. In the present work, a deep feedforward neural network (DNN) is used to learn the weights $\mathsfbi{W}$, biases $\boldsymbol{b}$ and features $\boldsymbol{\Theta}(\mathsfbi{A})$ of the dynamical systems investigated. Figure \ref{fig:1} shows a sample DNN arquitecture where the input $\mathsfbi{X}$ of the DNN is the matrix $\mathsfbi{A}$ and the target $\mathsfbi{Y}$ is the matrix $\mathsfbi{\dot{A}}$. The DNN calculation procedure is presented in the following algorithm form~\ref{alg:1}. In the current work, the open source machine learning framework Tensorflow~\citep{tensorflow2015-whitepaper} is used for training the DNN. 

\tikzset{%
	every neuron/.style={
		circle,
		draw,
		minimum size=1cm
	},
	neuron missing/.style={
		draw=none, 
		scale=4,
		text height=0.333cm,
		execute at begin node=\color{black}$\vdots$
	},
}

\begin{figure}
	\begin{center}	
		\tikzset{%
			every neuron/.style={
				circle,
				draw,
				minimum size=0.8cm
			},
			neuron missing/.style={
				draw=none, 
				scale=2.5,
				text height=0.333cm,
				execute at begin node=\color{black}$\vdots$
			},
		}
		
		\begin{tikzpicture}[x=1.3cm, y=1.22cm, >=stealth]
		
		\foreach \m/\l [count=\y] in {1,2,3,missing,4}
		\node [every neuron/.try, neuron \m/.try] (input-\m) at (0,2.5-\y) {};
		
		\foreach \m/\l [count=\y] in {1,missing,2,3,4,missing,5}
		\node [every neuron/.try, neuron \m/.try] (hidden-\m) at (1.6,3.5-\y) {};
		
		\foreach \m/\l [count=\y] in {1,missing,2,3,4,missing,5}
		\node [every neuron/.try, neuron \m/.try] (hidden1-\m) at (3.2,3.5-\y) {};
		
		\foreach \m/\l [count=\y] in {1,missing,2,3,4,missing,5}
		\node [every neuron/.try, neuron \m/.try] (hidden2-\m) at (4.8,3.5-\y) {};
		
		\foreach \m/\l [count=\y] in {1,missing,2,3,4,missing,5}
		\node [every neuron/.try, neuron \m/.try] (hidden3-\m) at (6.4,3.5-\y) {};
		
		\foreach \m [count=\y] in {1,2,3,missing,4}
		\node [every neuron/.try, neuron \m/.try ] (output-\m) at (8.0,2.5-\y) {};
		
		\foreach \l [count=\i] in {1,2,3,N}
		\draw [<-] (input-\i) -- ++(-1,0)
		node [above, midway] {$a_\l$};
	
		\foreach \l [count=\i] in {1,2,3,N}
		\draw [->] (output-\i) -- ++(1,0)
		node [above, midway] {$\dot{\hat{a}}_\l$};
		
		\foreach \i in {1,...,4}
		\foreach \j in {1,...,5}
		\draw [->] (input-\i) -- (hidden-\j);
		
		\foreach \i in {1,...,5} 
		\foreach \j in {1,...,5} 
		\draw [->] (hidden-\i) -- (hidden1-\j);
		
		\foreach \i in {1,...,5}
		\foreach \j in {1,...,5}
		\draw [->] (hidden2-\i) -- (hidden3-\j);
		
		\foreach \i in {1,...,5}
		\foreach \j in {1,...,4}
		\draw [->] (hidden3-\i) -- (output-\j);
		
		\foreach \l [count=\x from 0] in {Input, Hidden, Hidden,Hidden, Hidden, Ouput}
		\node [align=center, above] at (\x*1.6,3) {\l \\ layer};
		
		\node [scale = 2.5] at ($(hidden1-1)!.53!(hidden2-1)$) {$\dots$};
		\node [scale = 2.5] at ($(hidden1-2)!.53!(hidden2-2)$) {$\dots$};
		\node [scale = 2.5] at ($(hidden1-3)!.53!(hidden2-3)$) {$\dots$};
		\node [scale = 2.5] at ($(hidden1-4)!.53!(hidden2-4)$) {$\dots$};
		\node [scale = 2.5] at ($(hidden1-5)!.53!(hidden2-5)$) {$\dots$};
		
		\end{tikzpicture}
	\end{center}
	\caption{An example of a deep feedforward network with $N$ inputs, several hidden layers (HL), and one output layer with $N$ outputs.} \label{fig:1}
\end{figure}
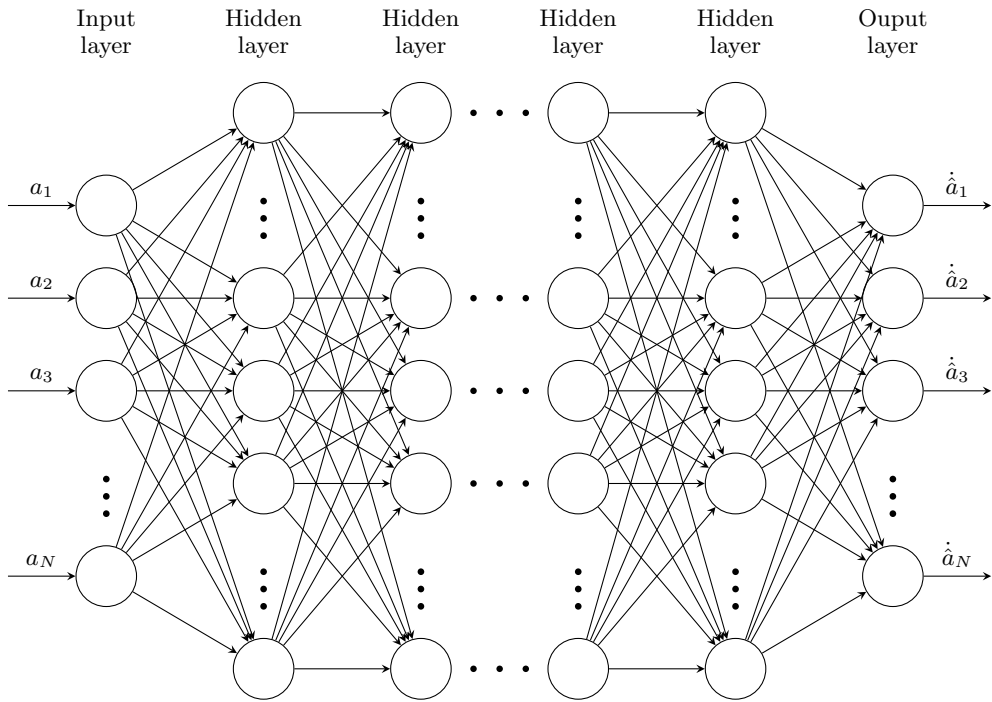

\begin{algorithm}
	\SetKwInOut{Input}{Input}
	\SetKwInOut{Output}{Output}
	
	\Input{Training data $\mathsfbi{X}$, Target $\mathsfbi{Y}$, network depth $L$, activation function $\sigma$, number of hidden units $n^{[l]}$ for layer $l$, learning rate $\alpha$, regularization parameter $\lambda$, maximum number of iterations $n_{iter}$, exponential decay rate for the 1st moment estimates $\beta_1$, exponential decay rate for the 2nd moment estimates $\beta_2$, and small constant for numerical stability $\epsilon$}
	\Output{$\mathsfbi{W}^{[l]}$, $\boldsymbol{b}^{[l]}$, $\mathsfbi{\hat{Y}}$}
	
	\textbf{Initialization of parameters}: All the weights $\mathsfbi{W}^{[l]}$ are initialized using Xavier's initialization \citep{Xavier}. The biases $\boldsymbol{b}^{[l]}$ are initialized to zero. The matrix $\mathsfbi{W}^{[l]}$ is of size $n^{[l+1]} \times n^{[l]}$ and the vector $\boldsymbol{b}^{[l]}$ is of size $n^{[l]} \times 1$\;
	
	\textbf{Initialization of Adam parameters}: Adam parameters $\mathsfbi{V}_{dW^{[l]}}$, $\boldsymbol{V}_{db^{[l]}}$, $\mathsfbi{S}_{dW^{[l]}}$ and $\boldsymbol{S}_{db^{[l]}}$ are initialized to zero. They have the same dimensions as $\mathsfbi{W}^{[l]}$ and $\boldsymbol{b}^{[l]}$\;
	
	\For{$iter = 1$ to $n_{iter}$ } 
	{
		\textbf{Forward propagation} \; 
		$\mathsfbi{G}^{[0]} = \mathsfbi{X}$ \;
		\For{$l = 1$ to $L$} 
		{
			$\mathsfbi{Z}^{[l]} = \mathsfbi{W}^{[l]}\mathsfbi{G}^{[l-1]} + \boldsymbol{b}^{[l]}$\;
			$\mathsfbi{G}^{[l]} = \sigma({\mathsfbi{Z}}^{[l]})$;  \Comment{for the last layer $L$:  $\mathsfbi{G}^{[L]} = \mathsfbi{Z}^{[L]}$}
		}
		$\mathsfbi{\hat{Y}} = \mathsfbi{Z}^{[L]}$\;
		\textbf{Cost function}: 	$J = \frac{1}{2N_T}\left[\sum\limits_{j=1}^N \sum\limits_{i=1}^{N_T}({\hat{Y}_{i,j}} - {Y}_{i,j})^{2} + \lambda\sum\limits_{l=1}^{L}\sum\limits_{k=1}^{n^{[l]}}\sum\limits_{j=1}^{n^{[l+1]}}{(W}^{[l]}_{j,k})^2\right]$\; 
		
		\textbf{Backward propagation} \;
		$d\mathsfbi{G}^{[L]} = d\mathsfbi{Z}^{[L]}$; \Comment{as $\mathsfbi{G}^{[L]} = \mathsfbi{Z}^{[L]}$ then $d\mathsfbi{Z}^{[L]} = \mathsfbi{\hat{Y}} - \mathsfbi{Y}$}
		${d\mathsfbi{W}}^{[L]} = \frac{1}{N_T} \left[d\mathsfbi{Z}^{[L]} \cdot (\mathsfbi{G}^{[L-1]})^{T}\right]$\;
		${d\boldsymbol{b}}^{[L]} = \frac{1}{N_T} \sum\limits_{i=1}^{N_T} d\mathsfbi{Z}^{[L]}$\;
		\For{$l = L-1$ to $1$} 
		{
			${d\mathsfbi{G}}^{[l-1]} = (\mathsfbi{W}^{[l+1]})^{T} \cdot (d\mathsfbi{Z}^{[l+1]})$\;
			${d\mathsfbi{Z}}^{[l]} = d\mathsfbi{G}^{[l]}*\sigma'(\mathsfbi{Z}^{[l]})$\;
			${d\mathsfbi{W}}^{[l]} = \frac{1}{N_T} \left[d\mathsfbi{Z}^{[l]} \cdot (\mathsfbi{G}^{[l-1]})^{T}\right]$\;
			${d\boldsymbol{b}}^{[l]} = \frac{1}{N_T} \sum\limits_{i=1}^{N_T} d\mathsfbi{Z}^{[l]}$\;
		}
		
		\textbf{Adam optimization \citep{Adam}}  \;
		\For{$l = 1$ to $L$}
		{
			$\mathsfbi{V}_{dW^{[l]}} = \beta_1 \mathsfbi{S}_{dW^{[l]}} + (1 - \beta_1)d\mathsfbi{W}^{[l]}$; $\qquad\,\,\,\,\,\,\,\boldsymbol{V}_{db^{[l]}} = \beta_1 \textbf{S}_{db^{[l]}} + (1 - \beta_1)d\boldsymbol{b}^{[l]}$\;
			$\mathsfbi{S}_{dW^{[l]}} = \beta_2 \mathsfbi{S}_{dW^{[l]}} + (1 - \beta_2)(d\mathsfbi{W}^{[l]})^2$; $\qquad{S}_{db^{[l]}} = \beta_2 \boldsymbol{S}_{db^{[l]}} + (1 - \beta_2)(d\boldsymbol{b}^{[l]})^2$\;
			$\mathsfbi{V}_{dW^{[l]}} := \frac{\mathsfbi{V}_{dW^{[l]}}}{1 - \beta^{iter}_1}$; $\,\,\,\,\,\boldsymbol{V}_{db^{[l]}} := \frac{\boldsymbol{V}_{db^{[l]}}}{1 - \beta^{iter}_1}$; 	$\,\,\,\,\,\mathsfbi{S}_{dW^{[l]}} := \frac{\mathsfbi{S}_{dW^{[l]}}}{1 - \beta^{iter}_2}$; $\,\,\,\,\,\boldsymbol{S}_{db^{[l]}} := \frac{\boldsymbol{S}_{db^{[l]}}}{1 - \beta^{iter}_2}$\;
			${\mathsfbi{W}}^{[l]} := {\mathsfbi{W}}^{[l]} - \alpha\frac{\mathsfbi{V}_{dW^{[l]}}}{\sqrt{\mathsfbi{S}_{dW^{[l]}}} + \epsilon}$\; 
			${\boldsymbol{b}}^{[l]} := {\boldsymbol{b}}^{[l]} - \alpha\frac{\boldsymbol{V}_{db^{[l]}}}{\sqrt{\boldsymbol{S}_{db^{[l]}}} + \epsilon}$\;
		}
	}
	
	\caption{Deep feedforward network (DNN).}
	\label{alg:1}
\end{algorithm}

Once we have learned the parameters $\mathsfbi{W}^{[l]}$ and $\boldsymbol{b}^{[l]}$ of our model \ref{eq:18}, we can use them to  predict the temporal coefficients ${a_i}$ given the initial conditions $\boldsymbol{a}(t_1)$. The system of coupled ODEs \ref{eq:12} can now be given by
\begin{equation} \label{eq:19}
\frac{d\boldsymbol{a}(t)}{dt} = \mathsfbi{W}^{[L]}\mathsfbi{G}^{[L-1]} + \boldsymbol{b}^{[L]} = \boldsymbol{F}(\boldsymbol{a}(t)) \mbox{ ,}
\end{equation}
where $L$ is the index of the last layer, $\mathsfbi{G}^{[L-1]}$ is the activation function matrix of the penultimate layer, $\mathsfbi{W}^{[L]}$ is the weight matrix of the last layer, and $\boldsymbol{b}^{[L]}$ is the bias vector of the last layer.

This system of coupled ODEs is integrated using an explicit 5 stage 4th-order Runge-Kutta scheme derived by \citet{Kennedy1999}. As we can see in algorithm~\ref{alg:1}, $\mathsfbi{G}^{[L-1]}$ depends on the weigths $\mathsfbi{W}$ and biases $\boldsymbol{b}$ from previous layers. Thus, for each stage of the 4th-order Runge Kutta, we need to perform forward propagation to obtain $\boldsymbol{F}(\boldsymbol{a}(t))$.
As we have the temporal coefficients $\boldsymbol{a}(t)$, one can reconstruct the flowfield using Eq. \ref{eq:3}. However, we are interested in using a reduced order model in circumstances other than simply reproducing the training data. The approach presented in this work allows to predict the flowfield beyond the training window as $\overline{\boldsymbol{q}}(\boldsymbol{x})$ and $\boldsymbol{\phi}(\boldsymbol{x})$ depend only on the spatial coordinates $\boldsymbol{x}$ and they are calculated using only the training data.    

\subsection{Data-driven ROMs}

As previously discussed in section \ref{sec:intro}, the construction of reduced order models is an active area of research and different methodologies for the development of ROMs are available. For example, POD-Galerkin based methods are directly related to the physics of the problem through the projection of the Navier Stokes equations into a system of ordinary differential equations. Galerkin projection methods require the treatment of the linear and non-linear spatial terms appearing in the full order model. Another issue with these methods relates to their expensive application for dynamical systems with strong non-linearities where one should employ, for example, hyper-reduction techniques \citep{Saifon:2010, Carlberg2011, Zimmermann:2016}.

In the current approach, we employ DNNs for the regression step in a context similar to the SINDy algorithm. Both the DNN and the original SINDy approaches are data-driven methods and, instead of having a direct connection with the physics of the problem, these techniques learn from data. An advantage of these methods is that the non-linearities of the problem are considered in the temporal derivatives of the primitive variables, which are obtained from the full order model. Therefore, neither the SINDy nor the DNN method requires the treatment of spatial derivatives.

In general, the resulting reduced order models constructed via DNNs are not physically interpretable due to the non-linearity of the matrix of features and their weights and biases, which are not sparse. On the other hand, in some cases, such as the flow past a cylinder, models constructed using sparse regression could lead to physically interpretable results \citep{Brunton3932}. As expected, the application of DNNs for the regression step adds a penalty cost compared to sparse regression. However, it is shown in this work that, for the cases analyzed, models obtained using neural networks present better long-term predictive capabilities compared to sparse regression. Both these issues will be discussed in the results section.

\section{Hyperparameter Optimization}

The performance of the DNN described in algorithm \ref{alg:1} depends dramatically on the selection of hyperparameters such as the network depth, $L$, number of hidden units for each layer,  $n^{[l]}$, regularization parameter, $\lambda$, and learning rate $\alpha$. Indeed, finding an optimal set of hyperparameters which minimizes the loss function over a hyperparameter space is a challenging task given the substantial number of free parameters involved.

The manual search, grid search, random search \citep{Bergstra} and Bayesian optimization \citep{bayesian} are the most widely used procedures for the hyperparameter optimization. The manual search consists in a direct human trial and error procedure in the search for an optimal configuration of hyperparameters. This procedure is entirely based on prior experience of the user and there is a high probability that an optimal set of hyperparameters is not found. However, manual search is still useful if the effect of a specific hyperparameter on the model performance can be monitored on the fly. 
In the grid search method, several combinations of hyperparameter values are tested in a range evenly spaced. This method is extremely time-consuming because the number of trials increases exponentially with the number of hyperparameters. In a random search, one randomly selects each hyperparameter from a defined range and evaluates the model performance. It is time-consuming when a high-dimensional hyperparameter space is analyzed, however, \citet{Bergstra} empirically show that random search outperforms a grid search for the hyperparameter optimization both in terms of computational time and model performance.

One of the recent strategies to find an optimal set of hyperparameters is the Bayesian optimization. It is a technique that involves constructing a probabilistic surrogate model to the data in order to determine the most promising hyperparameters to evaluate. \citet{Snoek} showed that Bayesian optimization was able to find optimal hyperparameters for a three-layer convolutional neural network considerably faster than previous approaches, and outperformed the state of the art performance at selecting the set of hyperparameters on the CIFAR-10 dataset \citep{Krizhevsky}. 

In this work, we use two hyperparameter optimization strategies: random search and Bayesian optimization. For random search, the model generation procedure is presented in the following algorithm form~\ref{alg:2}. Likewise, Bayesian optimization has the same inputs as random search. To report the performance of each model from a set of candidates, we compute the mean absolute error (MAE) over the training data. The candidate models with lower MAE values are most likely those which will provide the optimal parameters. It is possible, however, that the model with the lowest MAE suffers from overfitting. 

The current metric does not assess the generalization of the model. However, it is the only one available since we cannot split our data into training and validation sets. The validation dataset should provide an unbiased evaluation of the ROM. In our case, we employ the temporal coefficients of the POD modes for the training stage of the model. These modes are computed using a correlation of different snapshots and if we construct the ROM using POD modes obtained with data including the validation set, our model would use a biased set for the training stage. This occurs because the POD correlation matrix would be computed for snapshots of both the training and validation sets.
\begin{table}
	\begin{center}
		\begin{tabular}{  p{3cm}  p{3cm}  p{3cm}  p{3cm} }
			
			Hyperparameter & Improves performance if & Reason & Warning \\ \hline
			
			Number of hidden units, $n_{hidden}$ & Increased & Increasing the number of hidden units augments the capacity of the model to represent more complex functions & Increasing this parameter may cause overfitting to the training data \\ \hline
			
			Number of layers, $n_{layers}$ & Increased & Same as above  &  Same as above. $\,\,\,\,$ One should also be aware that a DNN with a large number of layers and a very small number of hidden units  will not work properly. \\ \hline
			
			Regularization parameter, $\lambda$ & Reduced & Reducing the regularization parameter allows larger weights for the model features. One should expect that some of these features are the most relevant for the model. & Reducing the regularization parameter causes the model to be more prone to overfitting to the training data  \\ \hline
			
			Learning rate, $\alpha$ & Tuned & If $\alpha$ is too small, the optimization process can be slow. If $\alpha$ is too high, the optimization method may lead to overshoot of local minima. This parameter is chosen by monitoring the learning curve. & If $\alpha$ is too large, the learning curve will show strong oscillations. If it is too small, the learning curve may stuck with a high value of the cost function.  \\ \hline
			
			Number of iterations, $n_{iter}$ & Tuned & The number of iterations is strictly related to the learning rate. It is chosen by monitoring the learning curve. & As the number of iterations increases, the model goes from underfitting to optimal and, then, to overfitting the training data. \\ \hline
			
			
		\end{tabular}
		\caption{Effects of hyperparameters on model performance.} 
		\label{effect_hyperparameters}
	\end{center}
\end{table}

An alternative to this issue is to use Akaike's information criterion (AIC) \citep{akaike} or the Bayesian information criterion (BIC) \citep{schwarz} as the model selection criteria. These methods try to balance the quality of the fit and model complexity and the main advantage is that there is no need for a validation set. The downside is that AIC and BIC impose a penalty for model complexity which is related to the number of parameters. This can be a problem when dealing with deep neural networks due to their large number of parameters. 

In our experiments, we noticed that a few of the hyperparameters employed in algorithm \ref{alg:1} need to be tuned for the best performance of the DNN. These are listed in algorithm \ref{alg:2}. The remaining hyperparameters are determined according to the following procedures in order to reduce the hyperparameter search space. For instance, hyperparameters related to the Adam optimization, such as the exponential decay rates $\beta_1$ and $\beta_2$, and the constant for numerical stability $\epsilon$, are set as described by \cite{Adam}. The learning rate $\alpha$ and the number of iterations $n_{iter}$ are chosen via manual search. Further details about this process can be found in \citet[p. 295]{Goodfellow-et-al}. As can be observed in the results section, we use the same values of the previous hyperparameters for all flow simulations studied in this work and these values should serve as references for other flows.
Following \citet{Goodfellow-et-al}, we provide table \ref{effect_hyperparameters} with information of the hyperparameters presented in algorithm \ref{alg:2} based on their effects on model performance. This table also presents a similar discussion for the hyperparameters chosen via manual search. We expect that it serves as a guideline for choosing the range of values to be explored for the hyperparameters presented in algorithm \ref{alg:2}.

Here, we consider that the activation function is also a hyperparameter since it plays a key role in the DNN performance. There are several available activation functions such as sigmoid, hyperbolic tangent (tanh), rectified linear unit (ReLU) \citep{RELU}, exponential linear unit (ELU) \citep{Clevert}, to name a few. For general regression problems, tanh and ELU are the most popular functions since they possess nonlinear properties and are continuously differentiable. \citet{Clevert} show that the ELU function reduces the vanishing gradient effect since its positive part returns the identity. Thus, in the positive part, the derivative is unitary and it is not contractive. On the other hand, tanh is contractive almost everywhere. Furthermore, in our experiments, the ELU function provided results with fewer iterations than a corresponding tanh network and, therefore,  we use ELU as the activation function.
\begin{algorithm}
	\SetKwInOut{Input}{Input}
	\SetKwInOut{Output}{Output}
	
	\Input{Number of candidate models $n_{models}$, minimum number of layers $n_{layers_{min}}$, maximum number of layers $n_{layers_{max}}$, number of inputs $n_{inputs}$ , minimum number of hidden units $n_{hidden_{min}}$, maximum number of hidden units  $n_{hidden_{max}}$, minimum order of magnitude of the regularization parameter $\lambda_{min}$, and maximum order of magnitude of the regularization parameter $\lambda_{max}$.}
	\Output{Network depth $L$, number of hidden units for each layer $n^{[l]}$ and regularization parameter $\lambda$}
	
	\For{$i = 1$ to $n_{models}$ } 
	{
		$\lambda_{i} = 10^{-rand(\lambda_{min},\lambda_{max})}$ \; 
		$L_{i} = int(rand(n_{layers_{min}},n_{layers_{max}}))$ ; \Comment{Network depth}
		$n^{[1]}_{i} = n_{inputs}$ ; $n^{[L]} = n_{inputs}$ ; \Comment{Number of units for the input and output layers}
		
		\For{$l = 2$ to $L_{i}-1$} 
		{
			$n^{[l]}_{i} = int(rand(n_{hidden_{min}},n_{hidden_{max}}))$ ; \Comment{Number of hidden units for layer l}
		}
		
	}
	
	\caption{Model generation procedure}
	\label{alg:2}
\end{algorithm}

\section{Results} 

In this section, the present method is first tested in the reconstruction of the dynamics of a damped cubic oscillator. Then, we evaluate the capability of the DNN-ROMs to reproduce the dynamics of a compressible flow past a cylinder including its noise generation. A comparison between the current DNN technique against sparse regression is also presented for the transient regime of an incompressible flow past a cylinder. Finally, the method is employed to create a ROM of a turbulent flow involving dynamic stall of a plunging airfoil. Again, the DNN solution is compared against that obtained by sparse regression. In each example, we show the ability and limitations of the methods to identify the dynamics of the different nonlinear systems comparing the ROM and full order model (FOM) solutions. 

\subsection{Nonlinear oscillator}

For this first example, we consider a canonical problem in system identification \citep{Brunton3932,Rudy2018}: the two-dimensional nonlinear oscillator. The system dynamics are given by
\begin{equation} \label{eq:20}
\frac{d}{dt} \begin{bmatrix}
x_1 \\
x_2 \\
\end{bmatrix}
=
\begin{bmatrix}
-0.1 & 2 \\
-2 & -0.1 \\
\end{bmatrix} 
\begin{bmatrix}
x_1^3 \\
x_2^3   \\
\end{bmatrix} \mbox{ .}\textbf{}
\end{equation}
with initial conditions $\mathopen{\begin{bmatrix}x_1 & x_2\end{bmatrix}}^T$  $=$ $\mathopen{\begin{bmatrix}2 & 0\end{bmatrix}}^T$.

\begin{figure}
	\centering
	\subfigure[State $x_1$.]{\includegraphics[width=.49\linewidth]{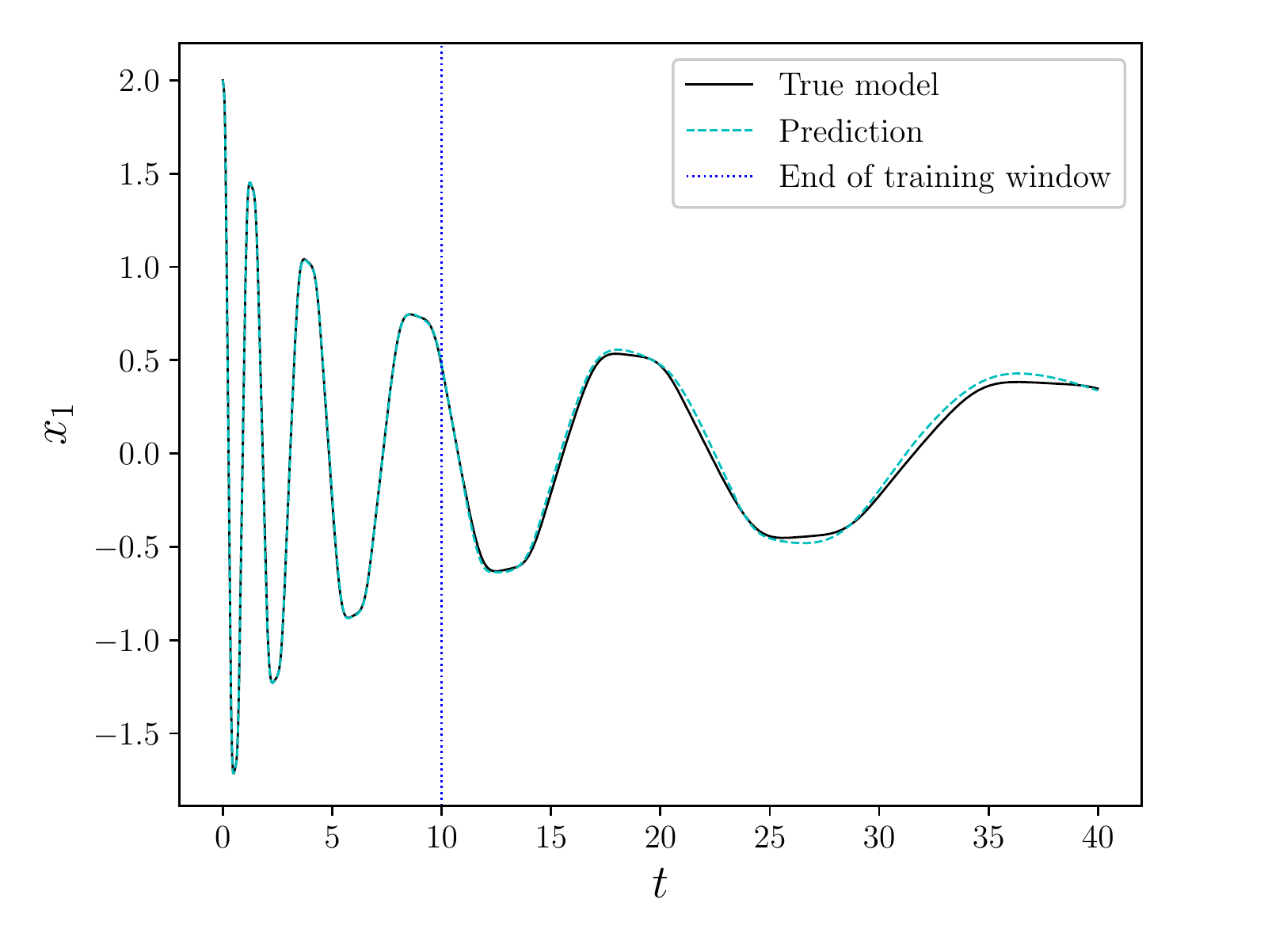}}
	\subfigure[State $x_2$.]{\includegraphics[width=.49\linewidth]{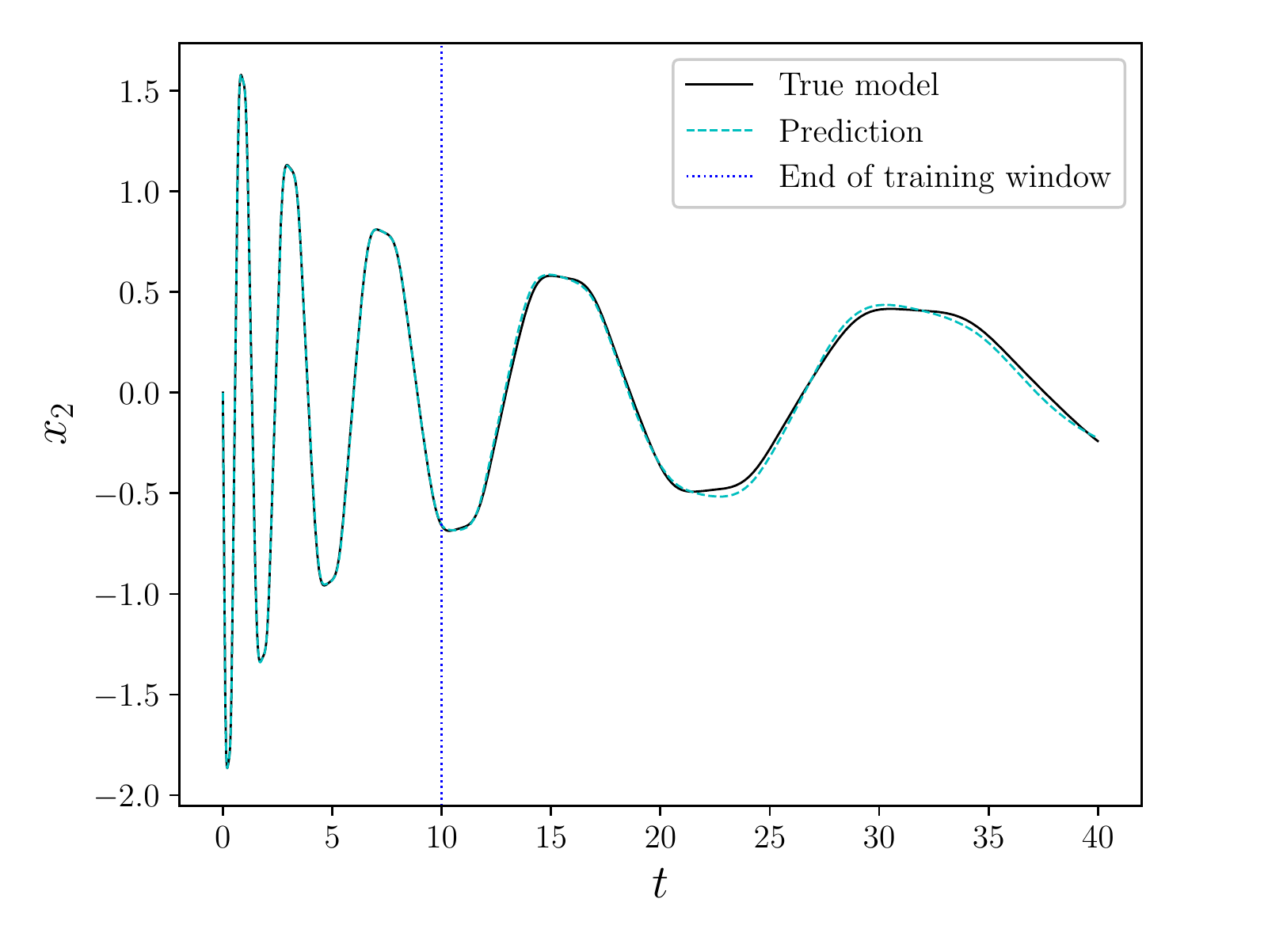}}
	\caption{Comparison between true model (solid black line) and ROM prediction (dashed cyan line) for nonlinear oscillator.}
	\label{fig:2}
\end{figure}

We generate 4000 snapshots from $t = 0$ to $t = 40$ by integrating equation \ref{eq:20} using an explicit 5 stage 4th-order Runge-Kutta scheme \citep{Kennedy1999} with a time step of $h = 0.01$. The training window spans the period $0 \leq t \leq 10$ and the remaining data is used as the test set. The system reconstruction is obtained following the procedures defined in section \ref{sec:formu}. For this first example, it is not necessary to use POD due to the low dimensionality of the system. The parameters employed in algorithm \ref{alg:2} are presented in table \ref{model_generation_oscillator} and the  best set of hyperparameters obtained via random search is listed in table \ref{hyperparamters_oscillator}. 
\begin{table}
	\begin{center}
		\begin{tabular}{ccccccc} 
			$n_{models}$ & $n_{layers_{min}}$ & $n_{layers_{max}}$ & $n_{hidden_{min}}$ & $n_{hidden_{max}}$ & $\lambda_{min}$ & $\lambda_{max}$ \\[3pt]  
			$300$ & $ 3$ & $10$ & $8$ & $36$ & $2$ & $6$\\
		\end{tabular}
		\caption{Model generation parameters for ROM of nonlinear oscillator} 
		\label{model_generation_oscillator}
	\end{center}
\end{table}
\begin{table}
	\begin{center}
		\begin{tabular}{cccccccc} 
			DNN architecture & $\sigma$ &$\alpha$ & $\lambda$ & $n_{iter}$ & $\beta_1$ & $\beta_2$ & $\epsilon$ \\[3pt] 
			$2-10-2$ & ELU & $0.001$ & $ 1.4062 \times 10^{-6}$ & $20000$ & $0.9$ & $0.999$ & $1.0 \times 10^{-8}$ \\
		\end{tabular} 
		\caption{Best set of hyperparameters for ROM of nonlinear oscillator}
		\label{hyperparamters_oscillator}
	\end{center}
\end{table}
Figure \ref{fig:2} presents the solutions obtained by the FOM (true model) and ROM for the damped cubic oscillator. Results show that the proposed algorithm accurately reproduces the system dynamics during the training window and beyond, for the test set.

\subsection{Construction of ROMs for fluid flow simulations} 
\label{sec:flow}

For the following investigations of the flow past a cylinder and the dynamic stall of a plunging SD7003 airfoil, the system dynamics are modeled by the compressible Navier-Stokes equations in 2D and 3D, respectively.
To simulate the airfoil undergoing a prescribed motion, the equations are solved in a non-inertial frame. In this form, source terms emerge from the grid curvature and frame movement. Here, all terms are solved in the full contravariant form to allow the use of a curvilinear coordinate system $\{\xi^1,\xi^2, \xi^3\}$. For a frame of reference with varying linear velocity, the equations reduce to
\begin{equation}
\frac{\partial}{\partial t}(\sqrt{g} \rho) + \frac{\partial}{\partial \xi^i} (\sqrt{g} \rho u^i) = 0
\mbox{ ,}
\label{eq:continuity_curv3}
\end{equation}
\begin{eqnarray}
\frac{\partial}{\partial t}(\sqrt{g} \rho u^i) +
\frac{\partial}{\partial \xi^j} \left[ \sqrt{g} \left( \rho u^i u^j - \tau^{ij} + g^{ij} p \right) \right] + \Christoffel{i}{jk} \sqrt{g} \Big( \rho u^k u^j + g^{jk} p - \tau^{kj} \Big) =   \sqrt{g} \rho \ddot{h}^i
\label{eq:momentum_curv3}
\mbox{ ,}
\end{eqnarray}
and
\begin{eqnarray}
\frac{\partial}{\partial t} (\sqrt{g} E) +
\frac{\partial }{\partial \xi^j} \left\{ \sqrt{g} \left[ (E+p) u^j  - \tau^{ij} g_{ik} u^k - \frac{ \mu }{Re \, Pr}  g^{ij} \frac{\partial T}{\partial \xi^i} \right] \right\} = \rho \sqrt{g} ( h^j + u^j ) g_{ji}\ddot{h}^i 
\mbox{ .}
\label{eq:energy_curv3}
\end{eqnarray}

The set of equations above represent the continuity, momentum and energy equations. In order to close the system of equations the following relations are employed
\begin{equation}
E = \frac{p}{\gamma - 1} + \frac{1}{2} \rho u^i g_{ij} u^j + \frac{1}{2} \rho \dot{h}^i  g_{ij} \dot{h}^i 
\mbox{ ,}
\end{equation}
\begin{equation}
\tau^{ij} = \frac{\mu}{Re} \left( g^{jk} u^i_{\ | k} + g^{ik} u^j_{\ |k} - \frac{2}{3} g^{ij} u^k_{\ |k} \right)
\mbox{ ,}
\end{equation}
and
\begin{equation}
h = h_o \sin(kt) \mbox{ .}
\end{equation}
Here, $\rho$ represents the density, $u^i$ the $i$-th component of the contravariant velocity vector and $p$ is the pressure. The term $h$ is the frame position (cross-stream motion of the plunging airfoil), $E$ is the total energy, $\mu$ is the dynamic viscosity and $T$ is the temperature. The dots represent temporal derivatives of the frame position, i.e., frame velocity and acceleration. The aforementioned terms are non-dimensionalized by freestream quantities such as density $\rho_{\infty}$ and speed of sound $c_{\infty}$. The length scales are made non-dimensional by the cylinder diameter or airfoil chord. In the above equations, $g_{ij}$ and $g^{ij}$ are the covariant and contravariant metric tensors, and $\Christoffel{i}{jk}$ represents the Christoffel symbols of the second kind. Further details about the present formulation can be found in \citet{Aris:1989}.

The numerical scheme for the spatial discretization of the equations is a high-resolution sixth-order accurate compact scheme \citep{Nagarajan2003} implemented on a staggered grid. A sixth-order compact interpolation scheme is also employed \citep{Lele1992} to obtain fluid properties on the different nodes of the staggered configuration.
Due to the non-dissipative characteristics of compact finite-difference schemes, numerical instabilities may arise from mesh non-uniformities, interpolations and boundary conditions and, hence, they have to be filtered to preserve stability of the numerical schemes. The high wavenumber compact filter presented by \citet{Lele1992} is applied in flow regions far away from solid boundaries at prescribed time intervals in order to control numerical instabilities.

The time integration of the fluid equations is carried out by the fully implicit second-order scheme of \citet{Beam1978} in the near-wall region in order to overcome the time step restriction due to the usual near-wall fine-grid numerical stiffness. A third-order Runge-Kutta scheme is used for time advancement of the equations in flow regions far away from solid boundaries. No-slip adiabatic wall boundary conditions are applied along the solid surfaces and characteristic plus sponge conditions are applied in the far field. For the study of dynamic stall, periodic boundary conditions are applied along the spanwise direction of the airfoil. In this case, we employ an implicit large eddy simulation to study the flow physics of deep dynamic stall over a plunging SD7003 airfoil profile, i.e., no subgrid scale model is used. The numerical tool has been previously validated for several simulations of unsteady compressible flows \citep{Wolf:2011,Wolf2012,Wolf:DU96,Brener:2019}.

\subsubsection{Flow past a cylinder}

In this case, the full order model (FOM) is obtained by solving the compressible Navier Stokes equations as detailed in section \ref{sec:flow}. The numerical simulations are conducted for Reynolds and Mach numbers $Re=150$ and $M= 0.4$, respectively. These non-dimensional parameters are computed based on freestream quantities.
The grid configuration consists of a body-fitted O-grid with 421 $\times$ 751 points in the streamwise and wall-normal directions, respectively.

The flow is recorded for 1120 snapshots with non-dimensional time steps of $h$ = 0.05. The snapshots are collected after an initial transient period of the simulation is discarded.
The reduced order model (ROM) is obtained following the procedure described in section \ref{sec:formu} using the pressure norm for the POD correlation matrix. It is important to mention that the use of a kinetic energy norm produced similar results for this case. The training data comprises the first 280 snapshots of the FOM and the remaining data is employed as the test set. The set of hyperparameters employed in algorithm \ref{alg:2} is listed in table \ref{model_generation_cylinder} and the best set of these parameters obtained via random search is presented in table \ref{hyperparamters_cylinder}. For this case, Bayesian optimization was also able to produce accurate models but at a higher computational cost compared to random search. One can see that the best architecture of the neural network for this case has 10 layers. However, we also found several other stable and accurate models with 6 layers, for example. The fact that the best model was found with 10 layers is a concidence and it occured because this particular model presented the lowest mean absolute error (MAE). The procedure to select the model could be improved if the MAE were computed from the error of the reconstructed flow as $||\boldsymbol{q}_{FOM} - \boldsymbol{q}_{ROM}||$. Although this procedure should be more expensive, it would be a better way to evaluate the models.
\begin{table}
	\begin{center}
		\begin{tabular}{ccccccc} 
			$n_{models}$ & $n_{layers_{min}}$ & $n_{layers_{max}}$ & $n_{hidden_{min}}$ & $n_{hidden_{max}}$ & $\lambda_{min}$ & $\lambda_{max}$ \\[3pt]
			$500$ & $ 6$ & $10$ & $10$ & $64$ & $2$ & $6$\\
		\end{tabular} 
		\caption{Model generation parameters for ROM of compressible flow past a cylinder}
		\label{model_generation_cylinder}
	\end{center}
\end{table}
\begin{table}
	\begin{center}
		\scalebox{0.92}{
			\begin{tabular}{cccccccc} 
				DNN architecture & $\sigma$ & $\alpha$ & $\lambda$ & $n_{iter}$ & $\beta_1$ & $\beta_2$ & $\epsilon$ \\[3pt] 
				$10-31-51-25-32-42-51-27-26-10$ & ELU & $0.001$ & $4.4085 \times 10^{-5}$ & $10000$ & $0.9$ & $0.999$ & $1.0 \times 10^{-8}$ \\
		\end{tabular}} 
		\caption{Best set of hyperparameters for ROM of compressible flow past a cylinder}
		\label{hyperparamters_cylinder}
	\end{center}
\end{table}

\begin{figure}
	\begin{center}
		{\includegraphics[trim = 1mm 1mm 1mm 1mm, clip, width=.75\textwidth]{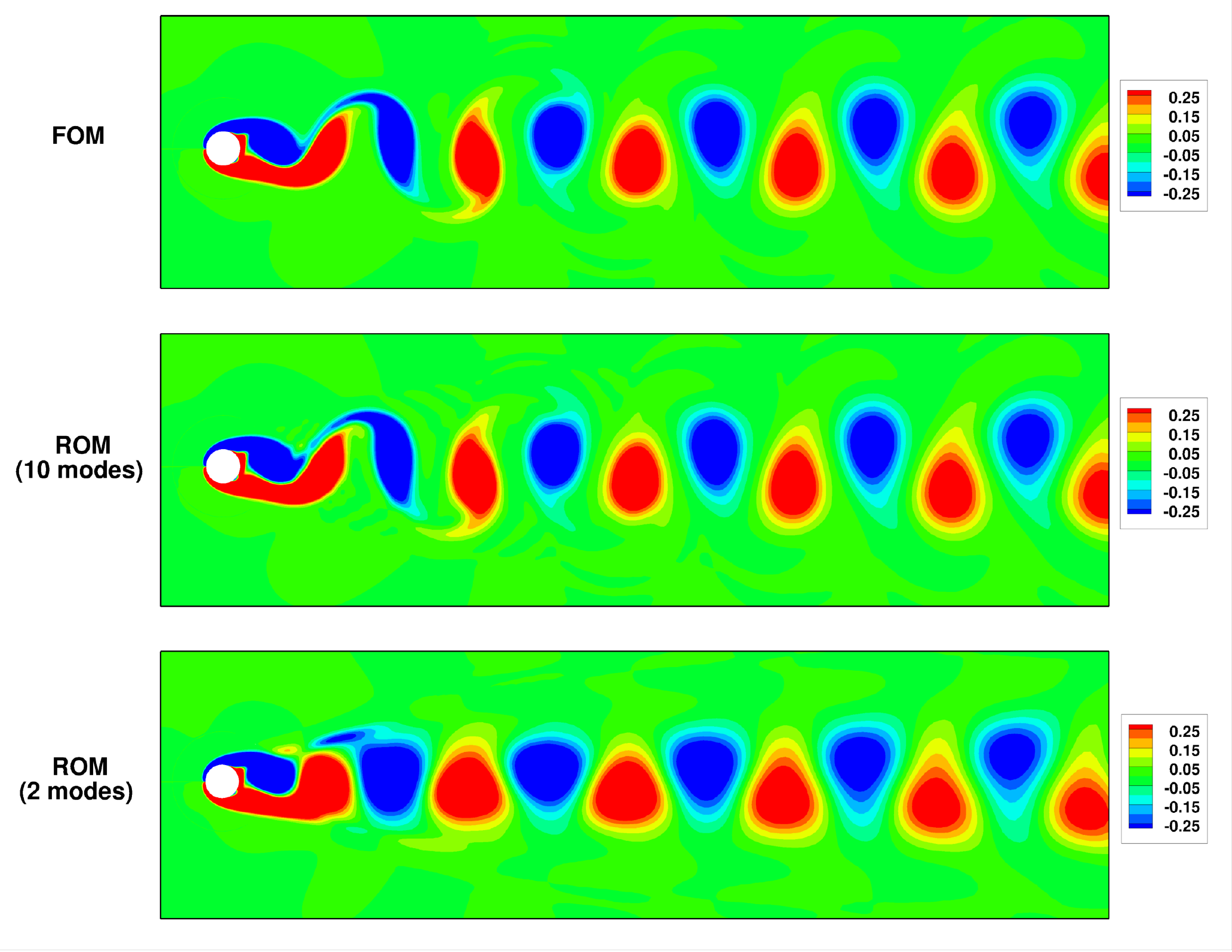}}
	\end{center}
	\caption{Contours of $\boldsymbol{z}$-vorticity, $t = 410$.}
	\label{fig:6}
\end{figure}

\begin{figure}
	\begin{center}
		{\includegraphics[trim = 1mm 1mm 1mm 1mm, clip, width=.85\textwidth]{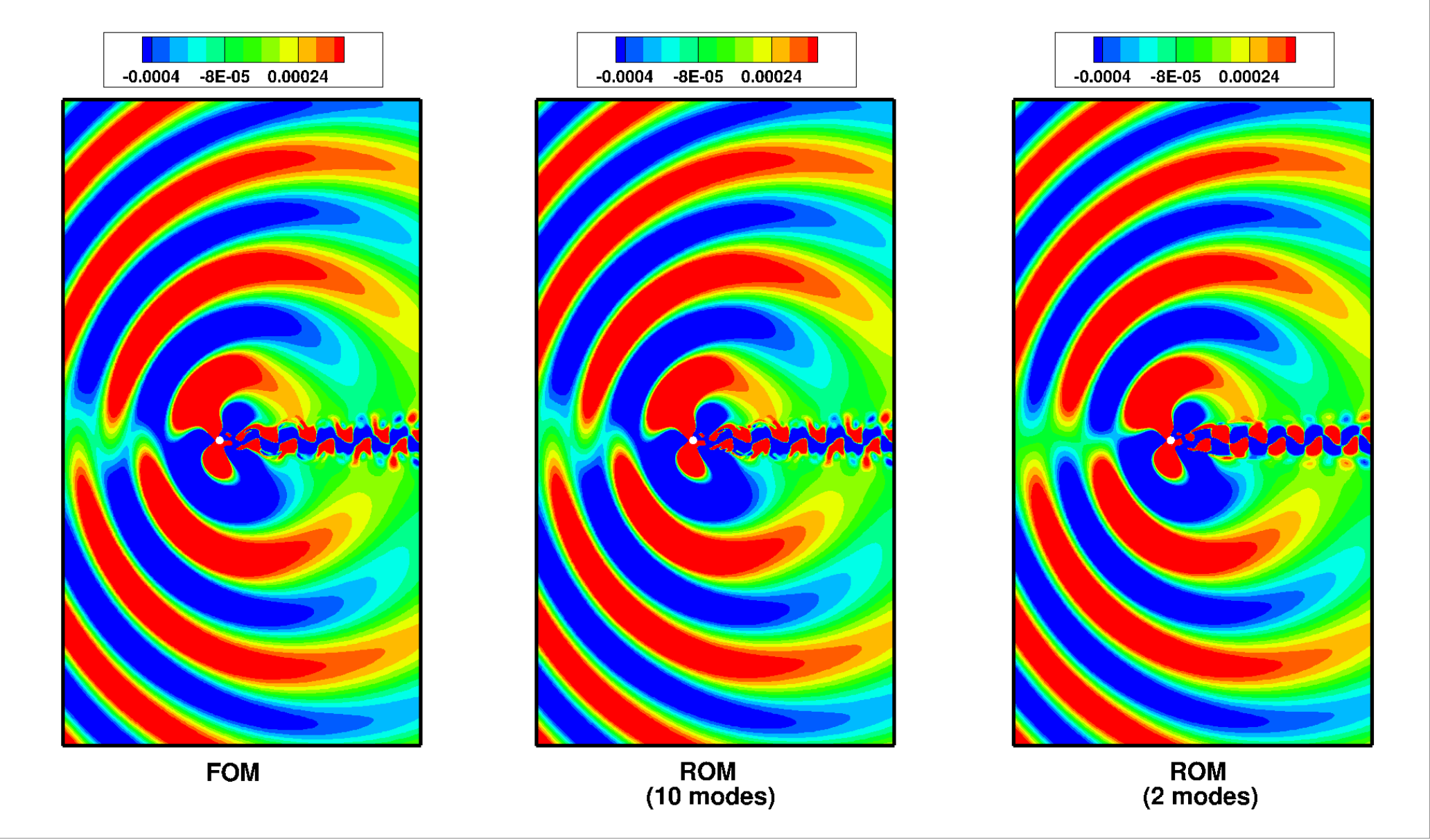}}
	\end{center}
	\caption{Contours of divergence of velocity, $t = 410$.}
	\label{fig:7}
\end{figure}

Figures \ref{fig:6} and \ref{fig:7} show contours of $z$-vorticity and divergence of velocity,  respectively, along the cylinder and wake regions at time $t = 410$, which is beyond the training window. The snapshots allow a comparison of the results between the FOM and ROM using 2 and 10 POD modes out of 280 modes. Hence, the flow is reconstructed using 0.7\% and 3.5\% of the total information available from the full order model. Although the current simulation is performed for a compressible flow at $M= 0.4$ and $Re=150$, we could verify that the POD spatial eigenfunctions are almost identical to those from \cite{Noack:2003}, which were obtained for an incompressible flow and $Re=100$. A video comparing the FOM and ROM solutions is provided together with the manuscript. Reconstruction of the individual flow variables with 2 POD modes could recover between 50 and 80\% of the total modal energy, depending on the variable. For example, density is reconstructed using 50\% of the total energy of the system dynamics while $y$-momentum is reconstructed with 80\% of the total modal energy. Here, we use the term ``modal energy'' to refer to the ratio between the sum of $N$ POD eigenvalues used in a particular reconstruction over the entire range $N_T$ of eigenvalues available, $\sum_{i=1}^{N} \lambda_{i}/\sum_{i=1}^{N_T} \lambda_{i}$. For 10 POD modes, the reconstructions could recover 99\% of the energy for all variables and, therefore, should lead to an accurate flow representation. For this particular case, the spectral proper orthogonal decomposition technique is not required since the modes are almost monochromatic and, hence, do not require filtering.

One can observe from the figures that the computations of the flow using the ROM framework show good agreement compared to those obtained by the FOM. For the current Reynolds number, the flow develops a typical von K\'arm\'an vortex street along the cylinder wake. The periodical pattern of the vortex shedding can be observed in the contours of $z$-vorticity. Noise generation also occurs in the current unsteady compressible flow simulation. In this case, pressure fluctuations along the cylinder surface are scattered to the far-field and can be observed in the contours of dilatation shown in figure~\ref{fig:7}. Both the near-field hydrodynamics and the far-field acoustics are recovered by the ROM. The reconstruction using 10 POD modes show an excellent agreement with the FOM. When 2 POD modes are employed in the flow reconstruction, discrepancies between the ROM and FOM are evident from the figures. However, the main features of the flow are still recovered by the model. One should remind that, despite the use of only 2 modes, the dynamical system is still stable beyond the training region.

\begin{figure}
	\centering
	\subfigure[Probe positioned just behind the cylinder.]{\includegraphics[width=.49\linewidth]{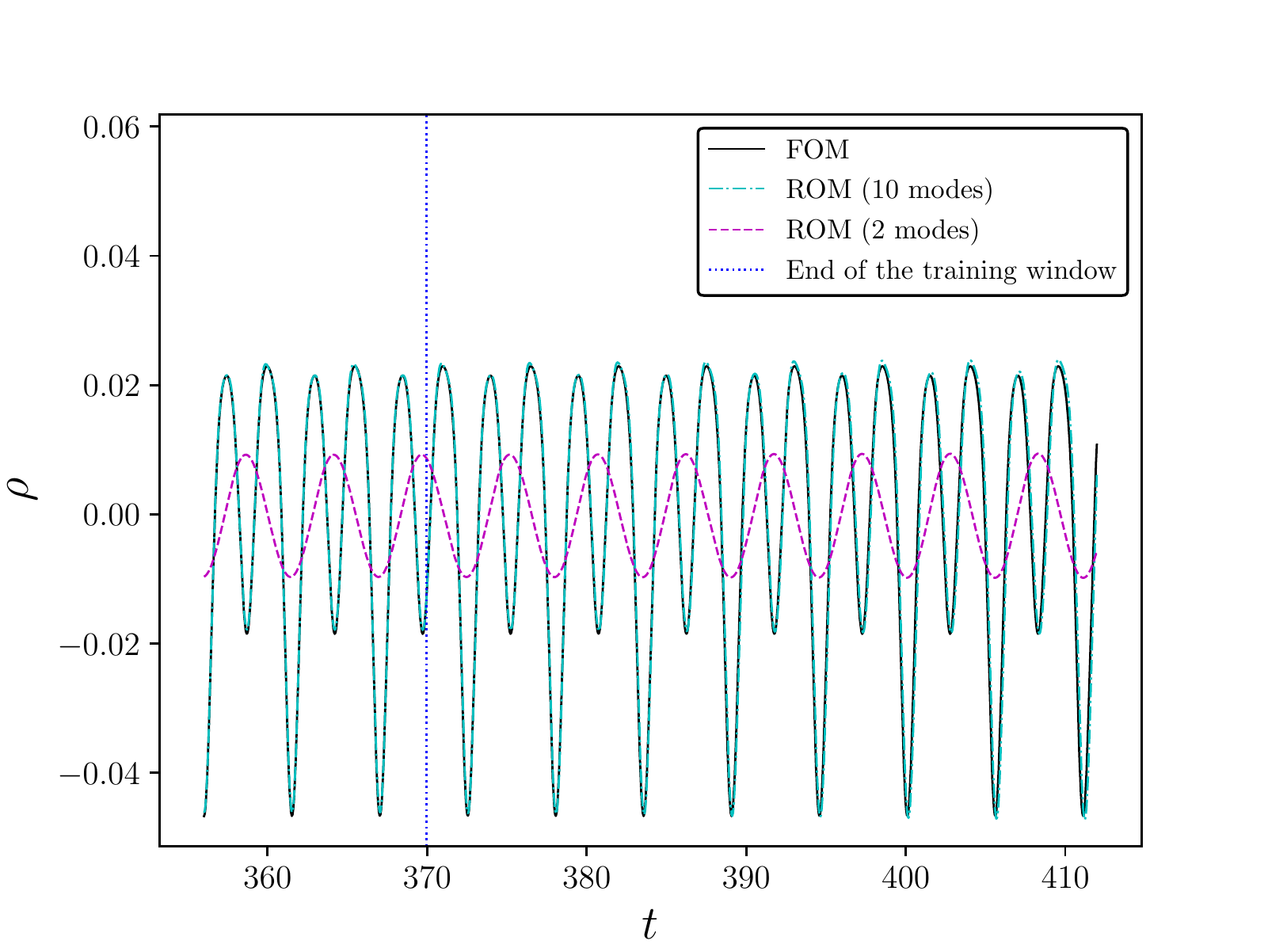}}
	\subfigure[Probe away from the cylinder, along the wake.]{\includegraphics[width=.49\linewidth]{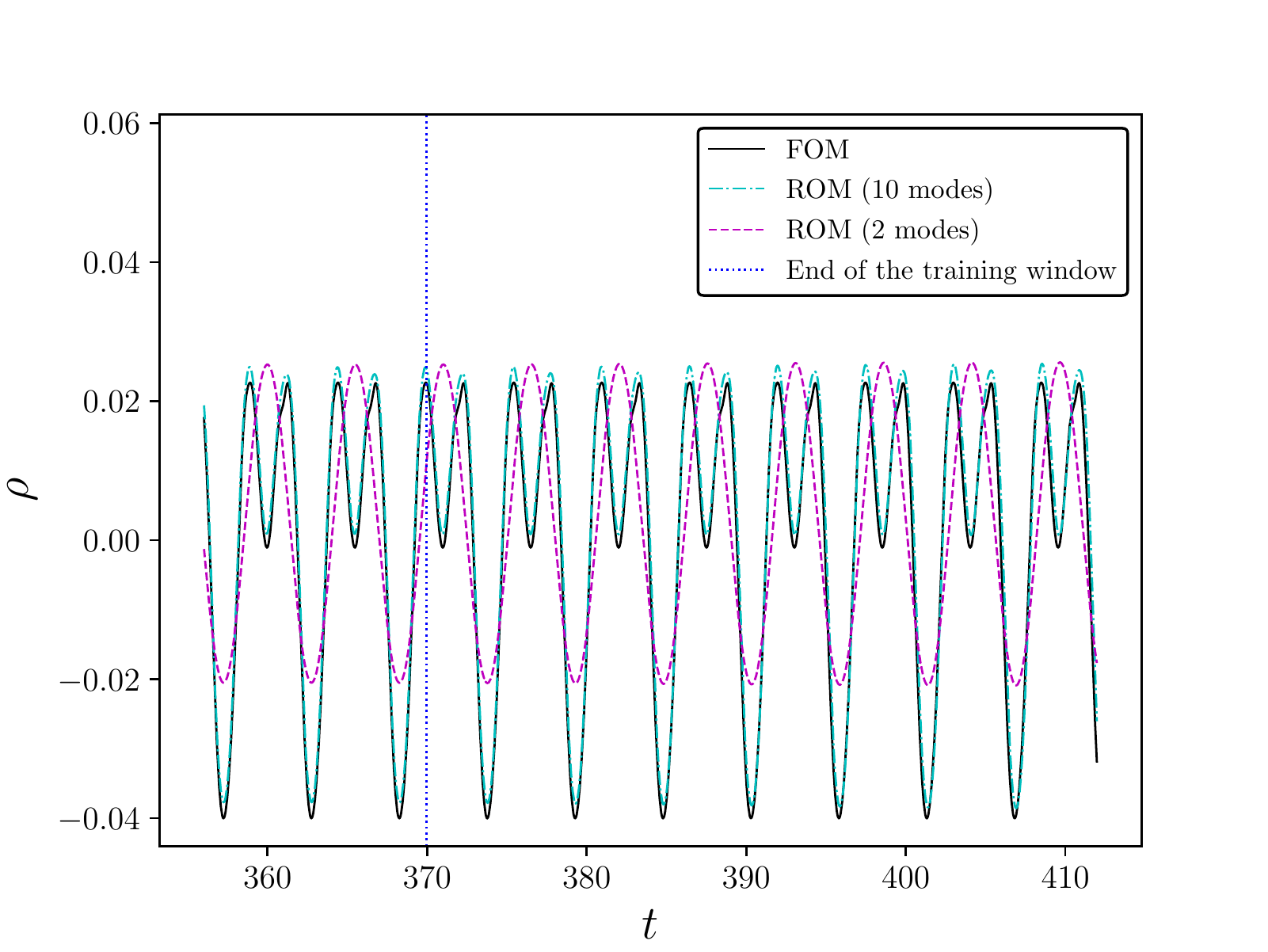}}
	\caption{Fluctuation time history of density.}
	\label{fig:8}
\end{figure}

\begin{figure}
	\centering
	\subfigure[Probe positioned just behind the cylinder.]{\includegraphics[width=.49\linewidth]{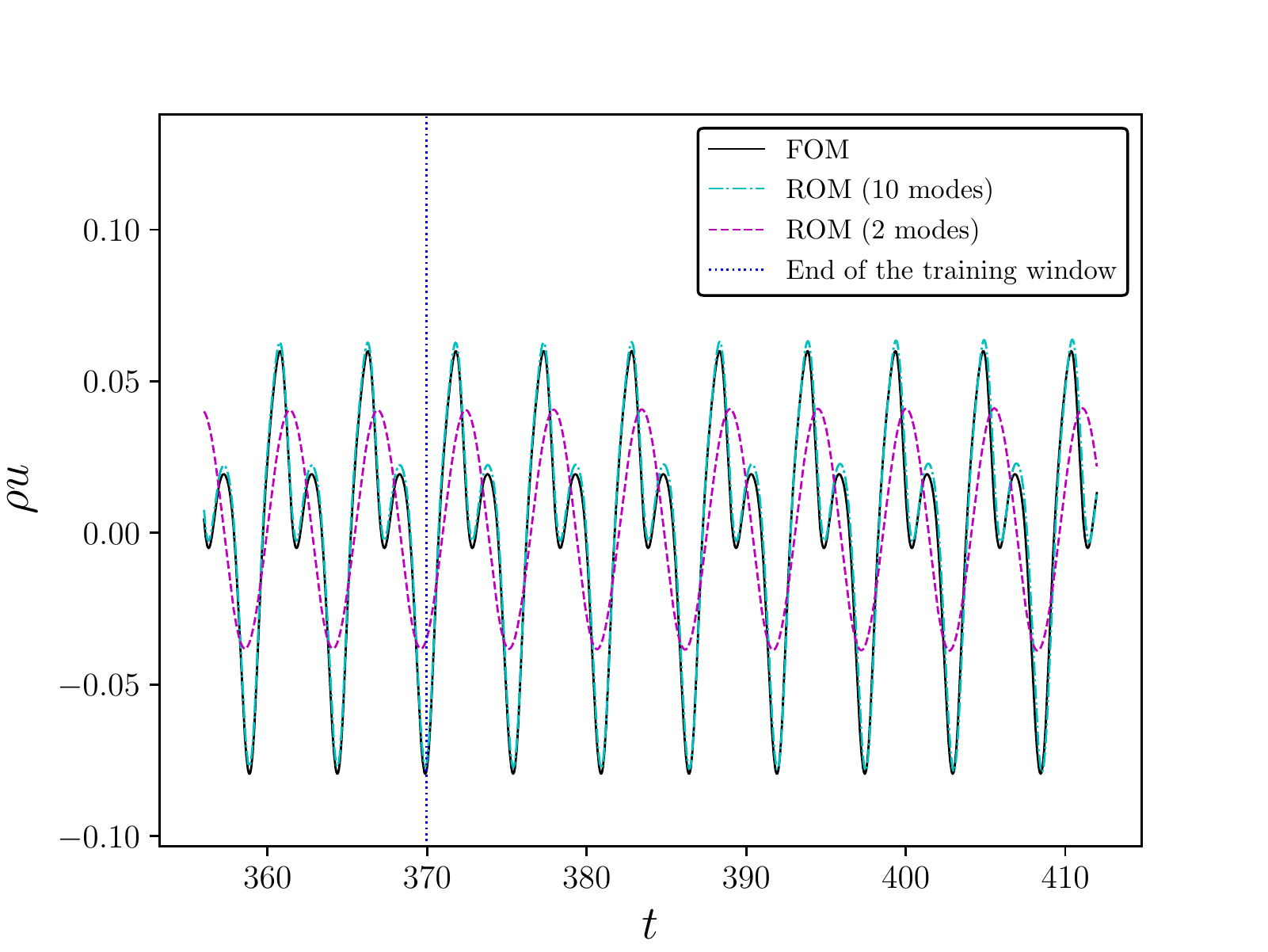}}
	\subfigure[Probe away from the cylinder, along the wake.]{\includegraphics[width=.49\linewidth]{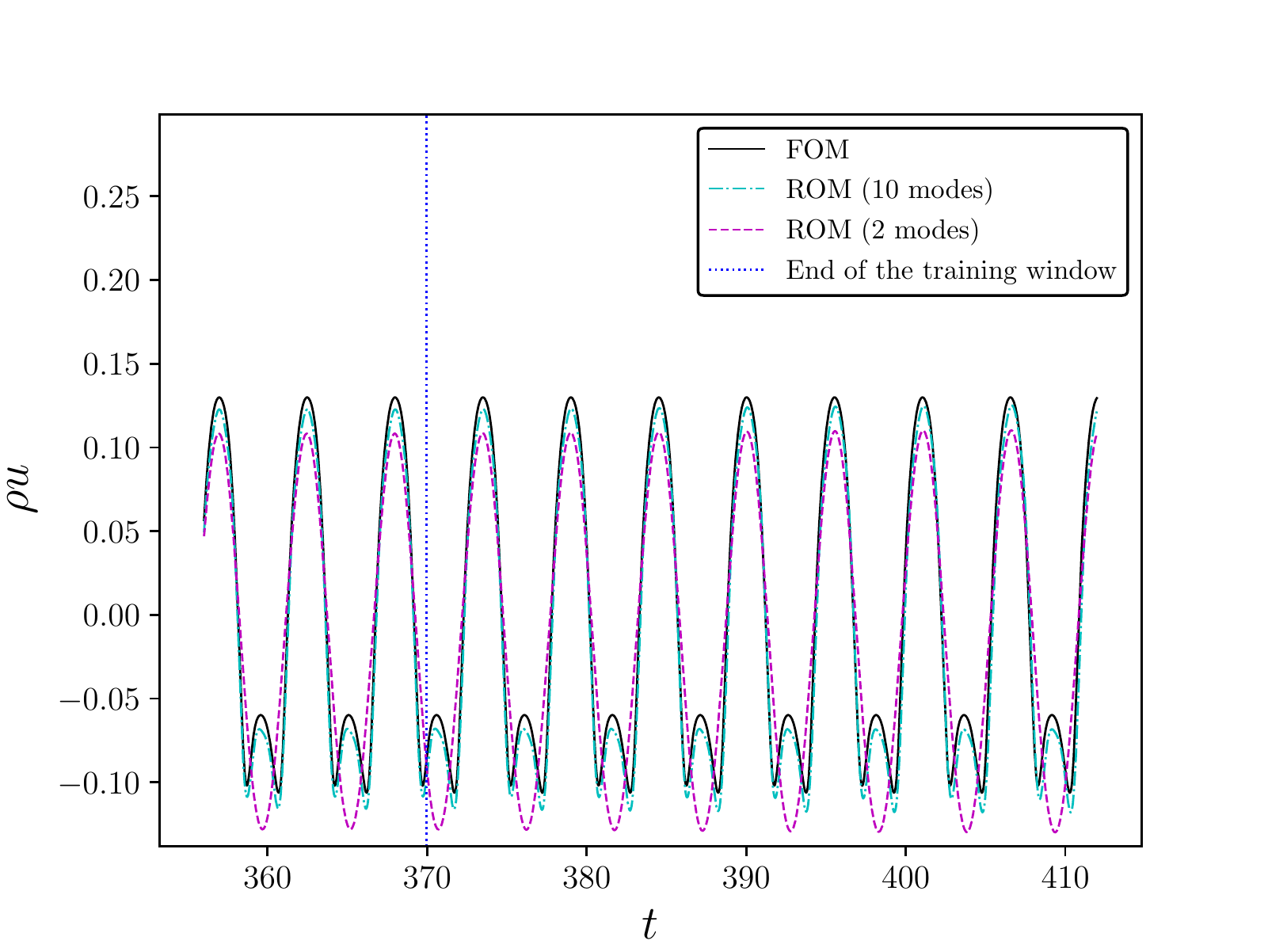}}
	\caption{Fluctuation time history of $x$-momentum.}
	\label{fig:9}
\end{figure}

In order to show a more qualitative evaluation of the model reconstructions, the density and $x$-momentum fluctuation time histories are presented for the FOM and ROMs in figures  \ref{fig:8} and \ref{fig:9}, respectively. The figures on the left column show results for a probe located just behind the cylinder, close to the surface, at $(x,y) = (0.55,-0.06)$. The cylinder has radius 0.5 and its center is positioned in the origin of the Cartesian system. On the right column, results are obtained for a probe downstream the cylinder wake, at $(x,y)=(1.1,-0.06)$. Results are shown for both the training window period and beyond. When 2 POD modes are employed, the solutions show a less accurate representation of the dynamics observed in the FOM. The density reconstruction is that with the highest discrepancy and that is attributed to the lower energetic content achieved by the first 2 POD modes. One can notice that the reduced order model accurately reproduces the full order model results during and beyond the training window when 10 POD modes are employed in the reconstruction.

\begin{figure}
	\centering
	\subfigure[Mode 1]{\includegraphics[trim = 0mm 0mm 10mm 0mm, clip,width=.49\linewidth]{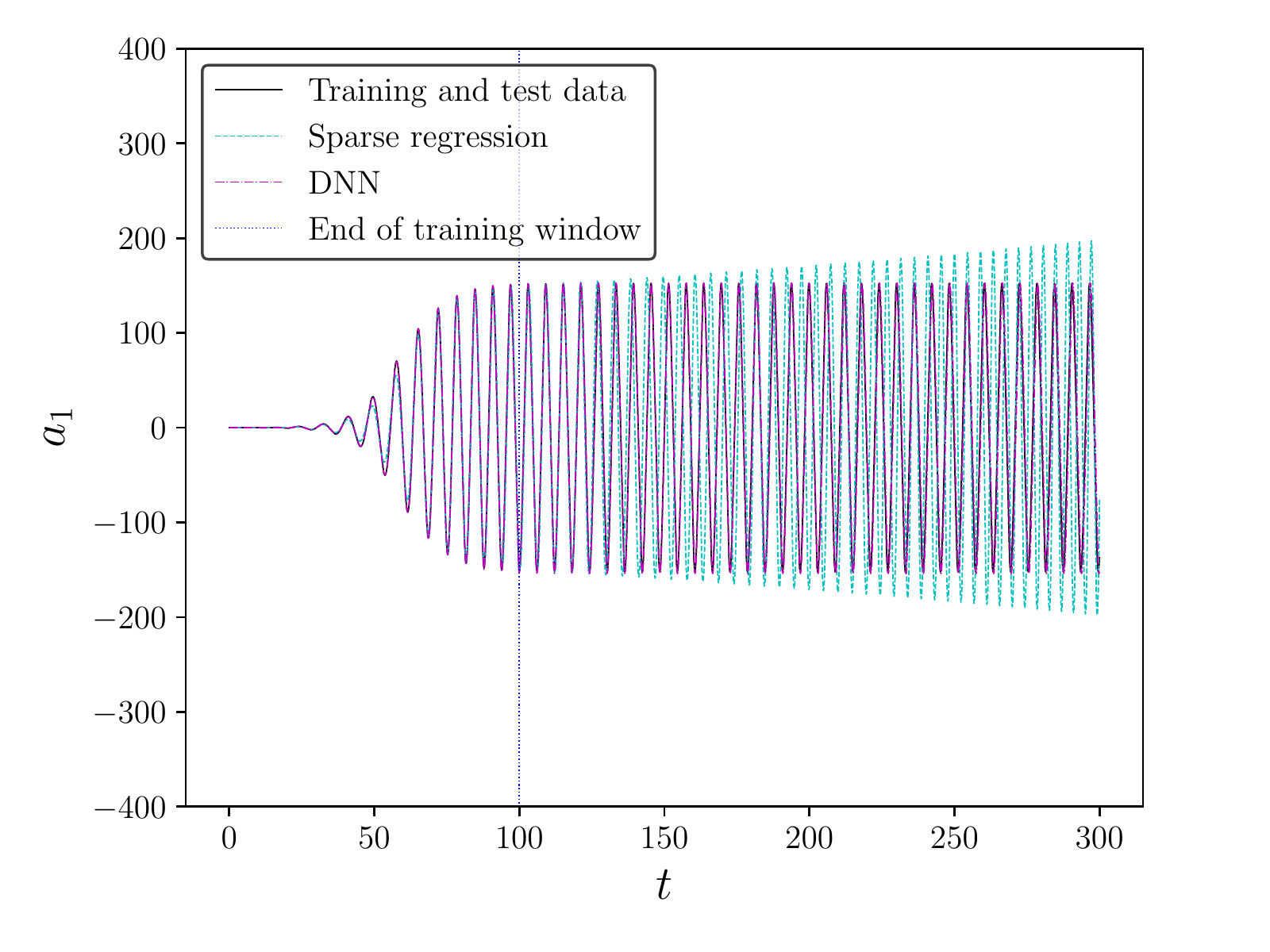}}
	\subfigure[Mode 2]{\includegraphics[trim = 0mm 0mm 10mm 0mm, clip,width=.49\linewidth]{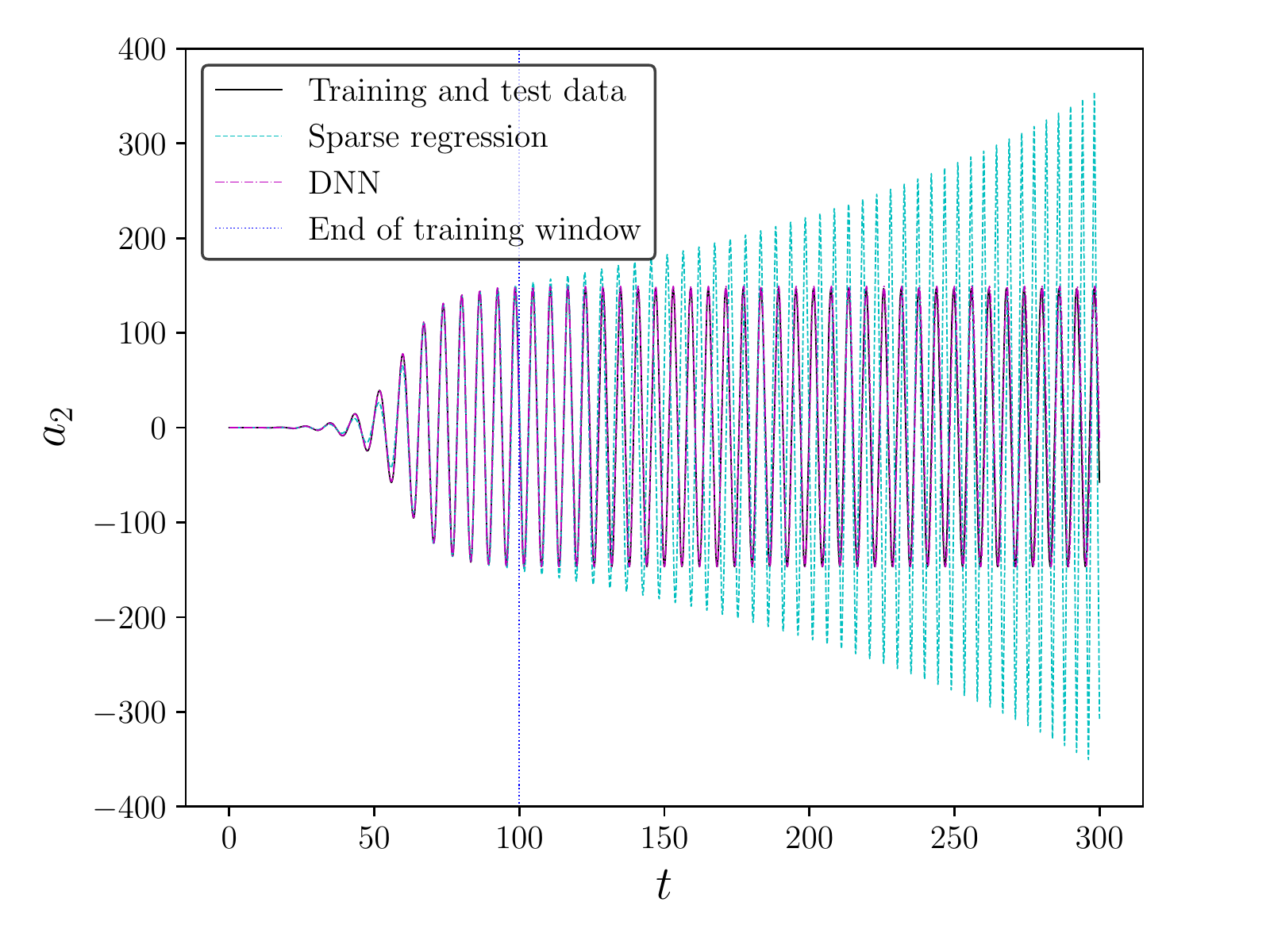}}
	\caption{Reconstruction of POD temporal modes for transient solution of incompressible flow past a cylinder.}
	\label{fig:transient}
\end{figure}

In order to test the robustness of the method, we employ the DNN approach for the reconstruction of the transient regime of an incompressible cylinder flow. For this study, the 2 most energetic POD modes containing both the transient and limit cycle dynamics of the flow are obtained from \cite{Brunton3932}. These modes contain both the transient and limit cycle dynamics of the flow. Figure \ref{fig:transient} shows the POD temporal modes used for training and testing the DNNs. The first 5000 temporal instants are used to train the models and one can see this dataset represented by the black line to the left of the vertical line marking the end of the training window. To the right of the training window, the black line represents the test data which is the correct solution for the temporal dynamics. Figure \ref{fig:transient} also shows the results obtained by the current DNN approach and by sparse regression. In the case of sparse regression, we employ the same model obtained by \cite{Brunton3932} in table 11 of the Supporting Information report. As one can observe, the DNN is able to accurately recover both the transient and limit cycle solutions of the test data for both POD modes. On the other hand, sparse regression reconstructs the transient portion of the dynamics but not the long term prediction of the limit cycle. Several models obtained by the DNNs presented similar results compared to those shown in figure \ref{fig:transient} and some had similar neural network architectures compared to that of table \ref{hyperparamters_cylinder}. These results show that the proposed DNN approach has good long-term predictive capabilities and can learn transient features of the flow.

\subsubsection{Deep dynamic stall of plunging airfoil}

The present study concerns a plunging SD7003 airfoil in deep dynamic stall. The flow conditions have a reduced frequency $k= \pi f L / U_{\infty} = 0.5$, where $f$ is the plunging frequency, $L$ is the chord of the airfoil and $U_{\infty}$ is the reference free stream velocity. The motion amplitude is set as $h_o/L = 0.5 $ with a static angle of attack $\alpha_0= 8$ deg. The chord Reynolds number based on the freestream velocity is $Re = 60,000$ and the freestream Mach number is $M=0.1$. This flow condition is relevant for micro air vehicle applications and this case is selected based on the availability of published results from other high fidelity simulations \citep{Visbal:2011} and particle image velocimetry data \citep{Baik:2009,Ol:2009}. In a previous work \citep{Brener:2019}, we performed a mesh refinement study and validation of the numerical solutions against the references above.

The present mesh configuration consists of a body-fitted O-grid with $441 \times 300 \times 64$ points in the streamwise, wall-normal and spanwise directions, respectively. The grid is generated with $70\%$ of the surface points located in the suction side of the airfoil to improve the capturing of finer flow scales developing along the turbulent region of the flow. Due to the favorable pressure gradients in the pressure side of the airfoil, the flow remains laminar along the entire cycle of the plunging motion. The trailing edge of the SD7003 airfoil is rounded in the current simulations with an arc of radius $r/L = 0.0008 $. This procedure is required for retaining the smoothness of the metric terms computed by the high order compact scheme. The spanwise domain is set as $z/c=0.4$ similarly to \citet{Visbal:2011} and the nondimensional time step of the simulation is set as $\Delta t^{*} = \dfrac{\Delta t U_{\infty}}{L}=0.00008$.

The plunge motion undergoes with an effective angle of attack in the range of $-6$ deg. $\leq \alpha \leq 22$ deg. Defining $\psi$ as the angular position in the plunging cycle, we say that at $\psi=0$ deg. the airfoil has no velocity in the $y$-direction and is at the top-most position of the plunging motion. At $\psi=90$ deg. it has the highest velocity in the $y$-direction downwards and, at $\psi=180$ deg., it has no velocity and is at the bottom-most position of the plunging motion. Finally, at $\psi=270$ deg. it has the highest velocity in the $y$-direction upwards. In summary, during the down-stroke, instabilities begin to form in the suction side of the airfoil, growing and eventually breaking into finer structures. While this takes place, the dynamic stall vortex forms along the leading edge and is transported through the airfoil suction side increasing the overall lift and creating a nose-down pitching moment. As the leading-edge vortex (LEV) approximates the trailing edge, a trailing-edge vortex (TEV) forms pushing the LEV away from the airfoil. A more complete discussion of the flow dynamics can be found in \citet{Visbal:2011} and \citet{Brener:2019}.

\begin{figure}
	\centering
	\subfigure[Flow at $\psi = 100$ deg. (FOM)]{\includegraphics[trim = 0mm 50mm 0mm 60mm, clip,width=.49\linewidth]{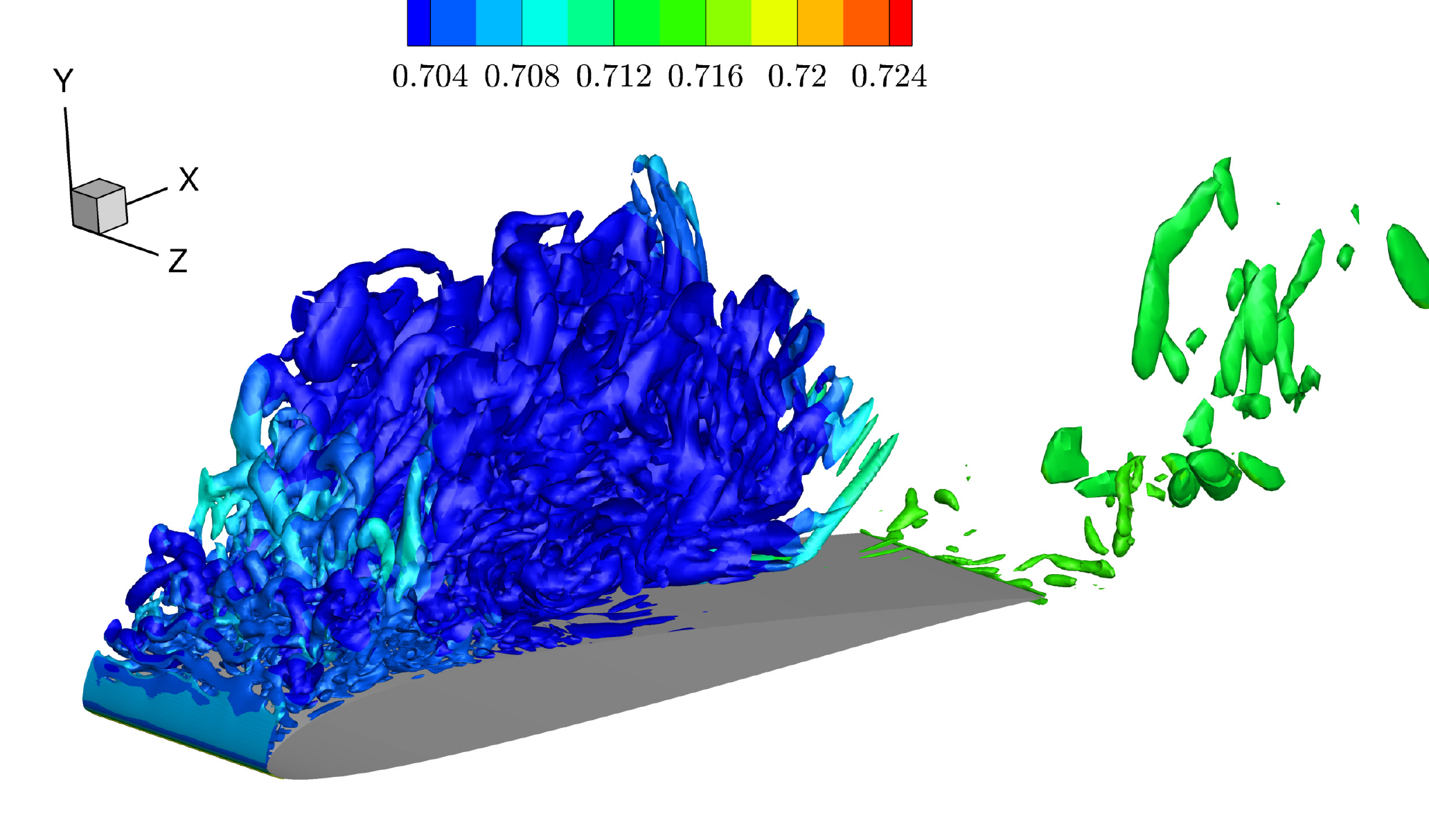}}
	\subfigure[Flow at $\psi = 100$ deg. (ROM)]{\includegraphics[trim = 0mm 50mm 0mm 60mm, clip,width=.49\linewidth]{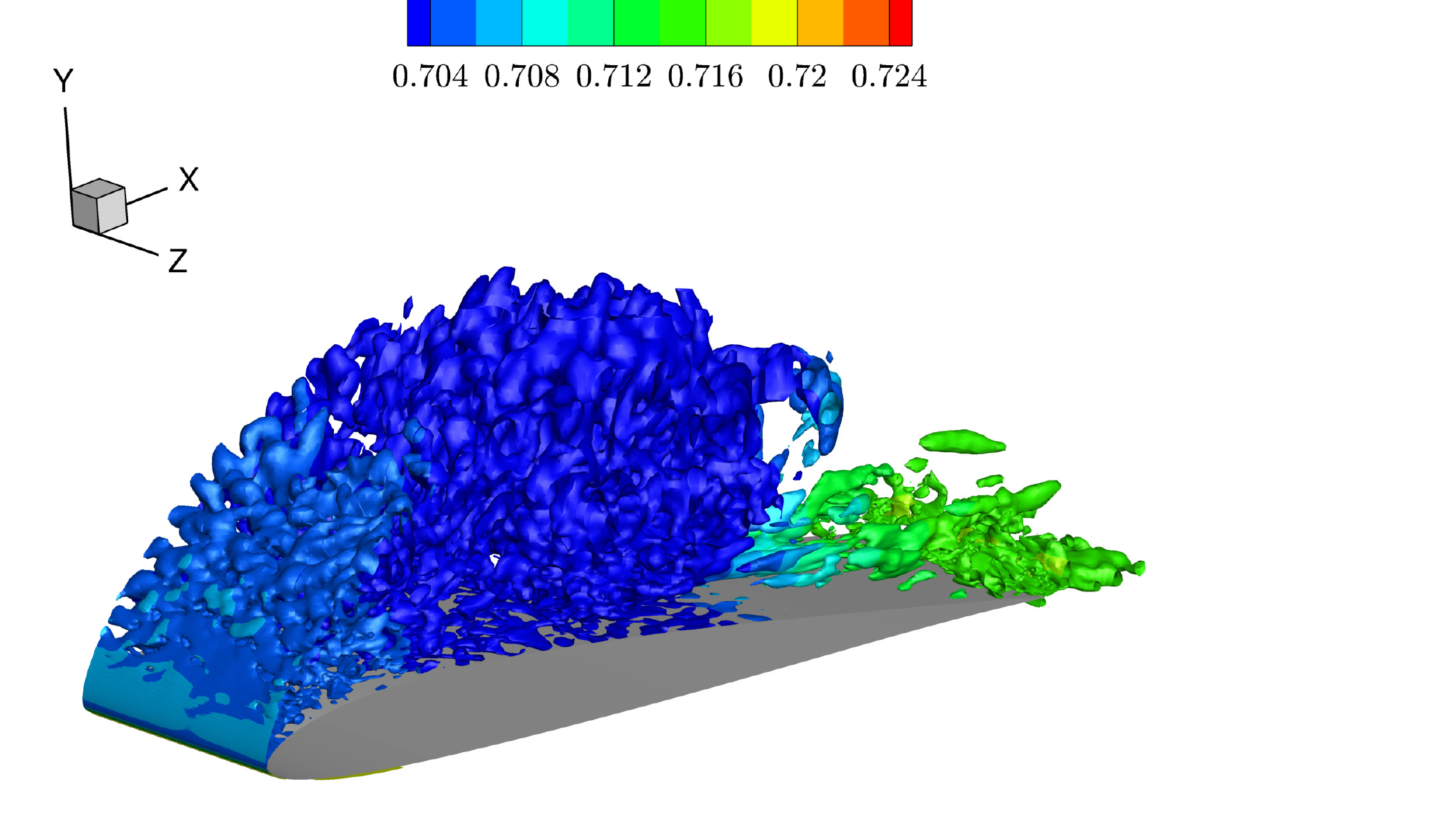}}
	\subfigure[Flow at $\psi = 160$ deg. (FOM)]{\includegraphics[trim = 0mm 0mm 0mm 60mm, clip,width=.49\linewidth]{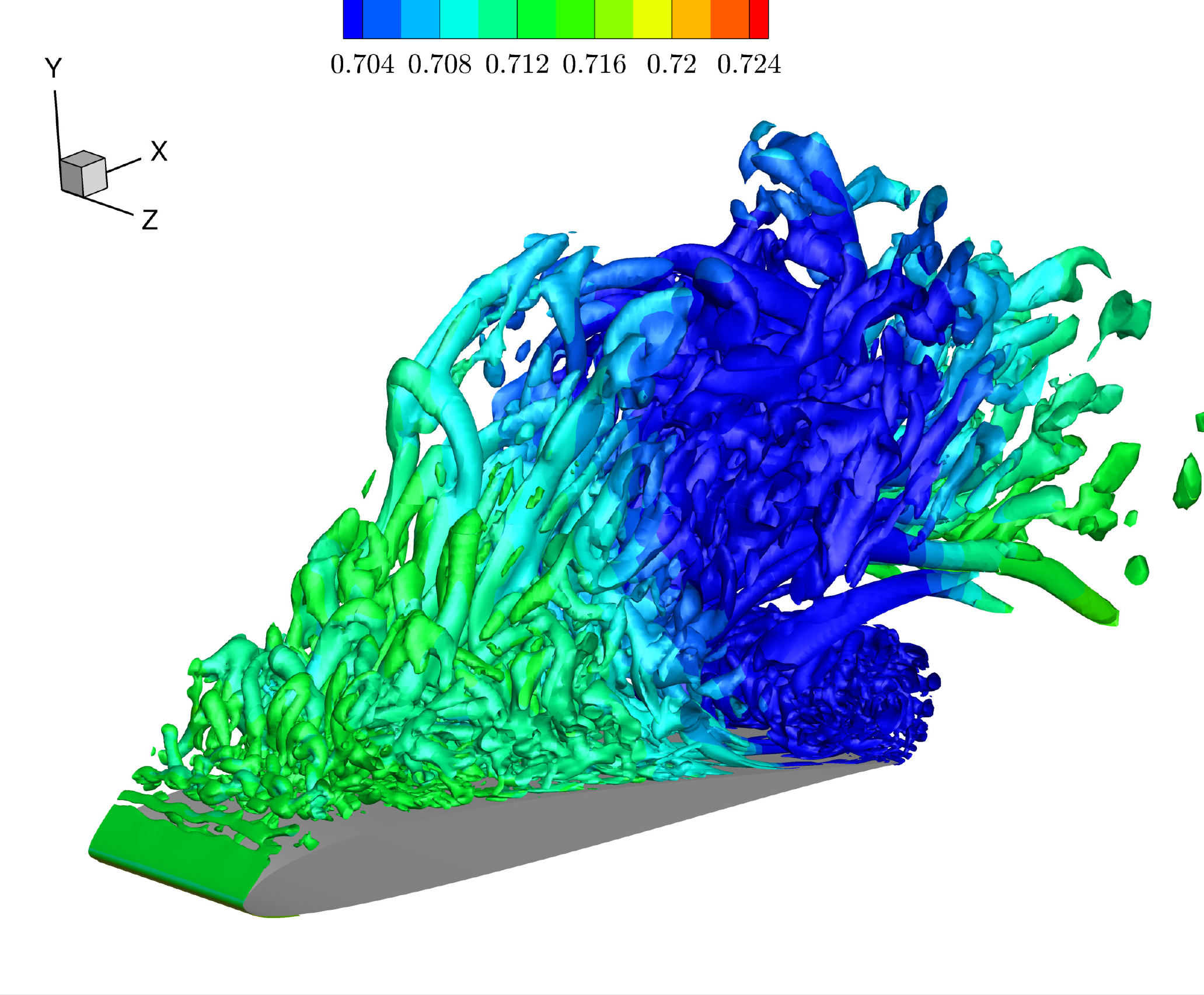}}
	\subfigure[Flow at $\psi = 160$ deg. (ROM)]{\includegraphics[trim = 0mm 0mm 0mm 60mm, clip,width=.49\linewidth]{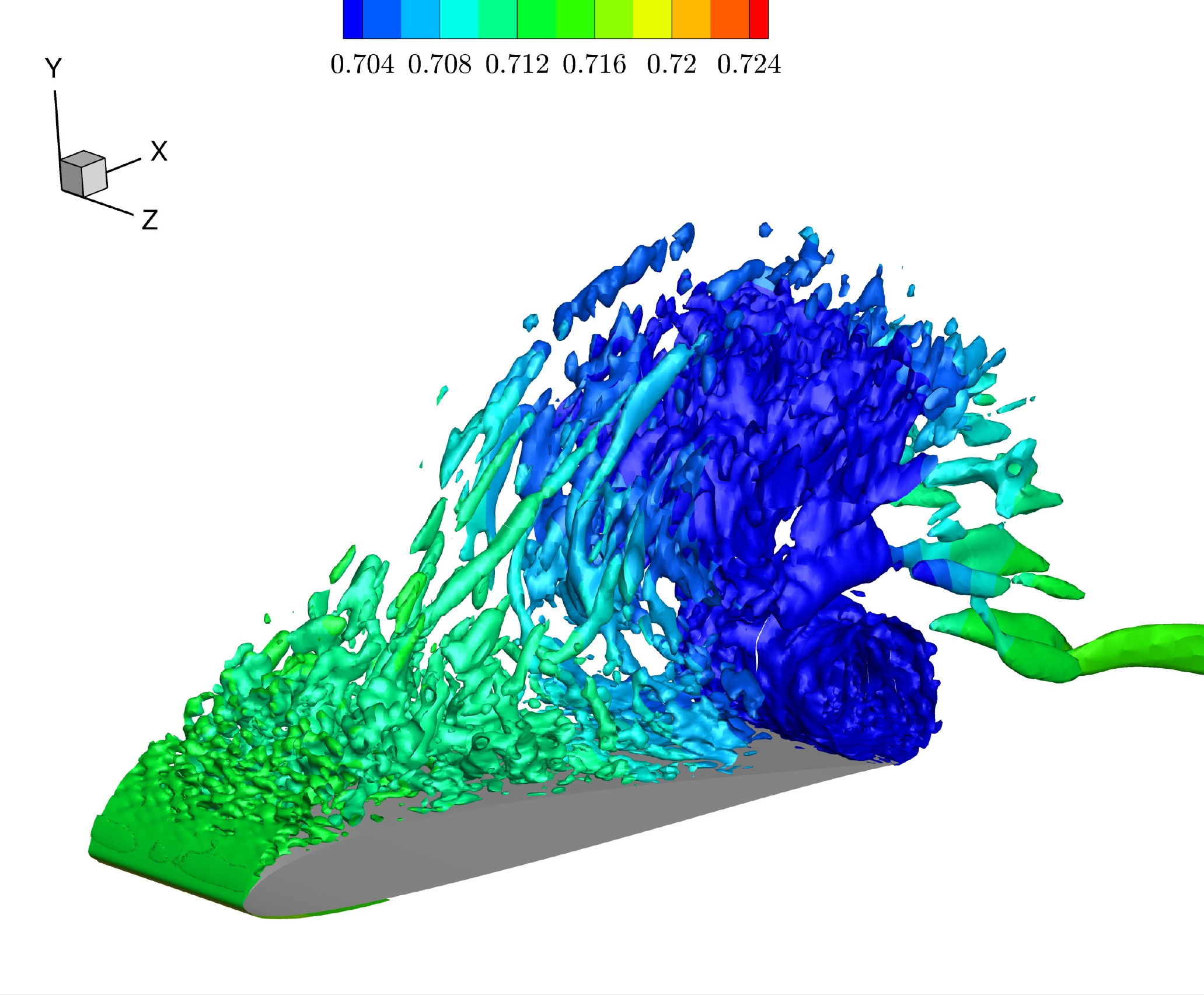}}
	\caption{Iso-surfaces of Q criterion colored by pressure at different instants of the plunge motion for the fourth cycle. The ROM was trained using only the first 2 cycles.}
	\label{fig:3D}
\end{figure}

Figure \ref{fig:3D} shows iso-surfaces of Q-criterion colored by pressure and it is possible to observe the main flow features described above. In figures \ref{fig:3D}(a) and (b), it is possible to compare the 3D solutions of the FOM and ROM, respectively, for the leading edge vortex formation. Figures \ref{fig:3D}(c) and (d) show a similar comparison  for the instant where the trailing edge vortex forms. For both instants, one can observe that the ROM is able to reconstruct the larger scale features of the 3D flow. A video comparing the FOM and ROM solutions for this flow configuration is provided together with the manuscript. The ROM is trained using the first 2 cycles of the plunging motion and the solutions presented in figure \ref{fig:3D} are computed for the fourth cycle, showing that the model is able to reproduce the 3D flow dynamics beyond the training window. For this study, the parameters employed in algorithm \ref{alg:2} are presented in table \ref{model_generation_dynamic_stall} and the best set of hyperparameters, obtained via random search, is listed in table \ref{hyperparamters_dynamic_stall_3D}. It is important to mention that, for this case, Bayesian optimization was unable to produce stable and accurate models.
\begin{table}
	\begin{center}
		\begin{tabular}{ccccccc} 
			$n_{models}$ & $n_{layers_{min}}$ & $n_{layers_{max}}$ & $n_{hidden_{min}}$ & $n_{hidden_{max}}$ & $\lambda_{min}$ & $\lambda_{max}$ \\[3pt]
			$700$ & $ 6$ & $12$ & $10$ & $64$ & $2$ & $6$\\
		\end{tabular} 
		\caption{Model generation parameters for ROM of plunging airfoil in deep dynamic stall}
		\label{model_generation_dynamic_stall}
	\end{center}
\end{table}
\begin{table}
	\begin{center}
		\begin{tabular}{cccccccc} 
			DNN architecture & $\sigma$ & $\alpha$ & $\lambda$ & $n_{iter}$ & $\beta_1$ & $\beta_2$ & $\epsilon$ \\[3pt]
			$16-19-52-50-31-16$ & ELU & $0.001$ & $8.0938 \times 10^{-5}$ & $10000$ & $0.9$ & $0.999$ & $1.0 \times 10^{-8}$ \\
		\end{tabular} 
		\caption{Best set of hyperparameters for ROM of plunging airfoil in deep dynamic stall (3D flow)}
		\label{hyperparamters_dynamic_stall_3D}
	\end{center}
\end{table}

The first 16 SPOD modes are employed to reduce the dimensionality of the input data in the 3D flow reconstruction. Other POD reconstructions were tested with a different number of modes and it was observed that the first 16 modes contained the main features of the leading and trailing edge vortex formation. For example, adding 10 more modes did not improve significantly the solution and, beyond mode 26, the SPOD temporal modes presented  complex behavior, being composed of several frequencies and difficulting the training stage of the DNNs. This issue could be improved running the simulation for a longer period with a lower time step to improve convergence of the POD modes. One should remind that the spectral proper orthogonal decomposition allows an energy shift of the modes to obtain frequency filtered temporal dynamics. In this sense, the high-frequency noise observed in the POD temporal modes is not discarded but is shifted to higher SPOD modes. This procedure allows a better pairing of POD modes if the coherent structures have periodicity. In the present study, due to the unsteady boundary conditions and the lack of symmetry in the flow, we do not expect pairing of the SPOD modes.
\begin{figure}
	\centering
	\subfigure[POD temporal mode 2.]{\includegraphics[trim = 1mm 1mm 1mm 1mm, clip,width=.49\linewidth]{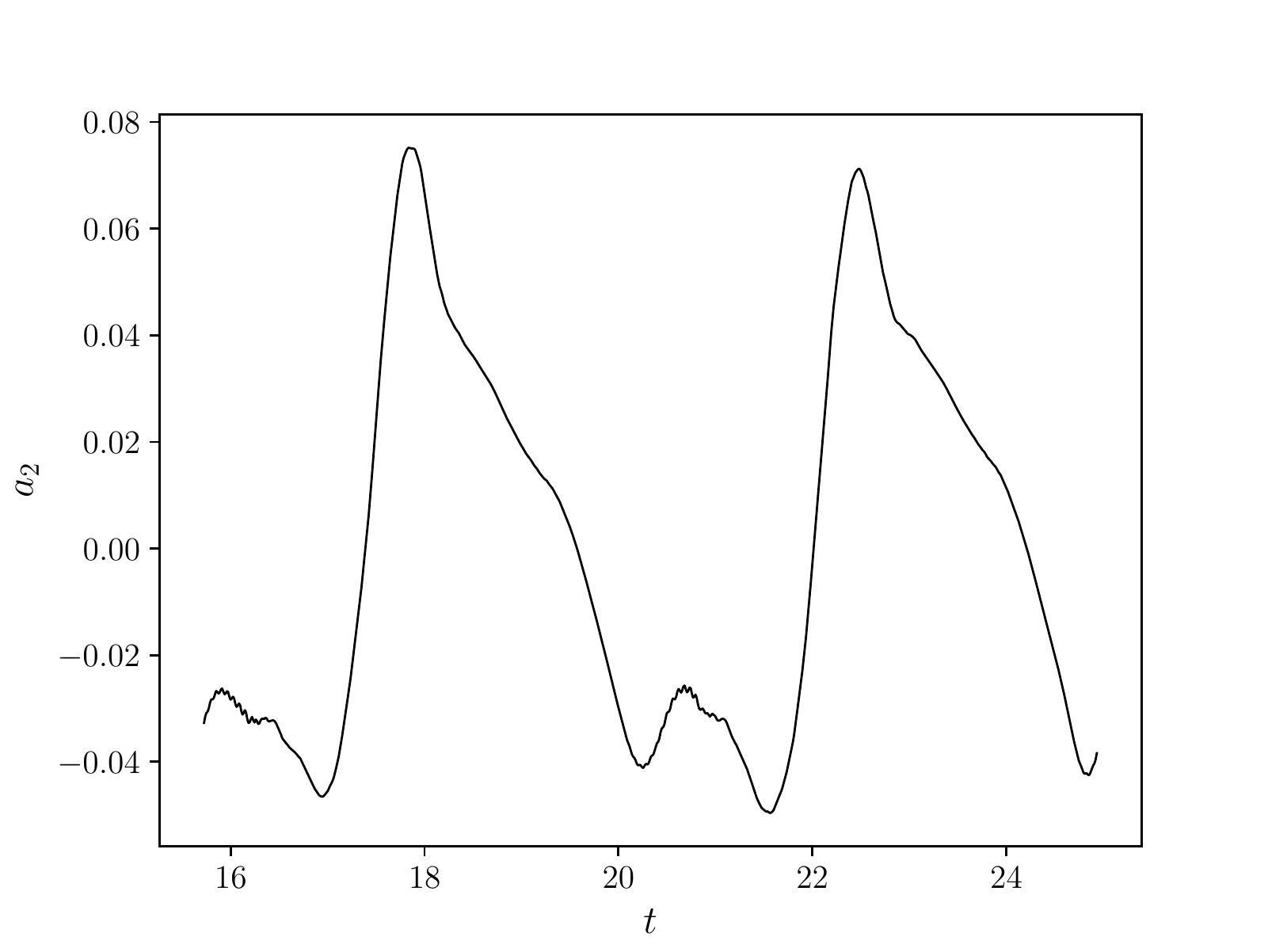}}
	\subfigure[SPOD temporal mode 2.]{\includegraphics[trim = 1mm 1mm 1mm 1mm, clip,width=.49\linewidth]{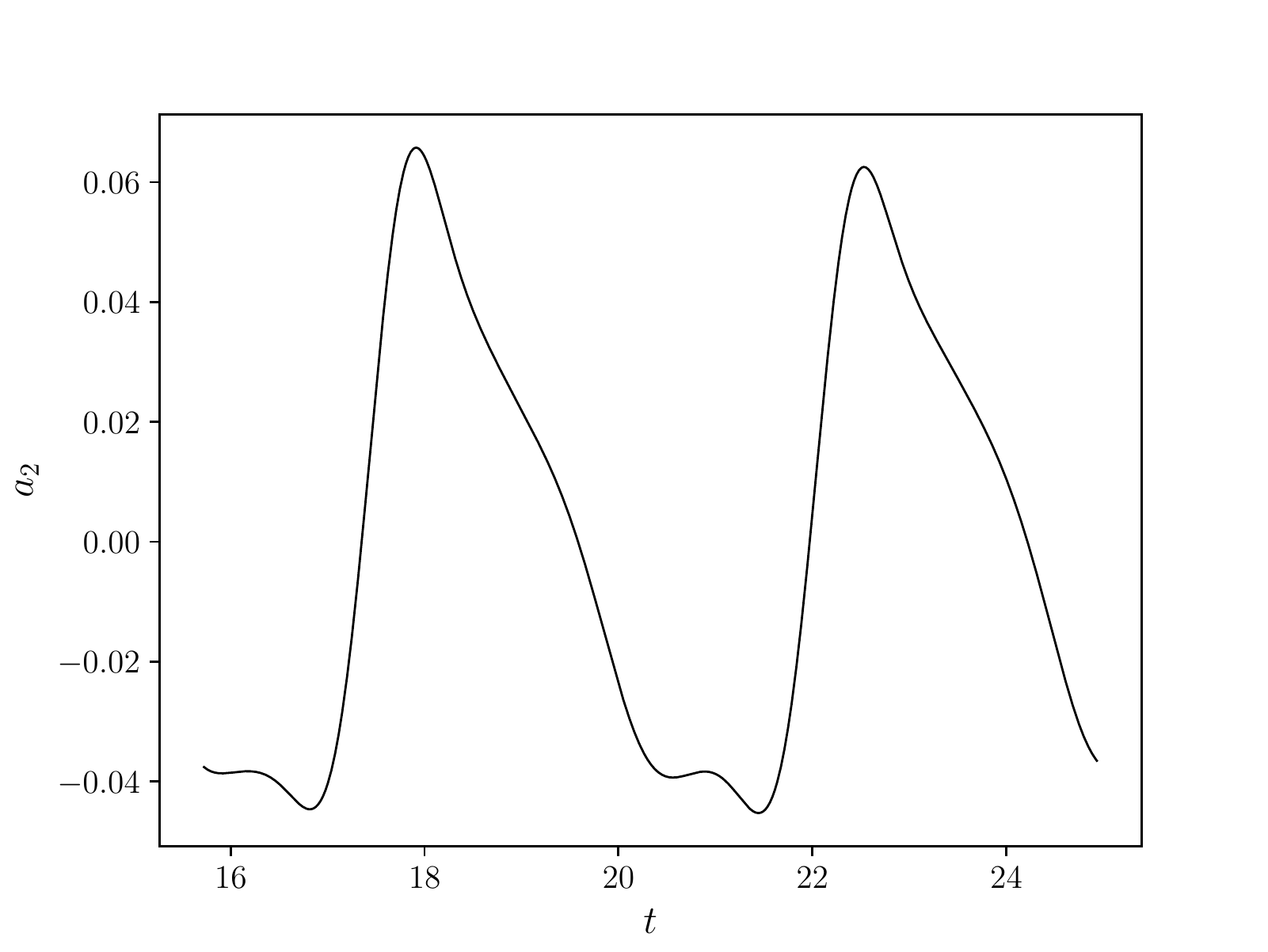}}
	\centering
	\subfigure[POD temporal mode 16.]{\includegraphics[trim = 0mm 1mm 1mm 1mm, clip,width=.49\linewidth]{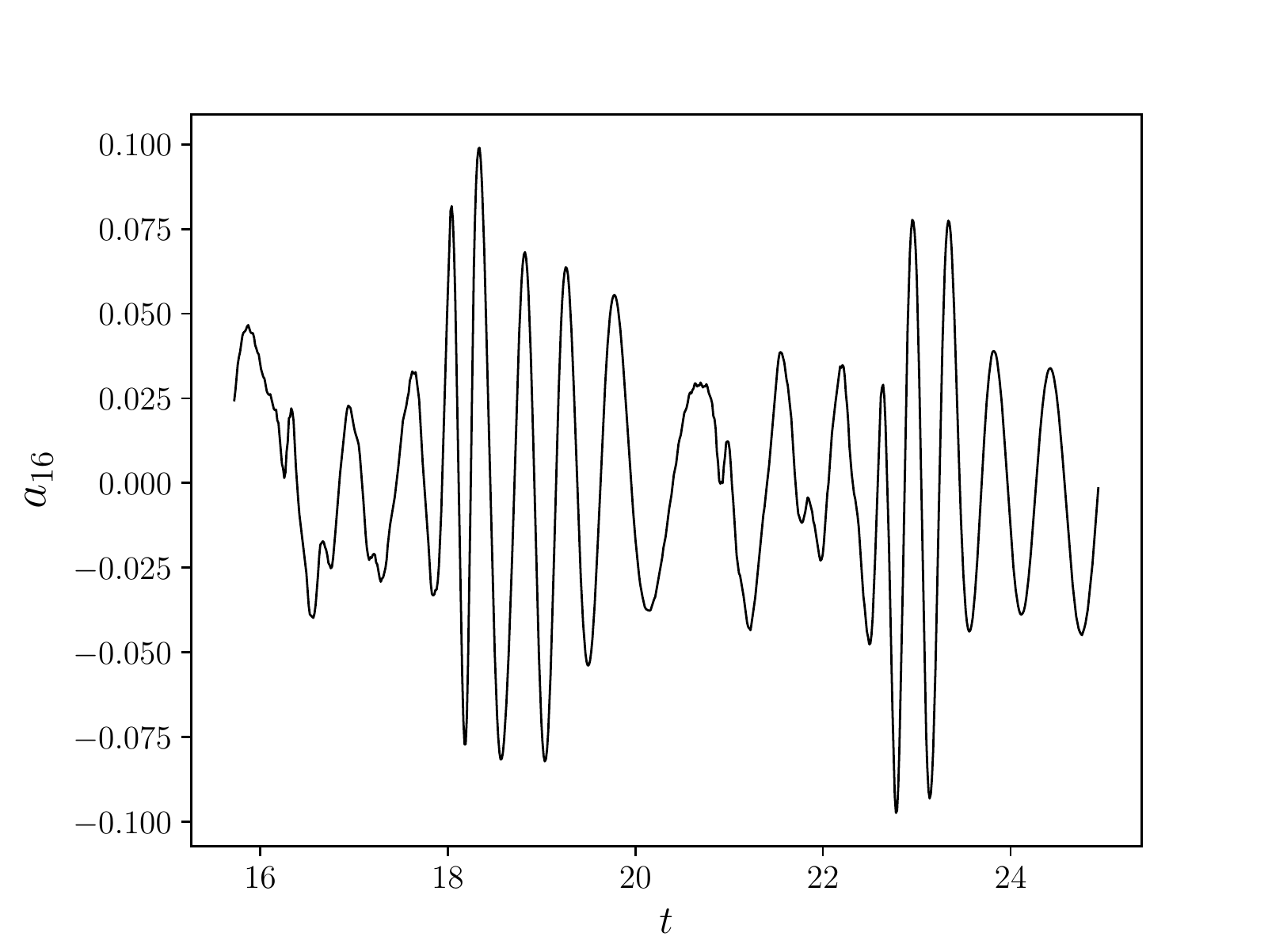}}
	\subfigure[SPOD temporal mode 16.]{\includegraphics[trim = 0mm 1mm 1mm 1mm, clip,width=.49\linewidth]{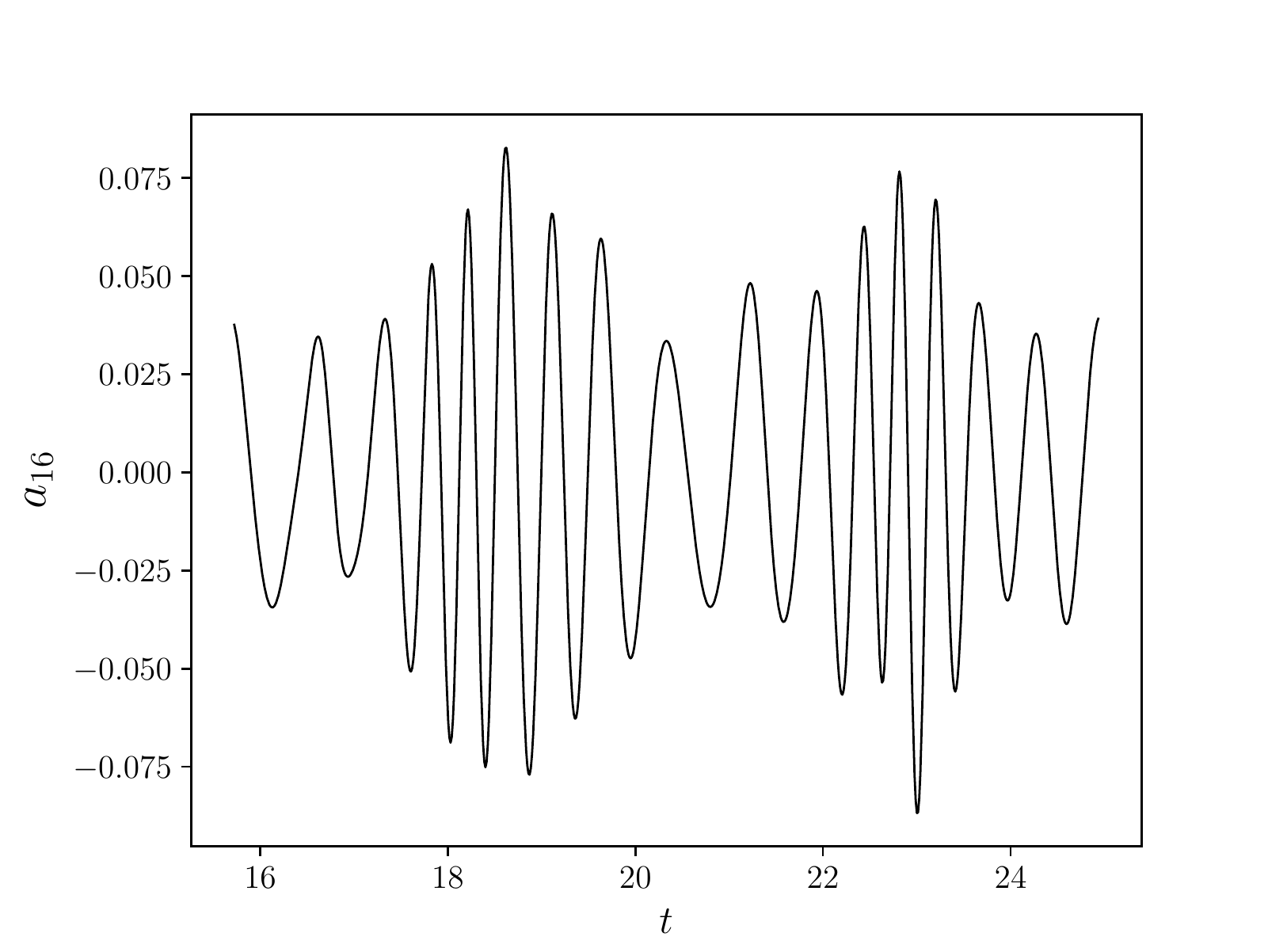}}
	\caption{Characteristics of POD and SPOD temporal modes for 3D flow past airfoil under plunging motion.}
	\label{fig:113D}
\end{figure}
	
For the present turbulent flow, the temporal modes of the standard snapshot POD are composed by several frequencies and, moreover, contain some noise that deteriorates the training of the DNNs. The SPOD provides the most energetic modes for specific frequency bands, allowing a better identification of the individual modes and smoothing out the temporal coefficients. Here, a box filter is applied to 25\% of the POD correlation matrix. We observed that filter values in the range of 20\% to 40\% provide good results for the model reconstructions. If a lower filter window is employed, noise can still be present in the temporal modes. On the other hand, higher filter windows may considerably modify the temporal modes transforming them into quasi-sinusoidal signals. This is undesirable since the relevant dynamics of the flow can be lost and too many SPOD modes may be required for the construction of the ROM. An example of the impact of SPOD on the characteristics of temporal modes 2 and 16 can be seen in figure~\ref{fig:113D}. In this example, the second POD mode exhibits high-frequency noise which is filtered by the SPOD. Meanwhile, POD mode 16 shows a complicated pattern due to the contribution of several frequencies. This temporal coefficient is considerably simplified by the application of SPOD as can be observed in the figure. It is clear that, differently from the cylinder case, for the plunging airfoil, the POD modes contain more complex dynamics which are composed of multiple Fourier modes.

In order to further evaluate the present DNN approch, the ROM reconstructions are also performed for spanwise-averaged solutions of the flow and the reduced order model is constructed following the procedures defined in section \ref{sec:formu}. We employ 6 plunging cycles (2244 snapshots) for computing the models, discarding initial flow transients. The training data contains the first two cycles (748 snapshots) of the FOM and the remaining data is used as the test set. The parameters employed in algorithm \ref{alg:2} are the same as those presented in table \ref{model_generation_dynamic_stall} and the best set of hyperparameters, obtained via random search, is listed in table \ref{hyperparamters_dynamic_stall}. Here, we apply the box filter to 25\% of the POD correlation matrix and employ the first 10 SPOD modes for the ROM construction, which was sufficient to recover the most important features of the dynamic stall solution for the spanwise-averaged study.
\begin{table}
	\begin{center}
		\begin{tabular}{cccccccc} 
			DNN architecture & $\sigma$ & $\alpha$ & $\lambda$ & $n_{iter}$ & $\beta_1$ & $\beta_2$ & $\epsilon$ \\[3pt]
			$10-46-23-17-19-10$ & ELU & $0.001$ & $2.3365 \times 10^{-5}$ & $10000$ & $0.9$ & $0.999$ & $1.0 \times 10^{-8}$ \\
		\end{tabular} 
		\caption{Best set of hyperparameters for ROM of plunging airfoil in deep dynamic stall (spanwise-averaged flow)}
		\label{hyperparamters_dynamic_stall}
	\end{center}
\end{table}

\begin{figure}
	\centering
	\subfigure[FOM, $\psi = 30^{\circ}$.]{\includegraphics[trim = 1mm 1mm 1mm 1mm, clip,width=.49\linewidth]{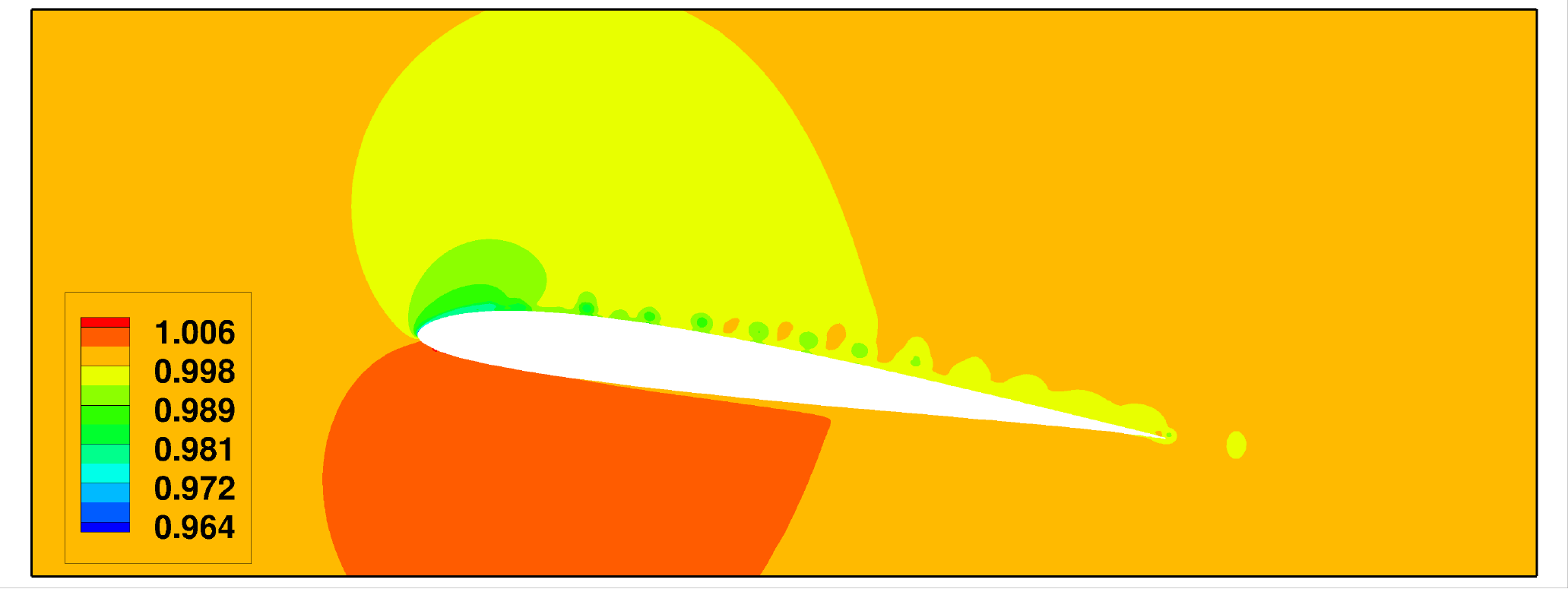}}
	\subfigure[ROM, $\psi = 30^{\circ}$.]{\includegraphics[trim = 1mm 1mm 1mm 1mm, clip,width=.49\linewidth]{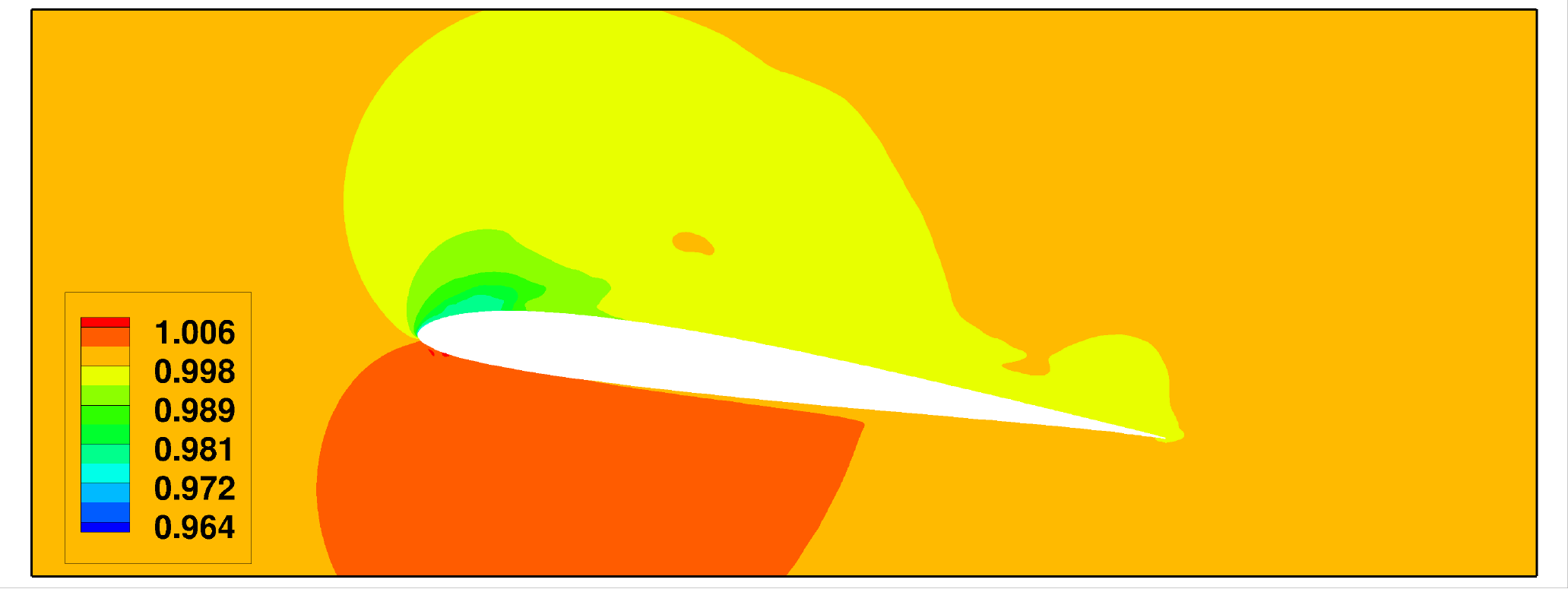}}
	\centering
	\subfigure[FOM, $\psi = 60^{\circ}$.]{\includegraphics[trim = 1mm 1mm 1mm 1mm, clip,width=.49\linewidth]{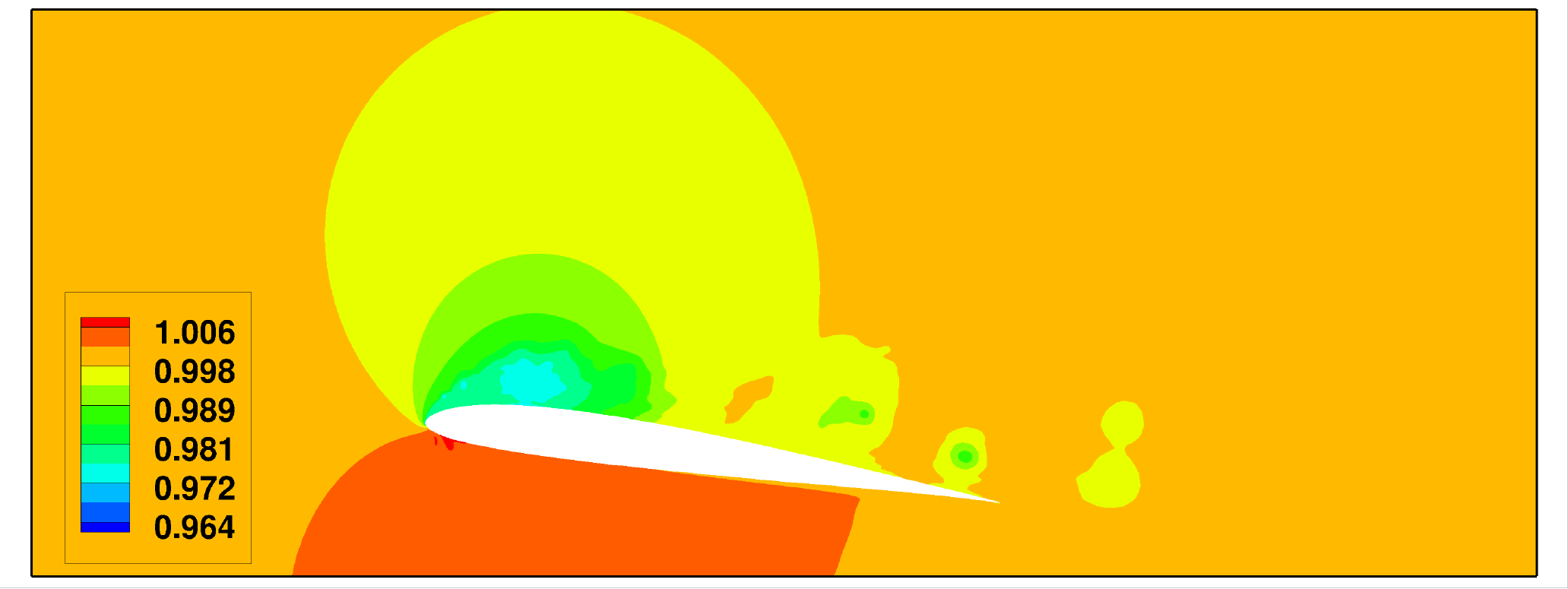}}
	\subfigure[ROM, $\psi = 60^{\circ}$.]{\includegraphics[trim = 1mm 1mm 1mm 1mm, clip,width=.49\linewidth]{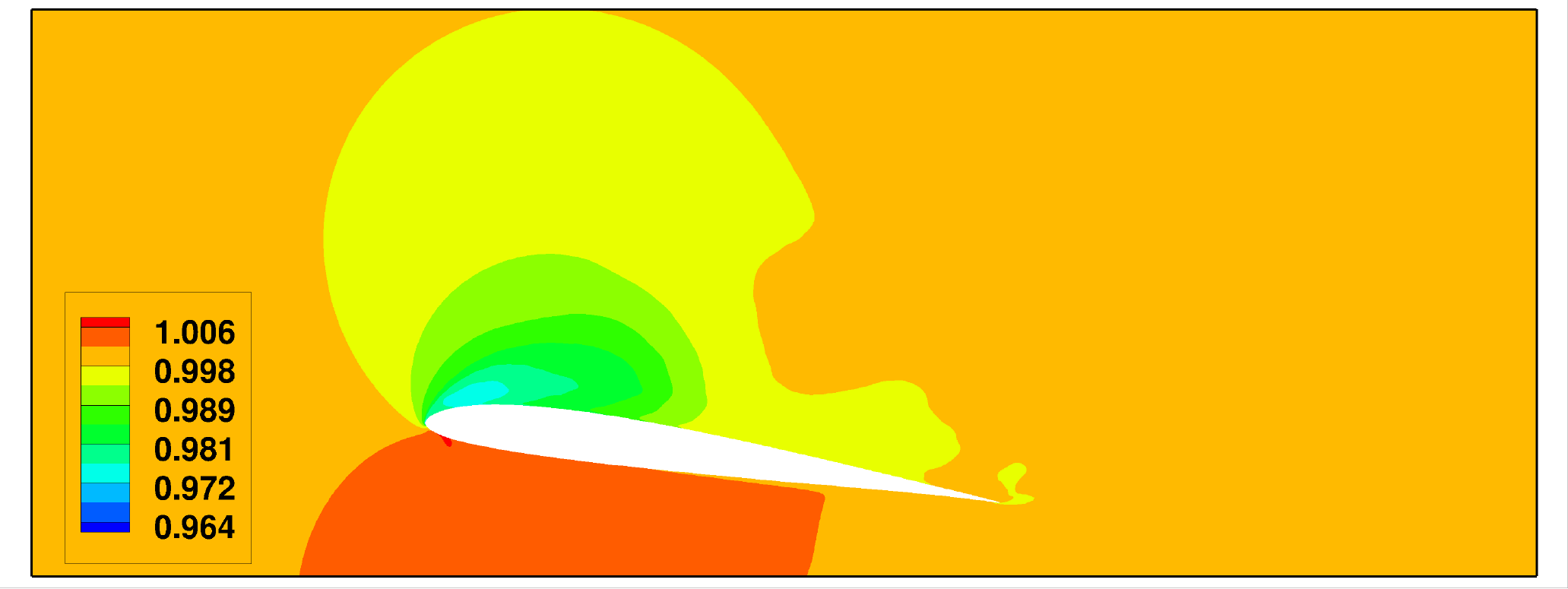}}
	\centering
	\subfigure[FOM, $\psi = 90^{\circ}$.]{\includegraphics[trim = 1mm 1mm 1mm 1mm, clip,width=.49\linewidth]{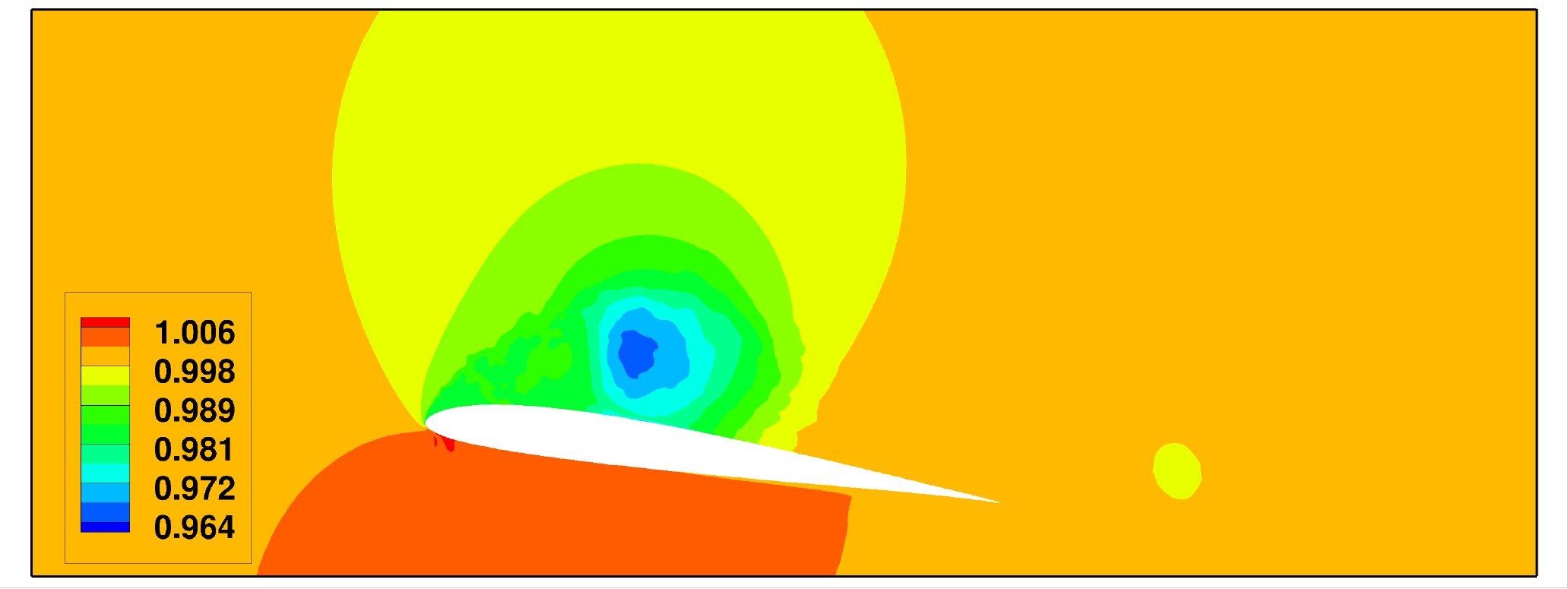}}
	\subfigure[ROM, $\psi = 90^{\circ}$.]{\includegraphics[trim = 1mm 1mm 1mm 1mm, clip,width=.49\linewidth]{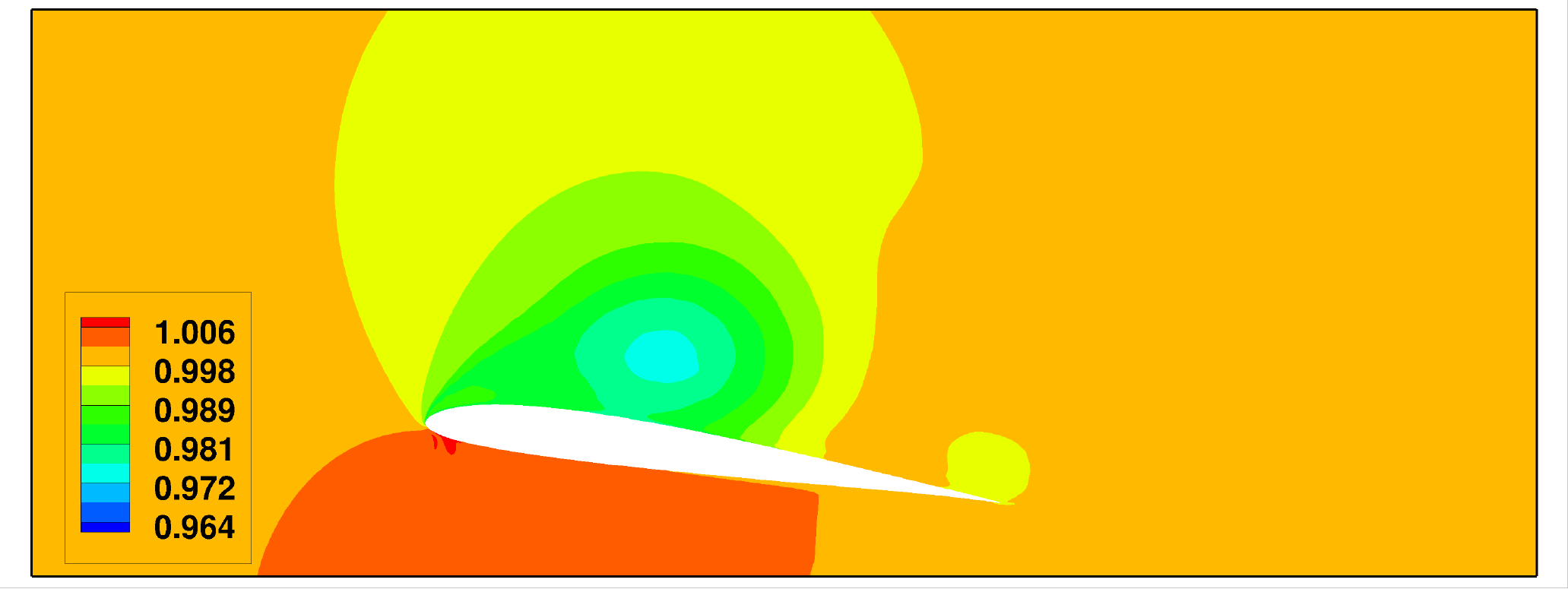}}
	\centering
	\subfigure[FOM, $\psi = 120^{\circ}$.]{\includegraphics[trim = 1mm 1mm 1mm 1mm, clip,width=.49\linewidth]{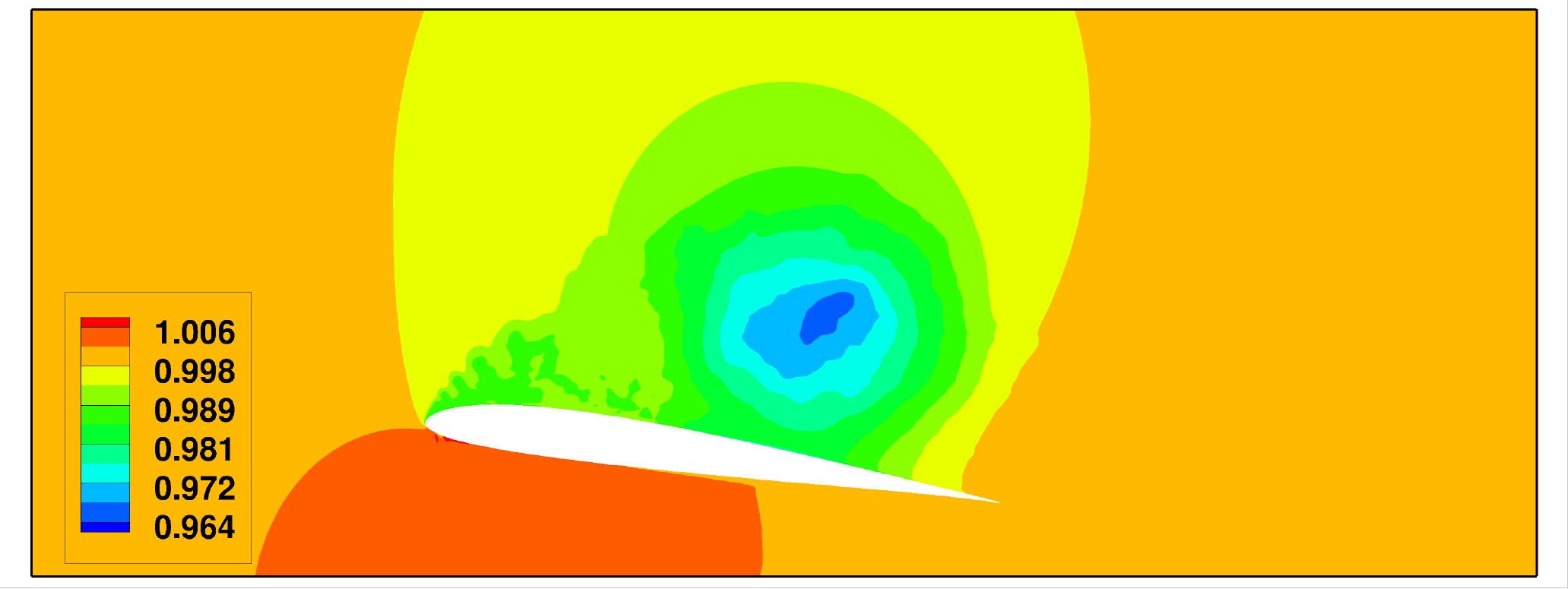}}
	\subfigure[ROM, $\psi = 120^{\circ}$.]{\includegraphics[trim = 1mm 1mm 1mm 1mm, clip,width=.49\linewidth]{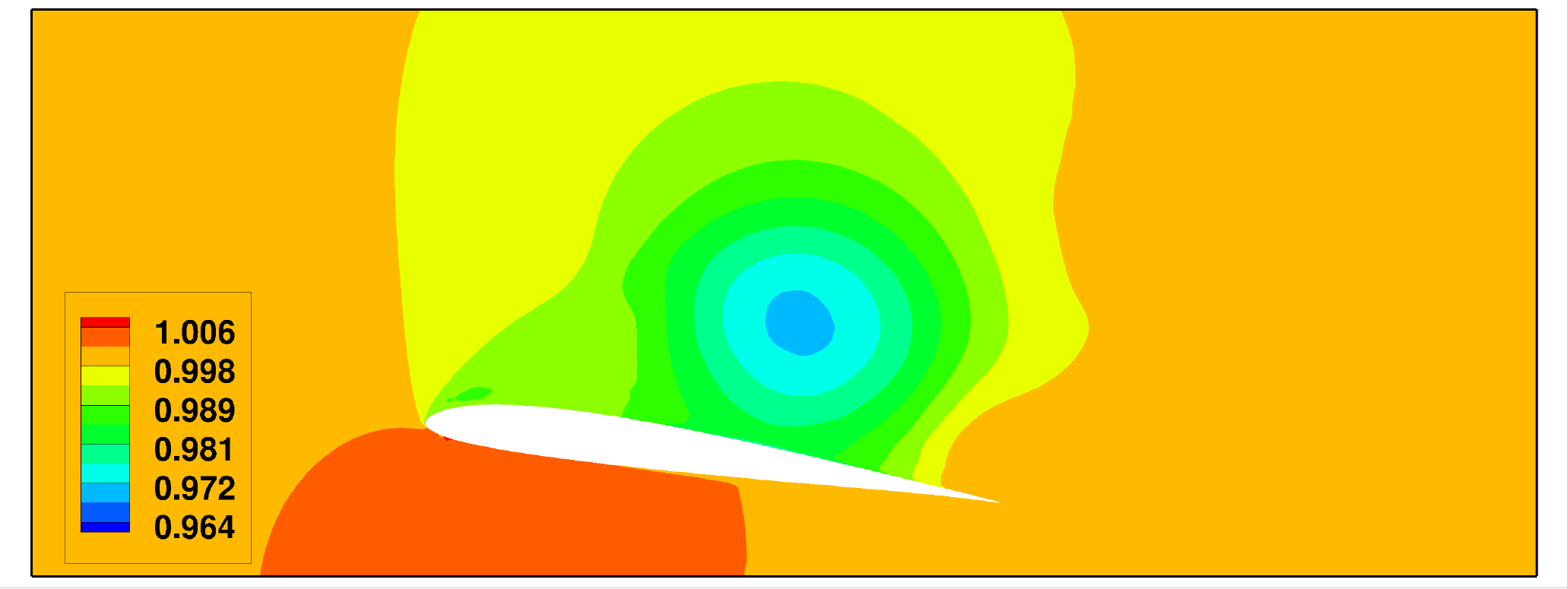}}
	\centering
	\subfigure[FOM, $\psi = 150^{\circ}$.]{\includegraphics[trim = 1mm 1mm 1mm 1mm, clip,width=.49\linewidth]{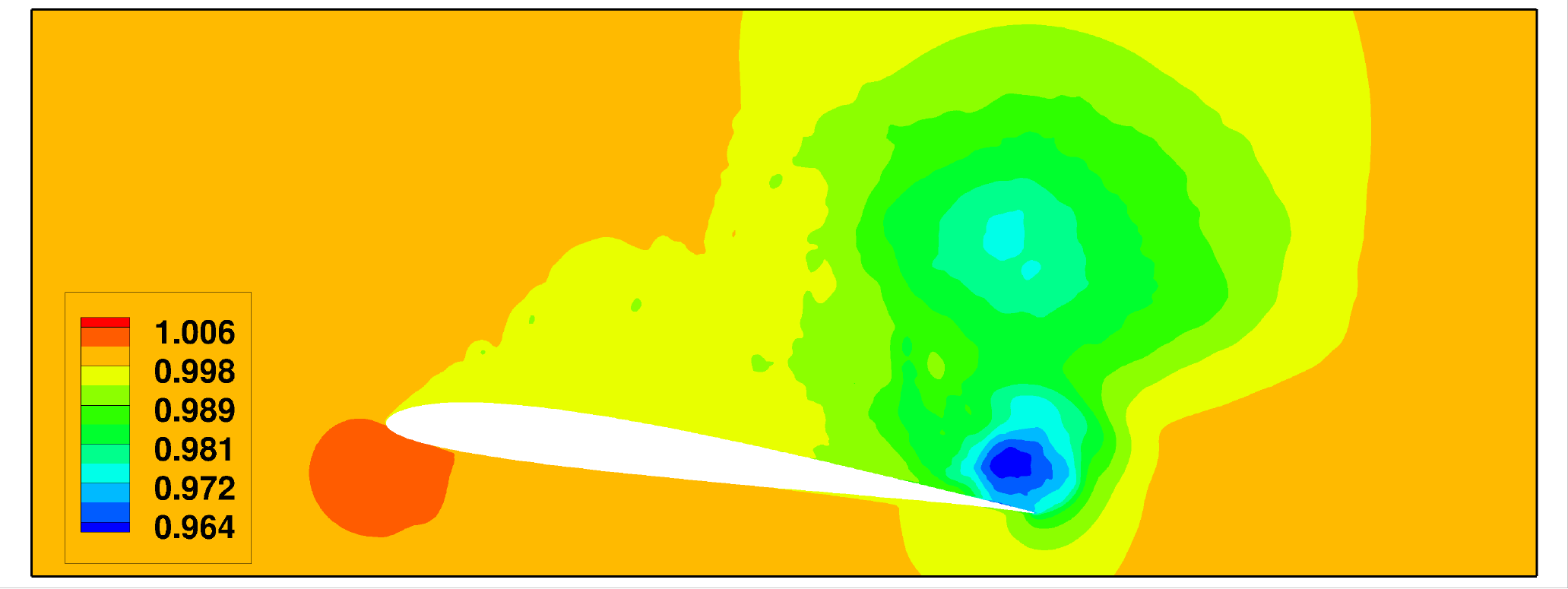}}
	\subfigure[ROM, $\psi = 150^{\circ}$.]{\includegraphics[trim = 1mm 1mm 1mm 1mm, clip,width=.49\linewidth]{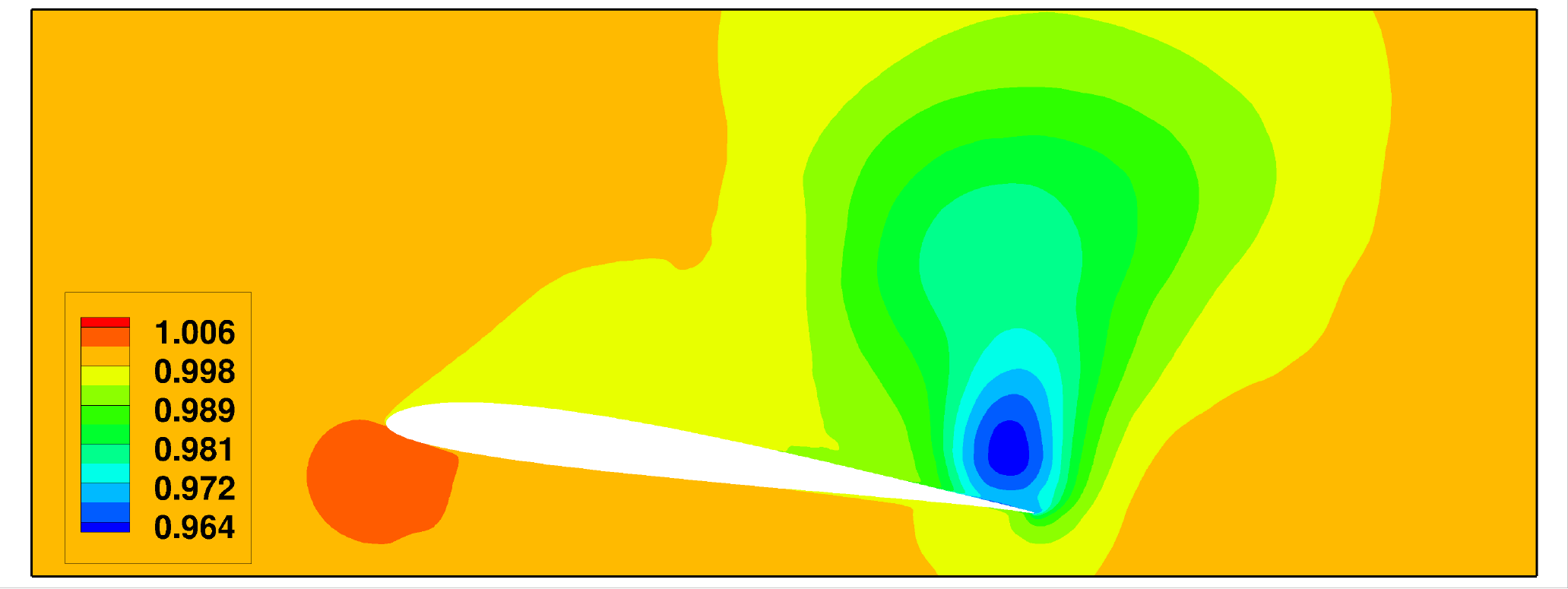}}
	\caption{Density contours at several phase angles for the fifth plunging cycle.}
	\label{fig:12}
\end{figure}

\begin{figure}
	\centering
	\subfigure[FOM, $\psi = 30^{\circ}$.]{\includegraphics[trim = 1mm 1mm 1mm 1mm, clip,width=.49\linewidth]{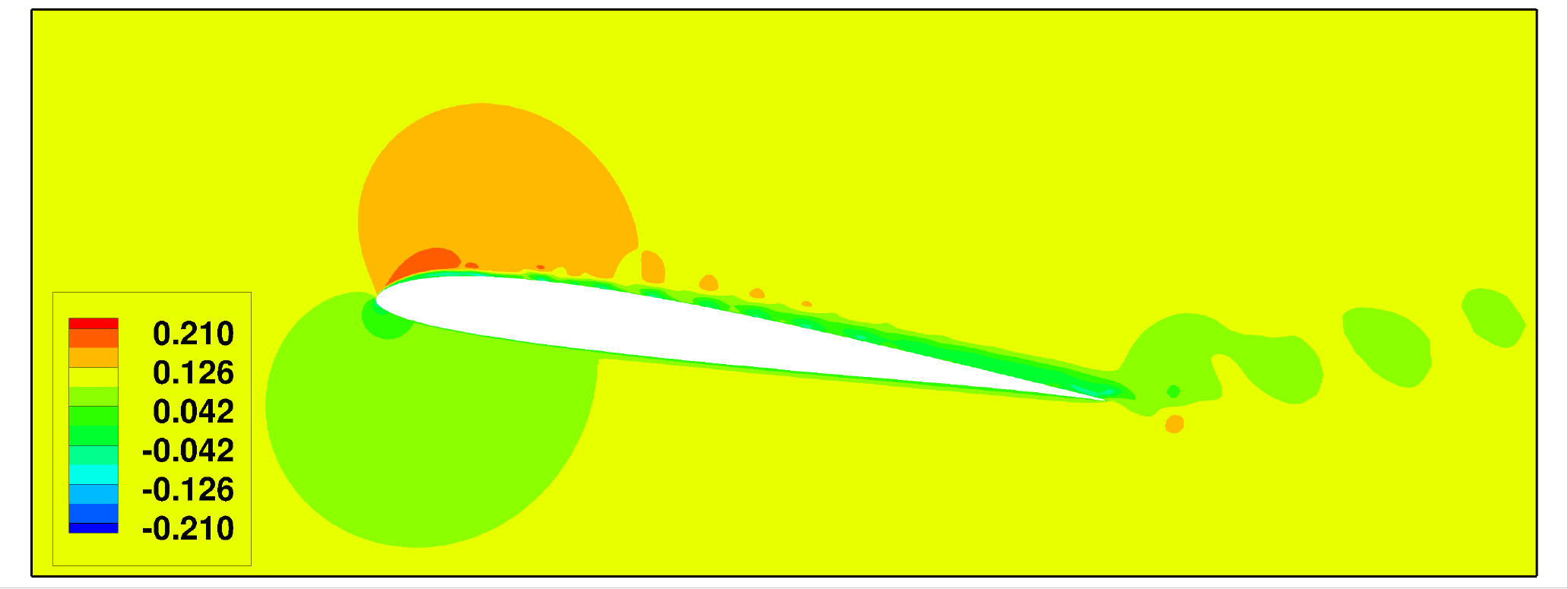}}
	\subfigure[ROM, $\psi = 30^{\circ}$.]{\includegraphics[trim = 1mm 1mm 1mm 1mm, clip,width=.49\linewidth]{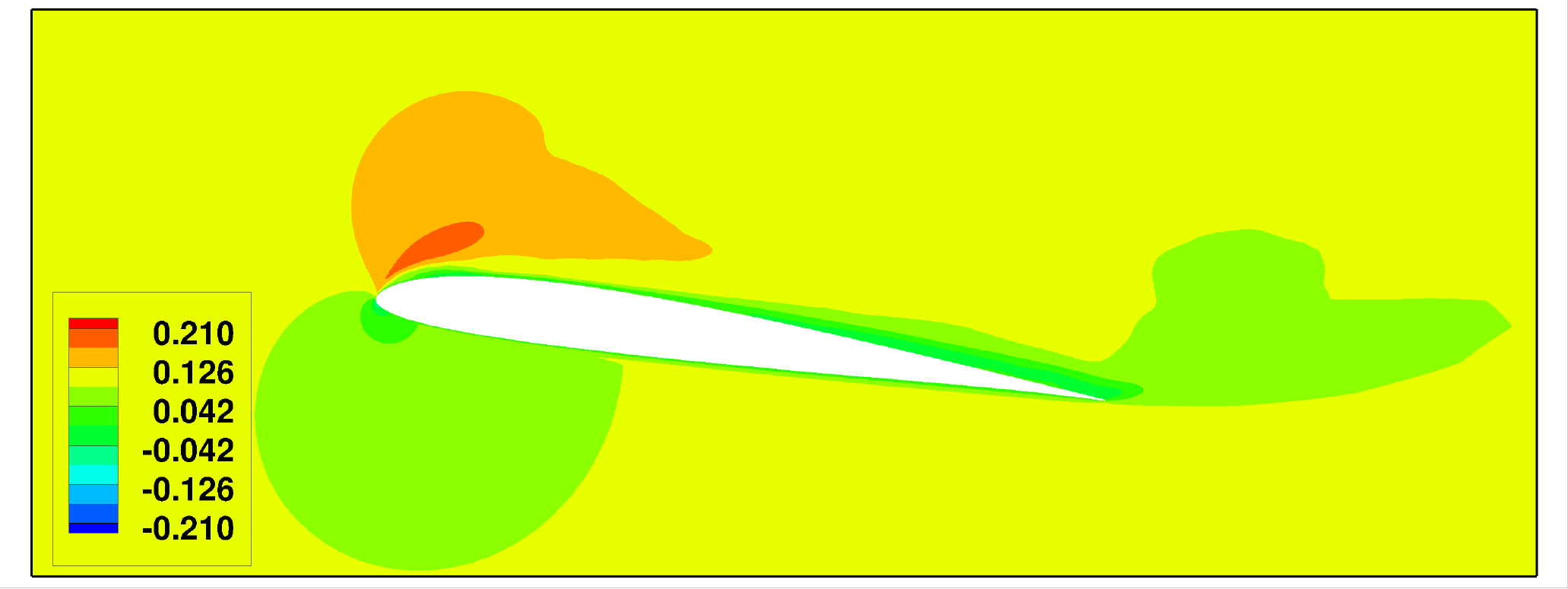}}
	\centering
	\subfigure[FOM, $\psi = 60^{\circ}$.]{\includegraphics[trim = 1mm 1mm 1mm 1mm, clip,width=.49\linewidth]{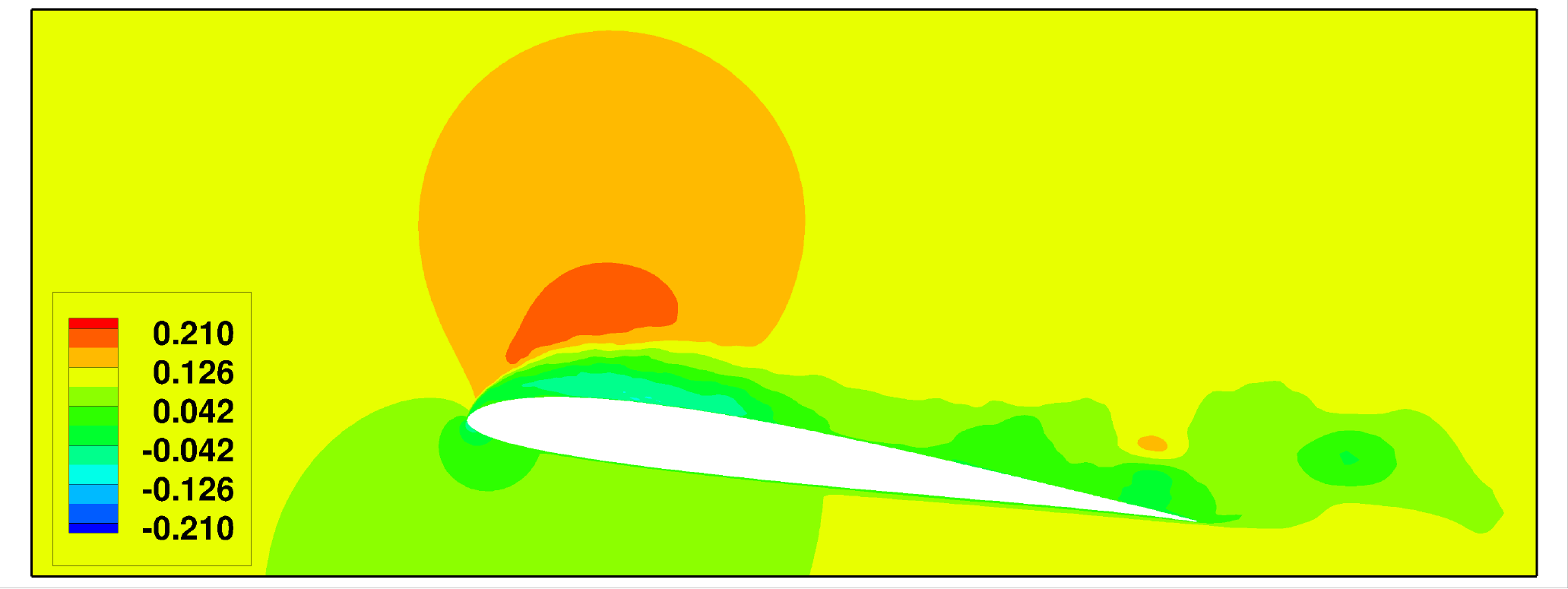}}
	\subfigure[ROM, $\psi = 60^{\circ}$.]{\includegraphics[trim = 1mm 1mm 1mm 1mm, clip,width=.49\linewidth]{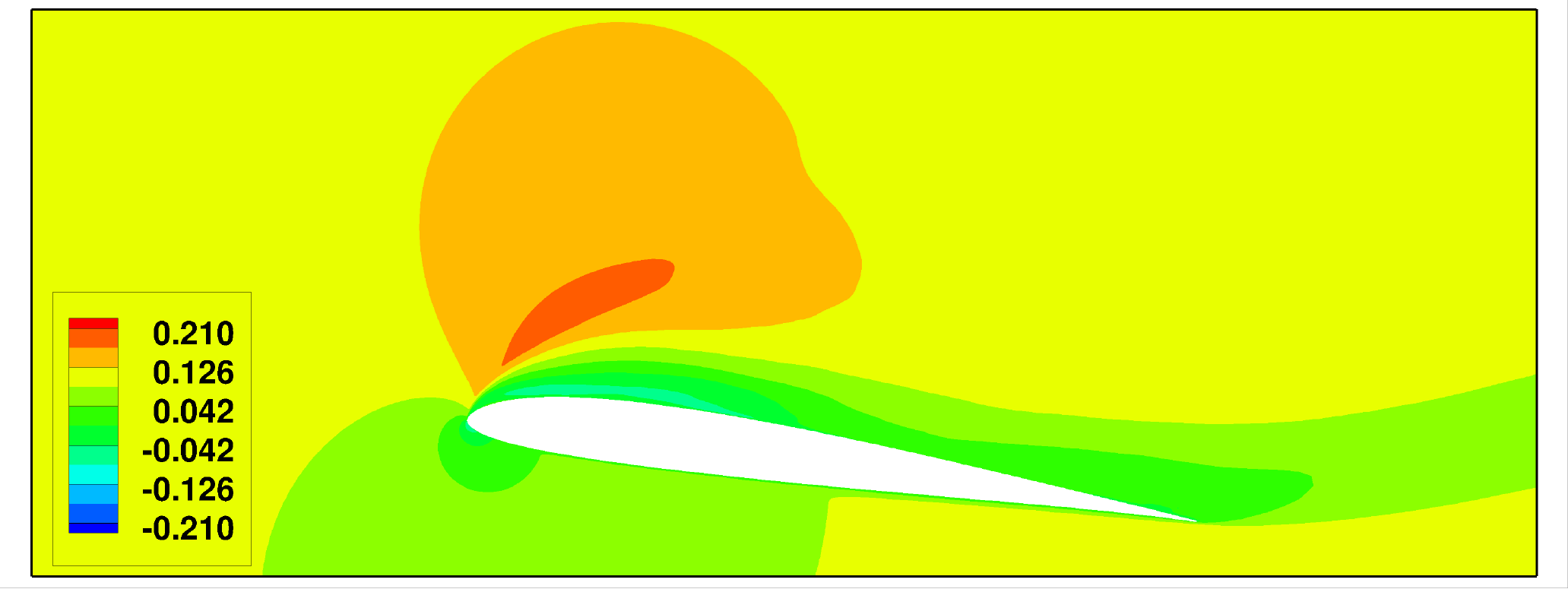}}
	\centering
	\subfigure[FOM, $\psi = 90^{\circ}$.]{\includegraphics[trim = 1mm 1mm 1mm 1mm, clip,width=.49\linewidth]{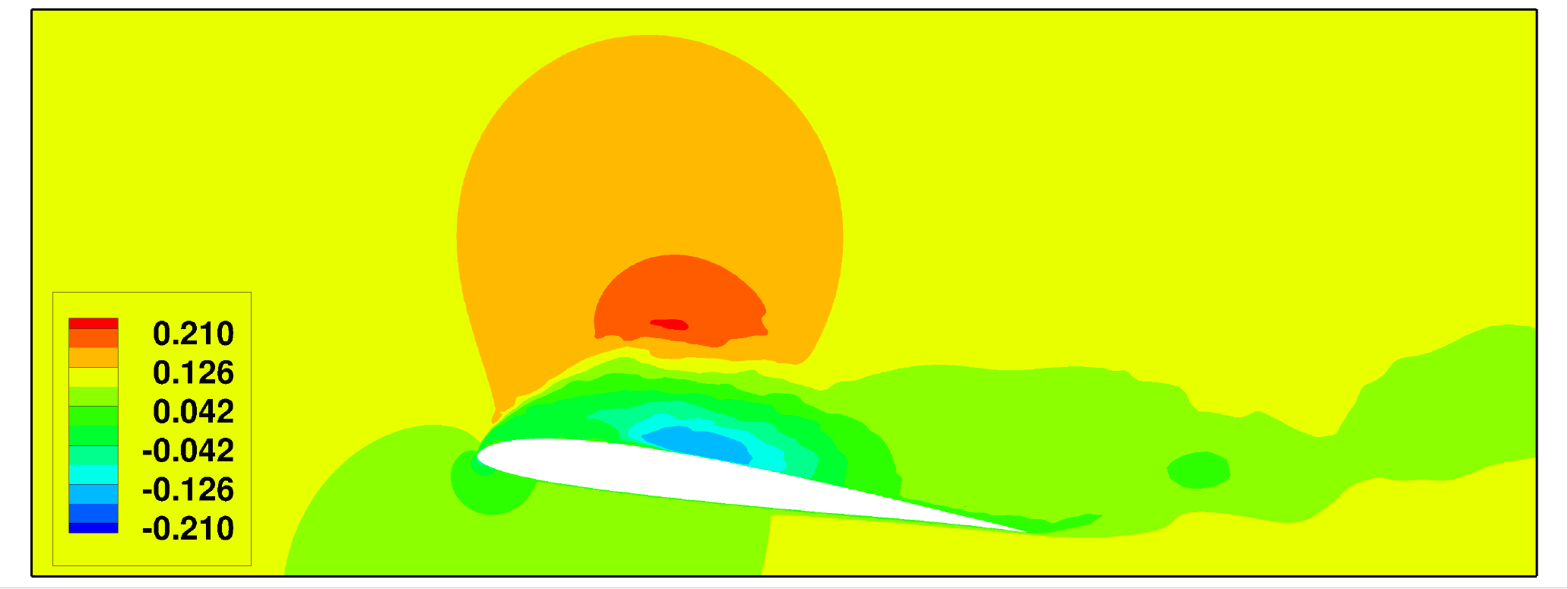}}
	\subfigure[ROM, $\psi = 90^{\circ}$.]{\includegraphics[trim = 1mm 1mm 1mm 1mm, clip,width=.49\linewidth]{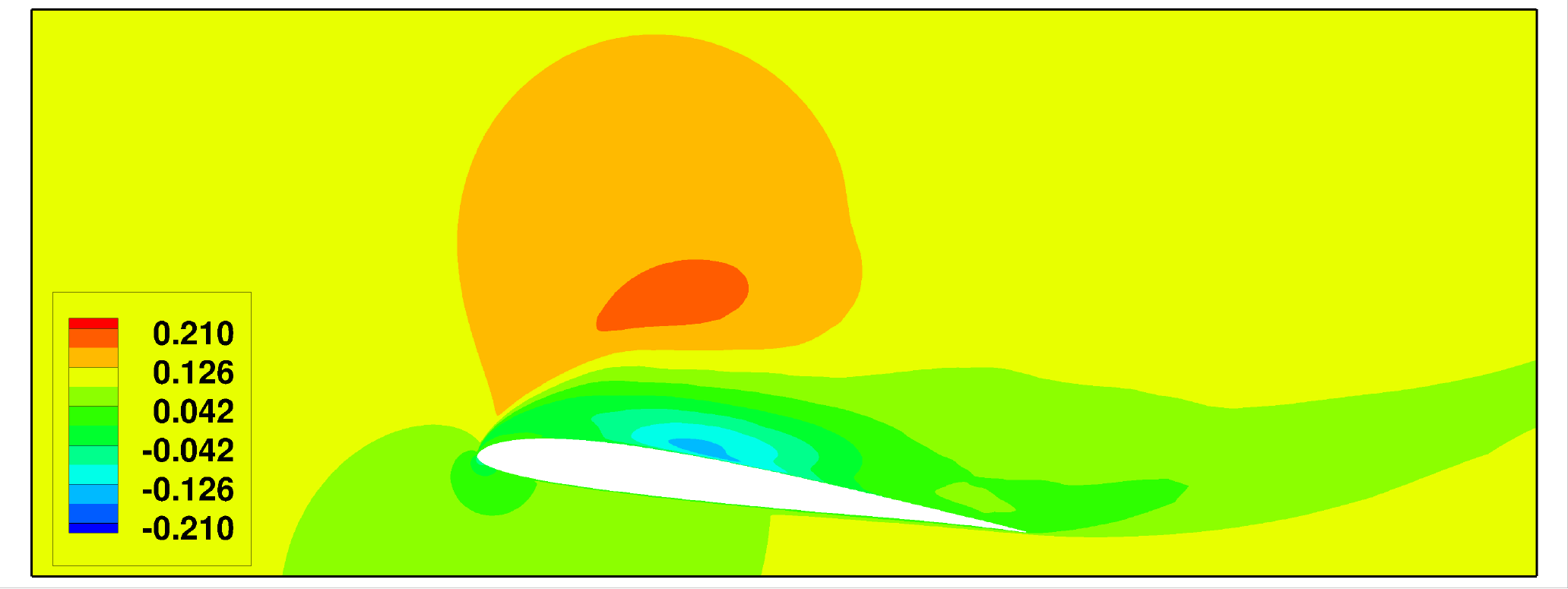}}
	\centering
	\subfigure[FOM, $\psi = 120^{\circ}$.]{\includegraphics[trim = 1mm 1mm 1mm 1mm, clip,width=.49\linewidth]{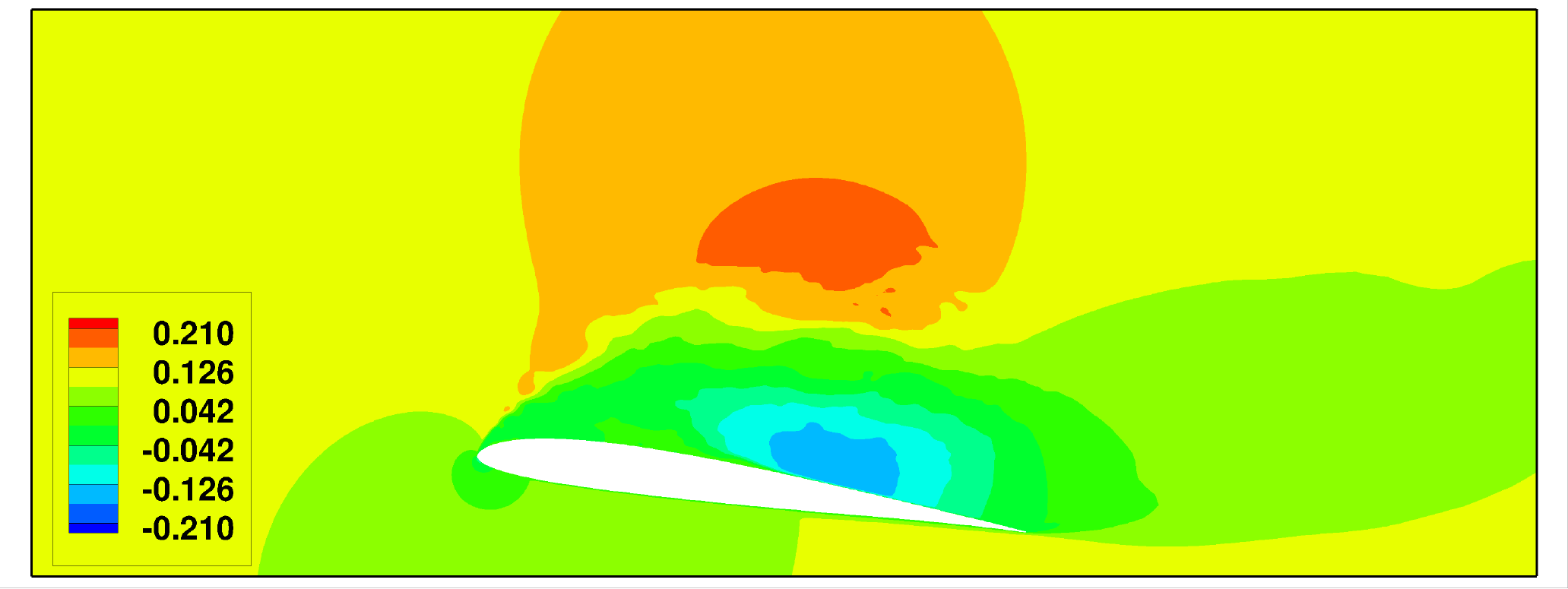}}
	\subfigure[ROM, $\psi = 120^{\circ}$.]{\includegraphics[trim = 1mm 1mm 1mm 1mm, clip,width=.49\linewidth]{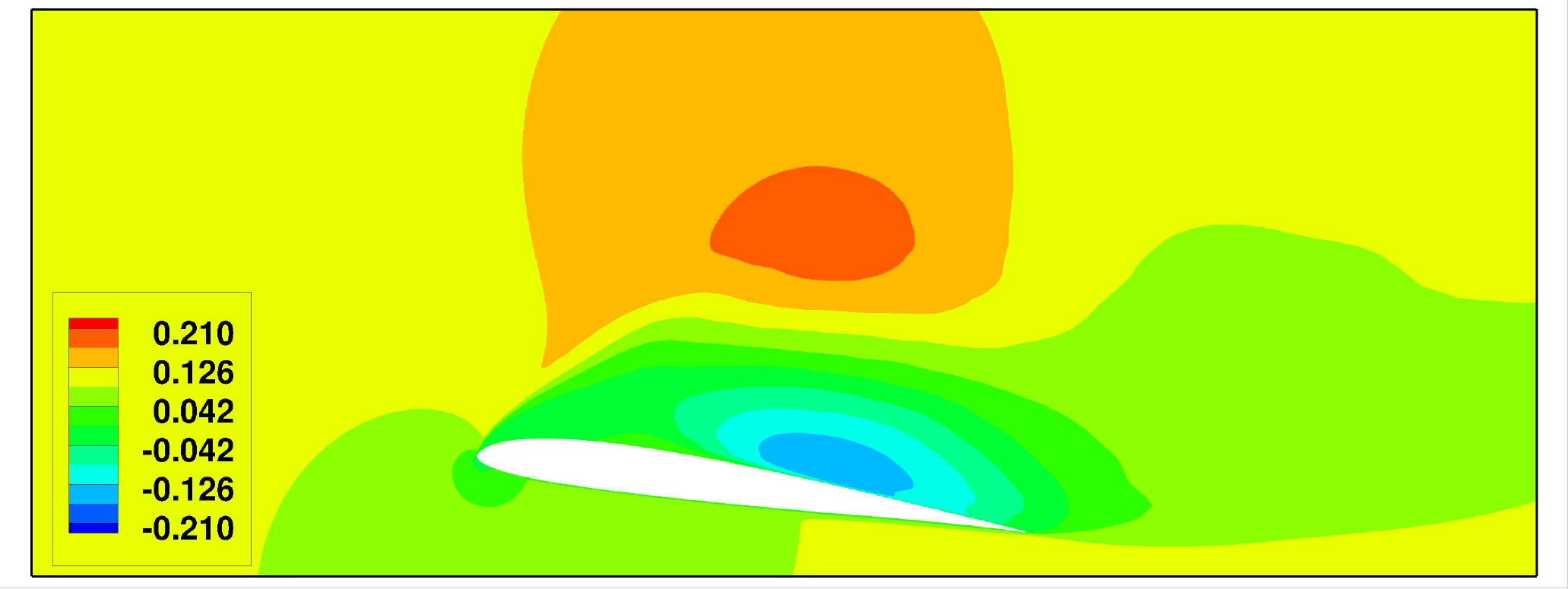}}
	\centering
	\subfigure[FOM, $\psi = 150^{\circ}$.]{\includegraphics[trim = 1mm 1mm 1mm 1mm, clip,width=.49\linewidth]{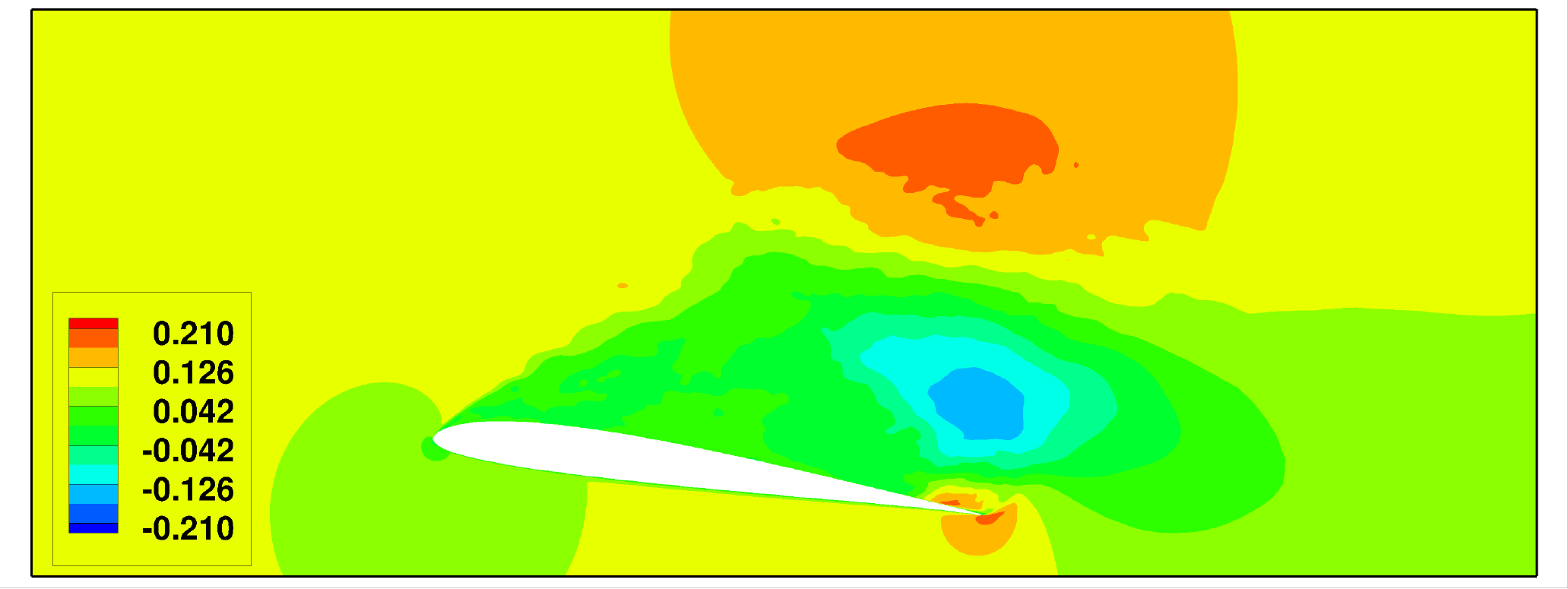}}
	\subfigure[ROM, $\psi = 150^{\circ}$.]{\includegraphics[trim = 1mm 1mm 1mm 1mm, clip,width=.49\linewidth]{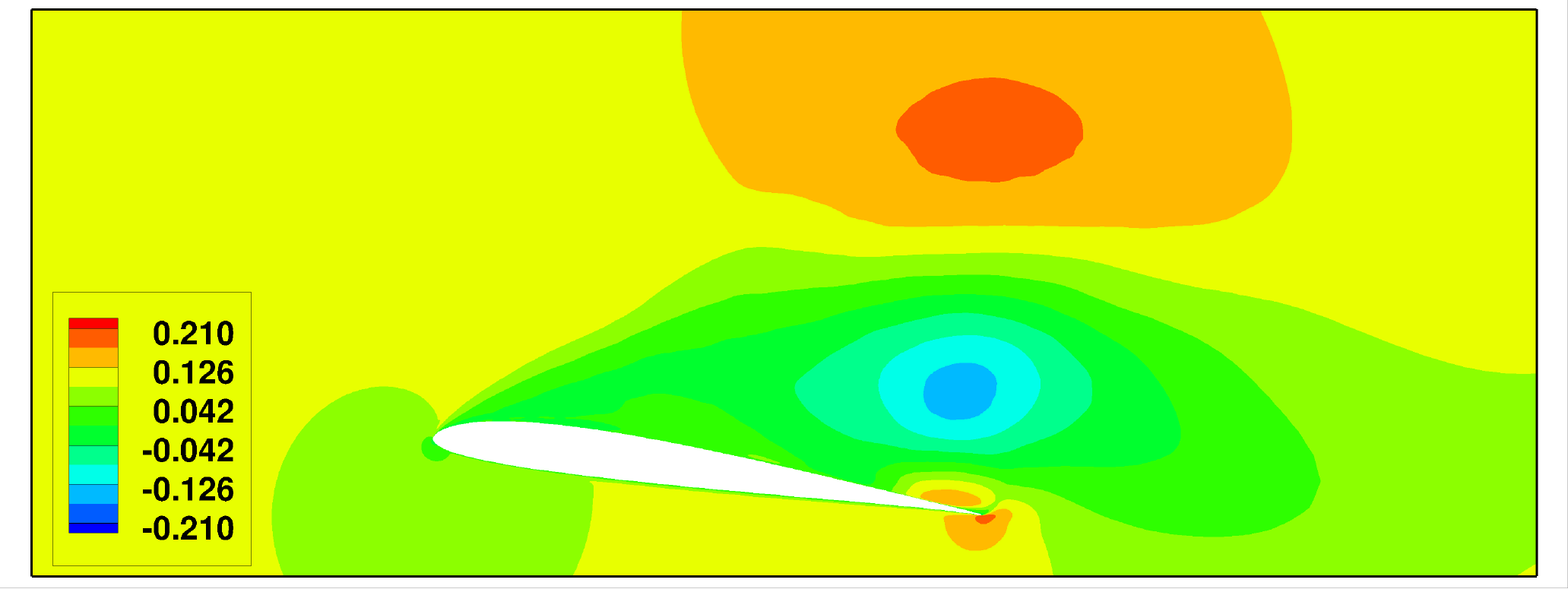}}
	\caption{Contours of $x$-component of momentum at several phase angles for the fifth plunging cycle.}
	\label{fig:13}
\end{figure}

Figures \ref{fig:12} and \ref{fig:13} present snapshots of density and x-component of momentum, respectively, for the FOM and ROM. A video comparing these solutions for the spanwise-averaged flow is also provided together with the manuscript. The flow features of the dynamic stall can be observed for different stages of the plunging cycle, given by different values of $\psi$ in the figures. It is possible to track the LEV over the suction side of the airfoil. The current ROM is able to recover the important dynamics of the leading edge vortex formation, its transport and ejection, besides the trailing edge vortex formation and ejection. One should remind that the results presented in these figures are obtained for a plunging cycle beyond the training region. Therefore, the current ROM is capable of reproducing the flow dynamics for the test set.
The high wavenumber features shown in the FOM are not present in the SPOD modes and, hence, they are not recovered by the ROM. If additional SPOD modes were employed in the model reconstruction, these features would appear. However, the overall cost of the simulations would considerably increase if accurate higher POD modes were required. For the present study, we observed that the fine scale flow dynamics are related to higher modes since most of the energetic content is related to the airfoil plunging motion represented by the first POD modes.

\begin{figure}
	\centering
	\subfigure[Density fluctuations.]{\includegraphics[width=.49\linewidth]{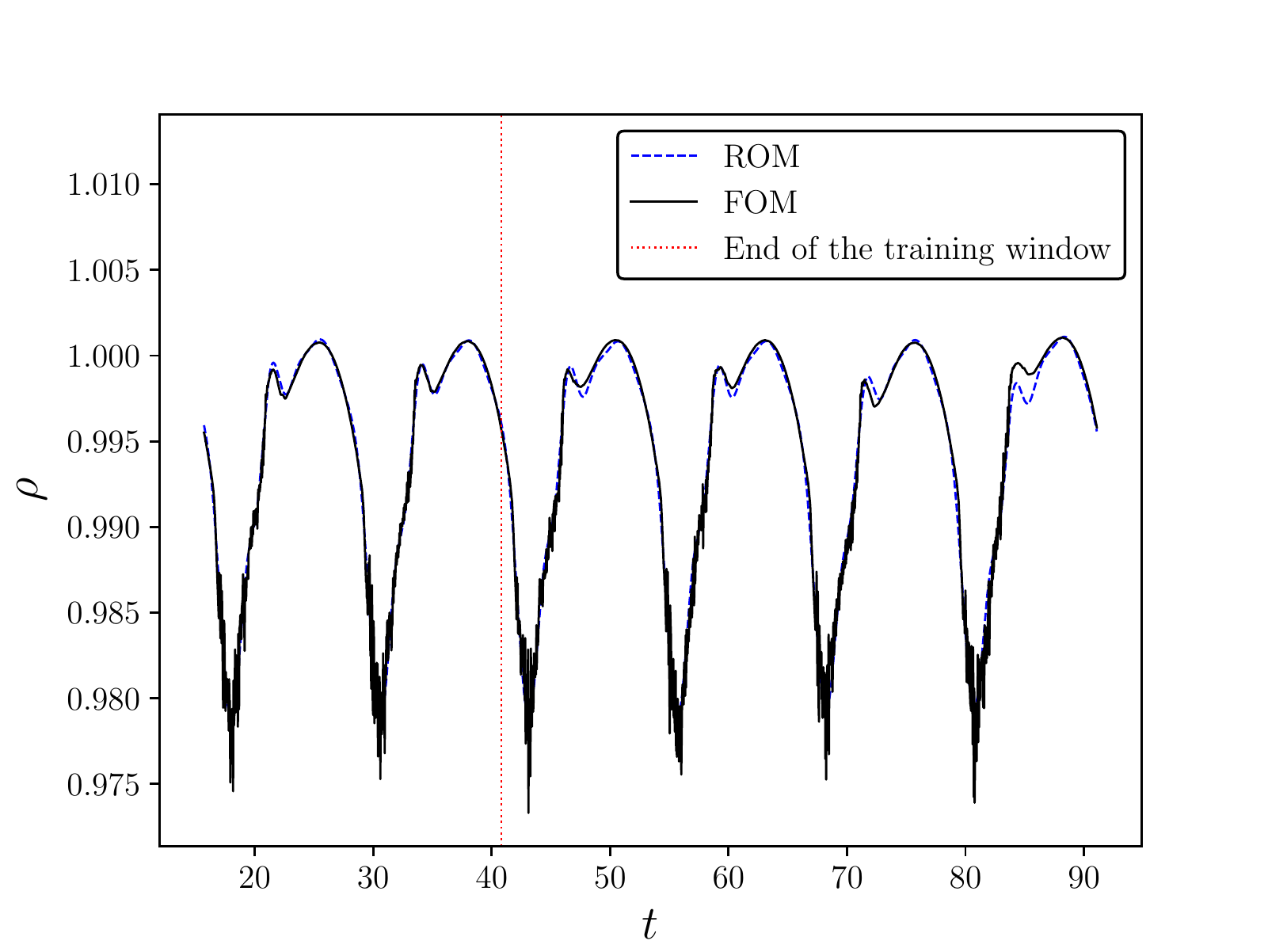}}
	\subfigure[Density fluctuations.]{\includegraphics[width=.49\linewidth]{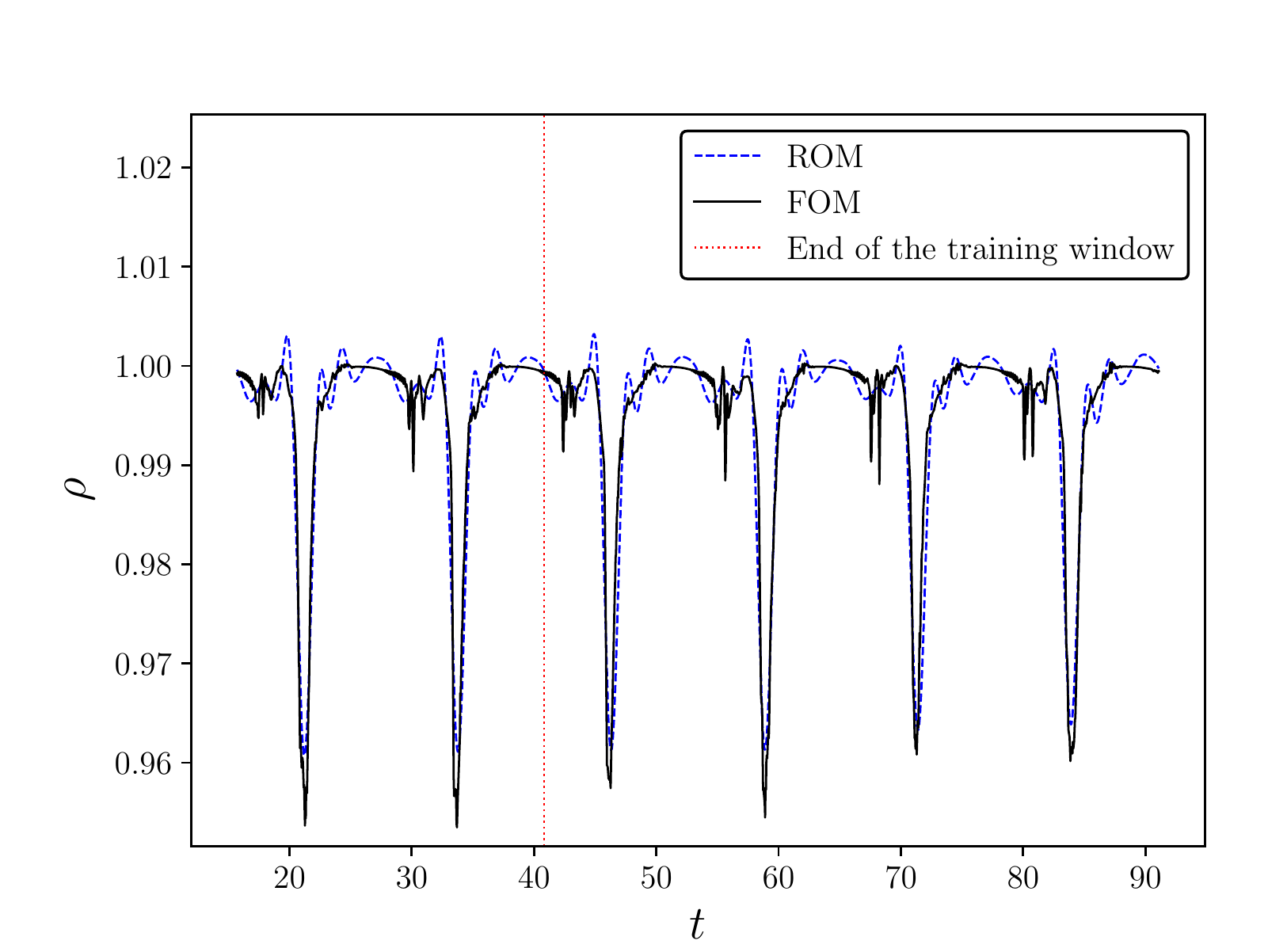}}
	\subfigure[Fluctuations of x-component of momentum.]{\includegraphics[width=.49\linewidth]{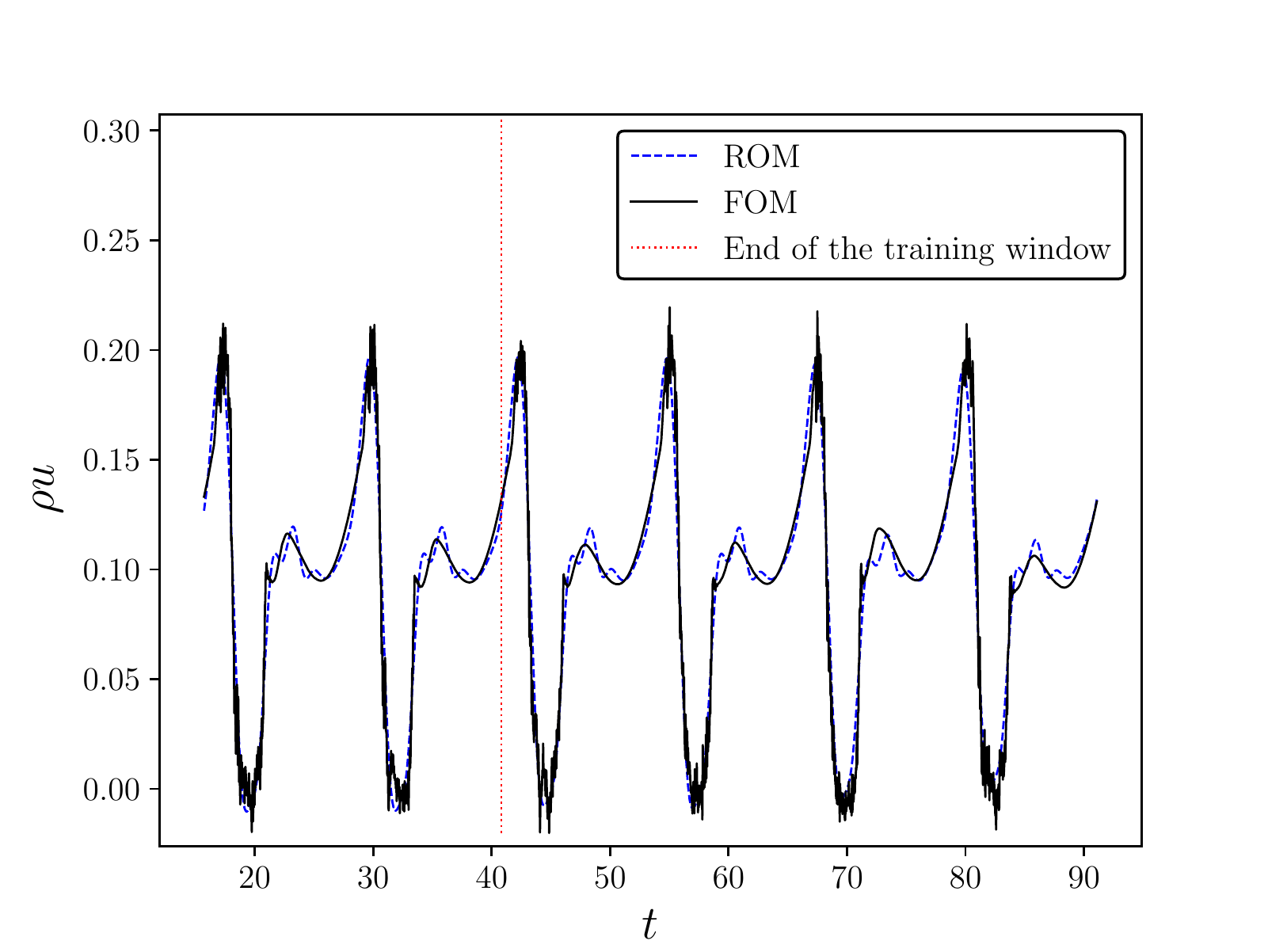}}
	\subfigure[Fluctuations of x-component of momentum.]{\includegraphics[width=.49\linewidth]{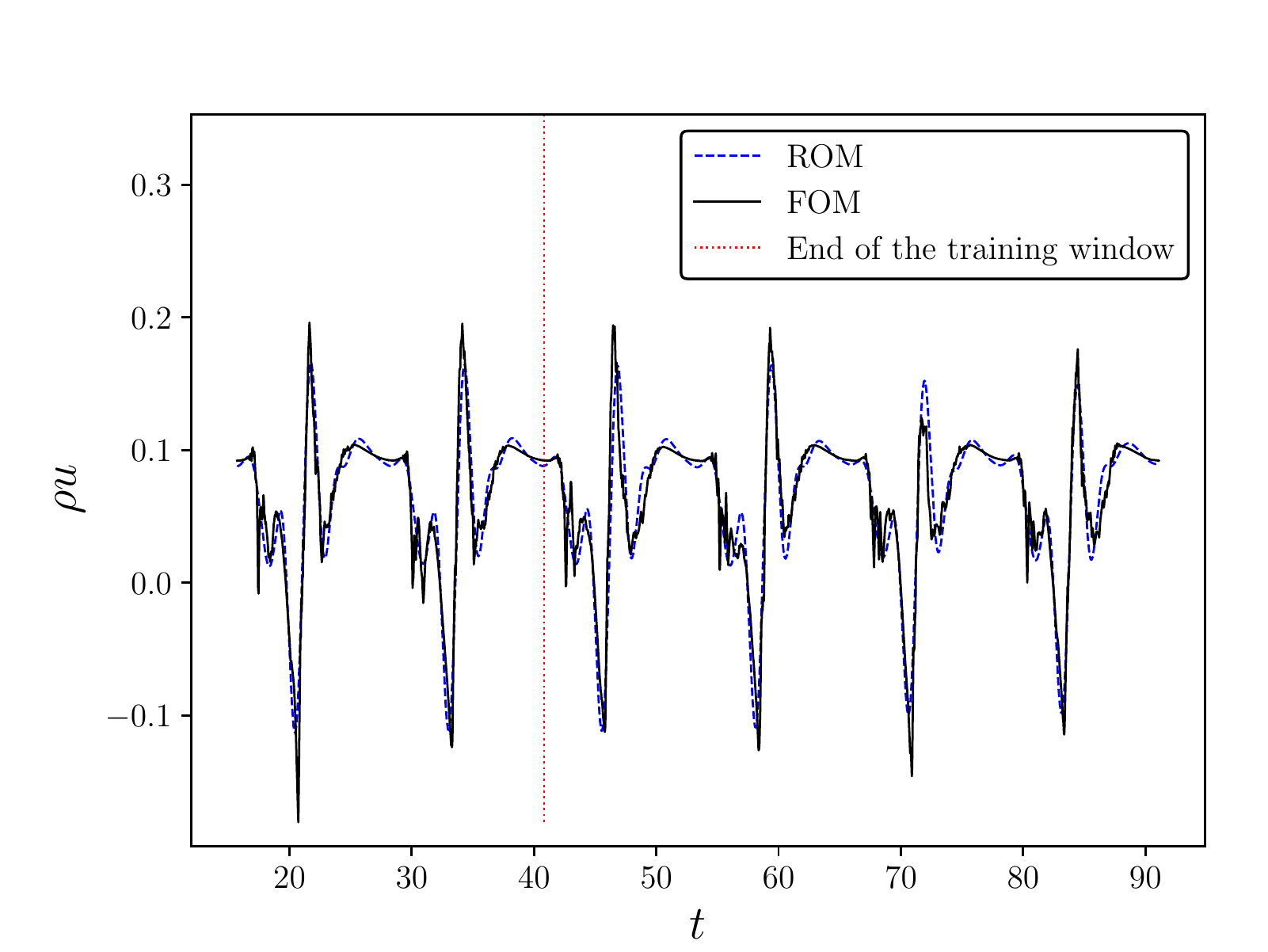}}
	\caption{Fluctuation time histories computed by the FOM and ROM for probe locations in the proximity of the leading edge (left column) and trailing edge (right column).}
	\label{fig:14}
\end{figure}

Fluctuation time histories are presented in figure \ref{fig:14} for density and $x$-momentum. These properties are computed by the FOM and ROM at probe locations in the proximity of the leading and trailing edges, at $(x,y)=(0.07,0.07)$ and $(x,y)=(0.97,0.07)$, respectively. It is important to mention that the airfoil leading edge is at the origin of the Cartesian system. The plots show the training and test regions demonstrating that the reduced order model is stable beyond the training set and can accurately reproduce the main dynamics of the flow. The fast oscillations resolved by the full order model are not represented due to the SPOD basis truncation. However, all the relevant features of the flow fluctuations are captured by the DNN model obtained with the first 10 SPOD modes. 

\begin{figure}
	\centering
	\subfigure[Mode 1.]{\includegraphics[width=.49\linewidth]{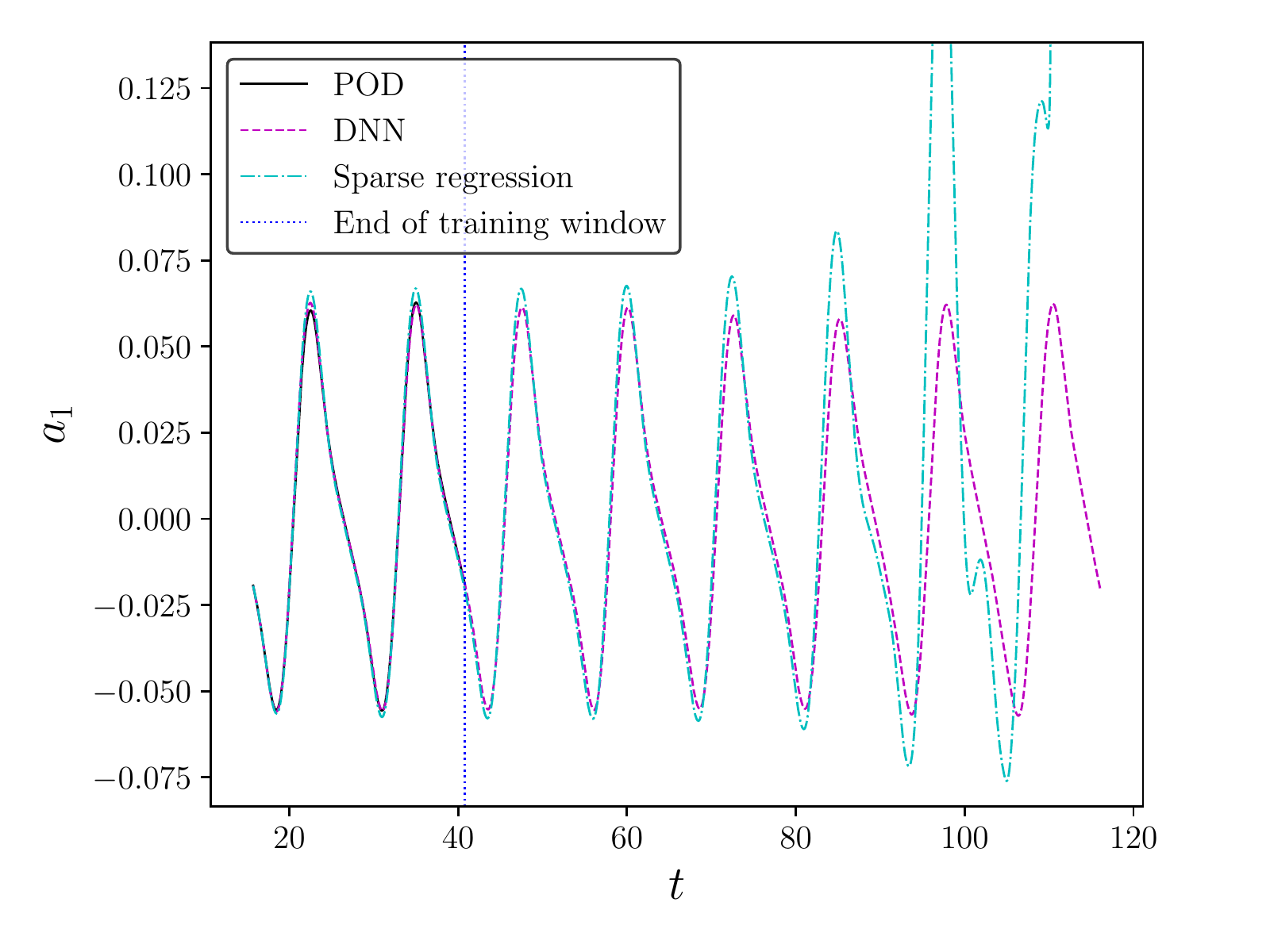}}
	\subfigure[Mode 8.]{\includegraphics[width=.49\linewidth]{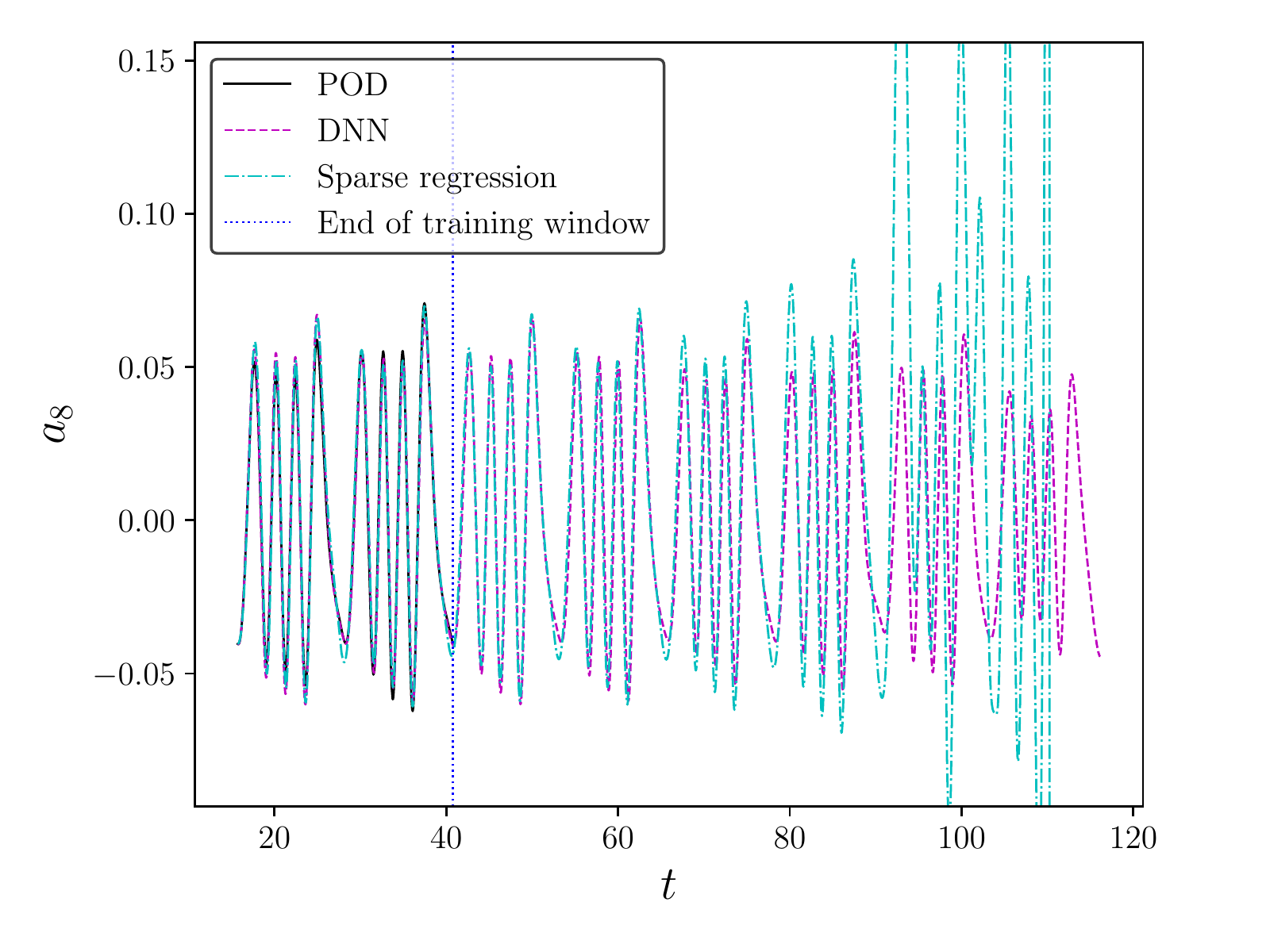}}
	\caption{Reconstruction of SPOD temporal modes using the current DNN approach and the original SINDy algorithm with LASSO.}
	\label{fig:comparison1}
\end{figure}

In figure \ref{fig:comparison1}, one can observe a comparison between reduced order models built using the current DNN approach and sparse regression. For this case, both the LASSO and Ridge regression techniques were tested with the original SINDy algorithm. Although the LASSO presented a higher computational cost, it provided better models compared to the Ridge regression. For this case, the computational cost of the DNN regression is, on average, 40 times higher than that of SINDy. However, as shown for SPOD modes 1 and 8, the SINDy algorithm was not able to provide stable models. The same can be said for the other SPOD modes. Sparse regression can learn the dynamics during the training window and also for a few cycles in the test set but its solution is not stable for long-time predictions. On the other hand, despite the higher cost, the DNN approach presents good long-time predictive capabilities with stable and accurate solutions beyond the training window.

\begin{figure}
	\centering
	\subfigure[Mode 1 obtained using $\Delta t$ as in the simulation.]{\includegraphics[width=.49\linewidth]{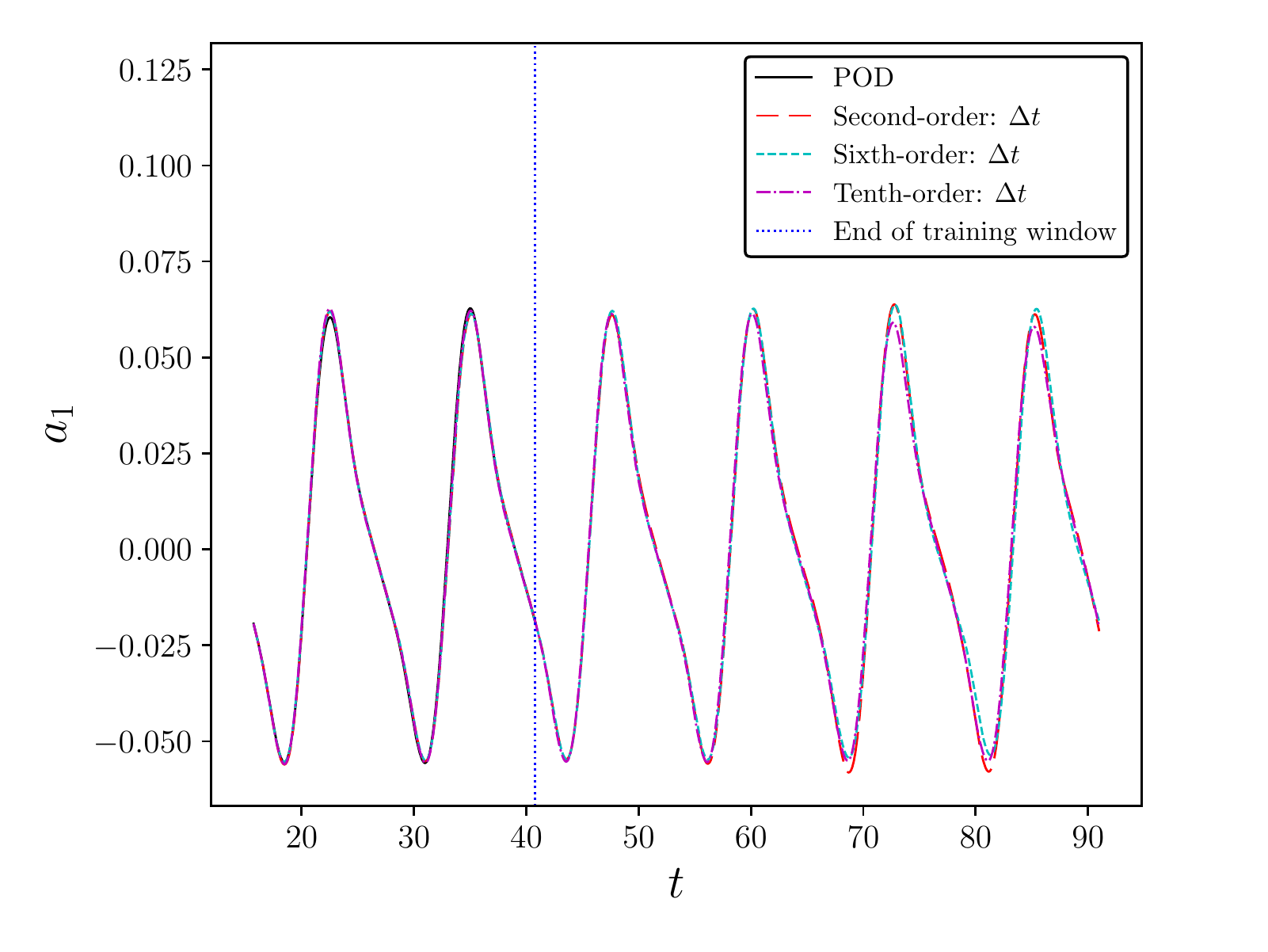}}
	\subfigure[Mode 1 obtained using $5 \times \Delta t$.]{\includegraphics[width=.49\linewidth]{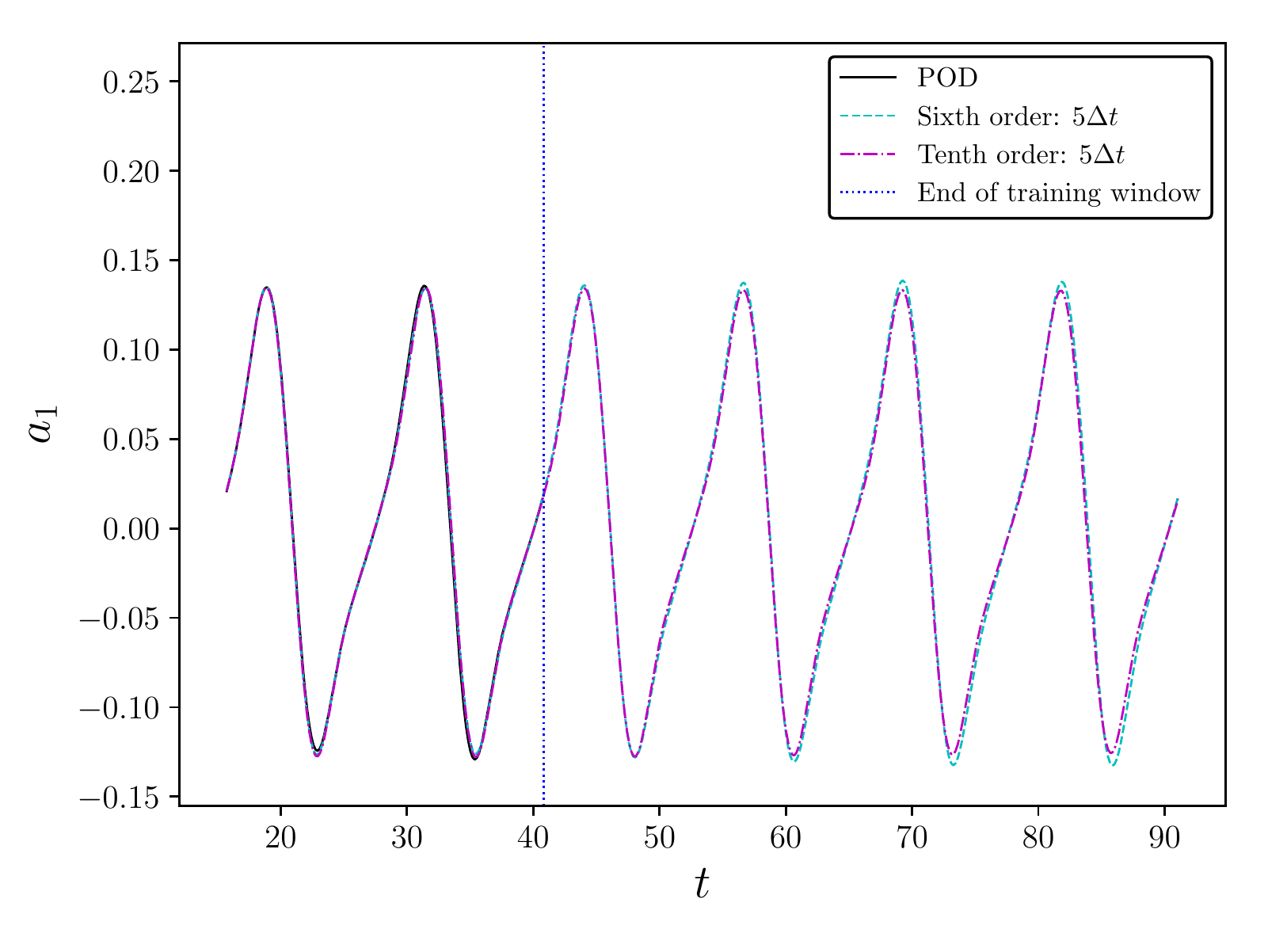}}
	\subfigure[Mode 8 obtained using $\Delta t$ as in the simulation.]{\includegraphics[width=.49\linewidth]{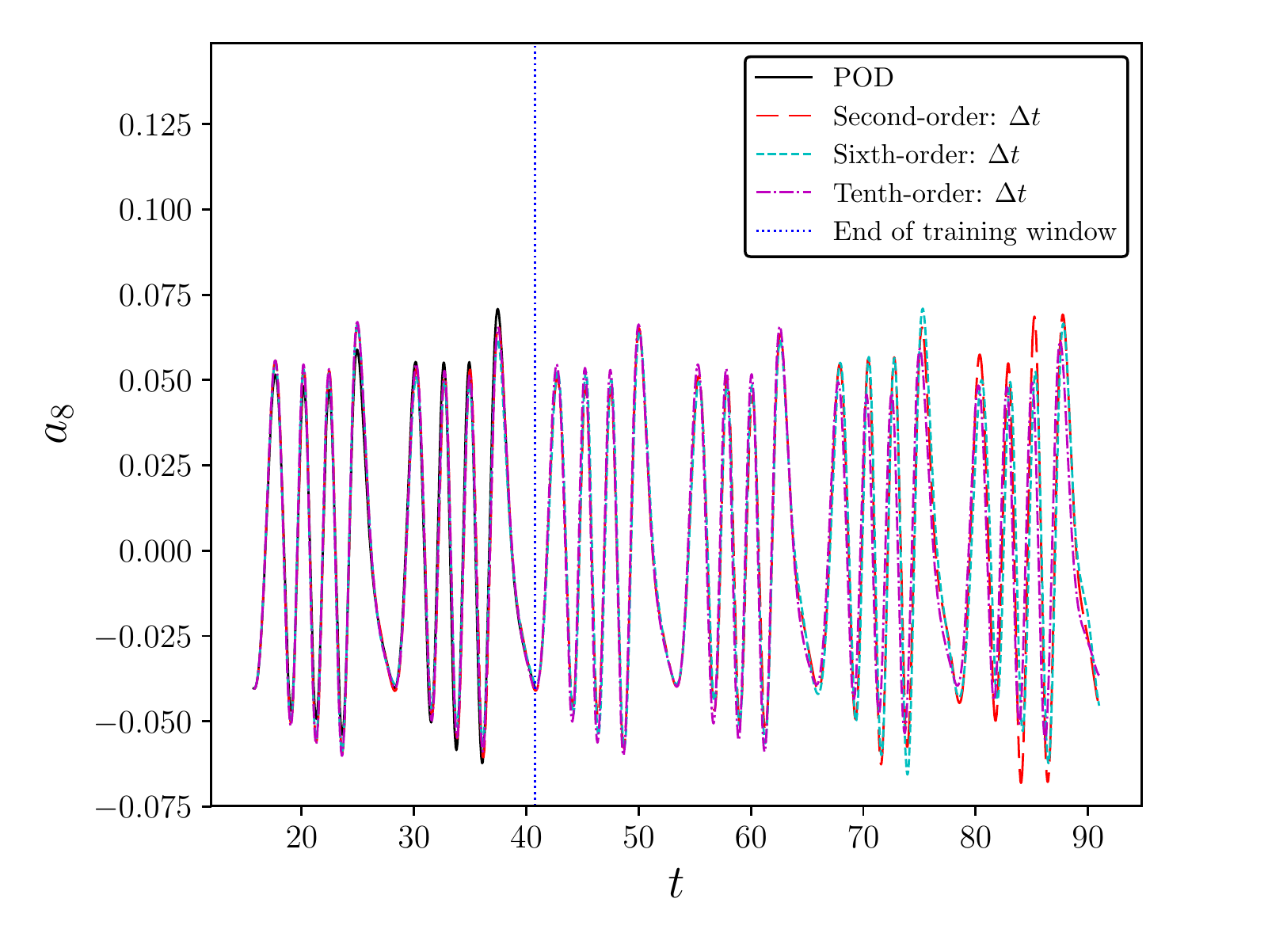}}
	\subfigure[Mode 8 obtained using $5 \times \Delta t$.]{\includegraphics[width=.49\linewidth]{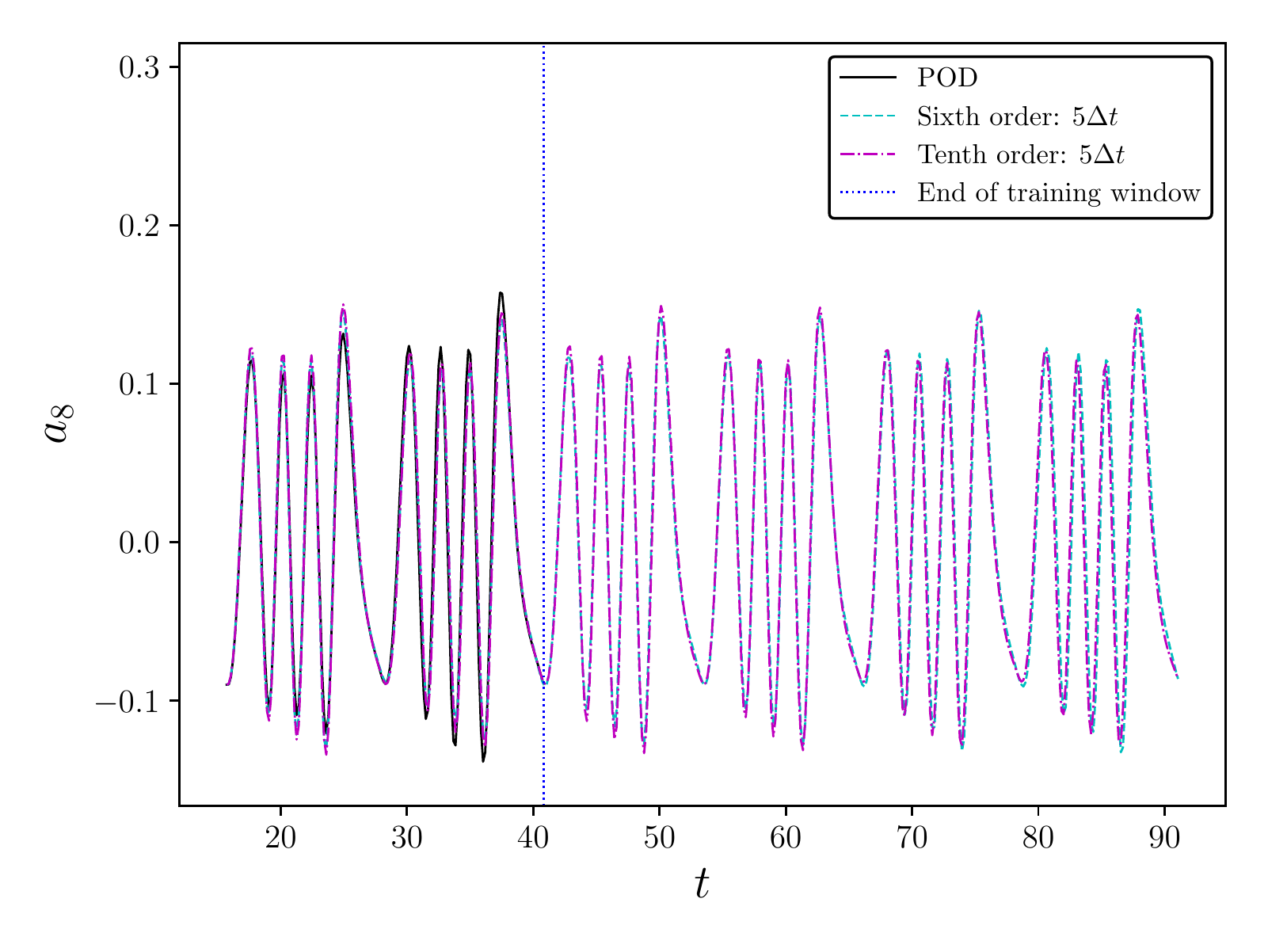}}
	\caption{Reconstruction of SPOD temporal modes using different finite difference schemes and time steps.}
	\label{fig:15}
\end{figure}

In order to evaluate the robustness of the present framework, we test the capability of the ROM in reconstructing the flowfield using a time step different than that of the simulation. Figure \ref{fig:15} shows the reconstruction of SPOD temporal modes using different finite difference schemes and time steps. The derivatives of the temporal modes are computed via an explicit second order scheme and by sixth and tenth order compact schemes. When the reduced order model is constructed using the same time step of the numerical simulation, the first 10 SPOD modes are highly resolved in time. For this case, all three schemes for computing the temporal derivatives provide accurate results and the random search is able to find a set of hyperparameters which results in an accurate ROM. However, when the DNN is trained using SPOD modes obtained by a time step five times larger than that of the simulation, the resolution of the higher POD temporal modes can become compromised. Then, it is important to use the compact schemes to obtain accurate representations of the derivatives. In this case, the random search could only find suitable ROMs using the compact schemes. As shown in figure \ref{fig:15}, the models are still stable and reproduce the same dynamics observed with lower time steps. This shows that the current ROM framework is robust and can be constructed using a subset of the data of the full order model, with lower temporal resolution.

\begin{figure}
	\centering
	\subfigure[Lift coefficient $C_L$.]{\includegraphics[trim = 1mm 1mm 1mm 1mm, clip,width=.49\linewidth]{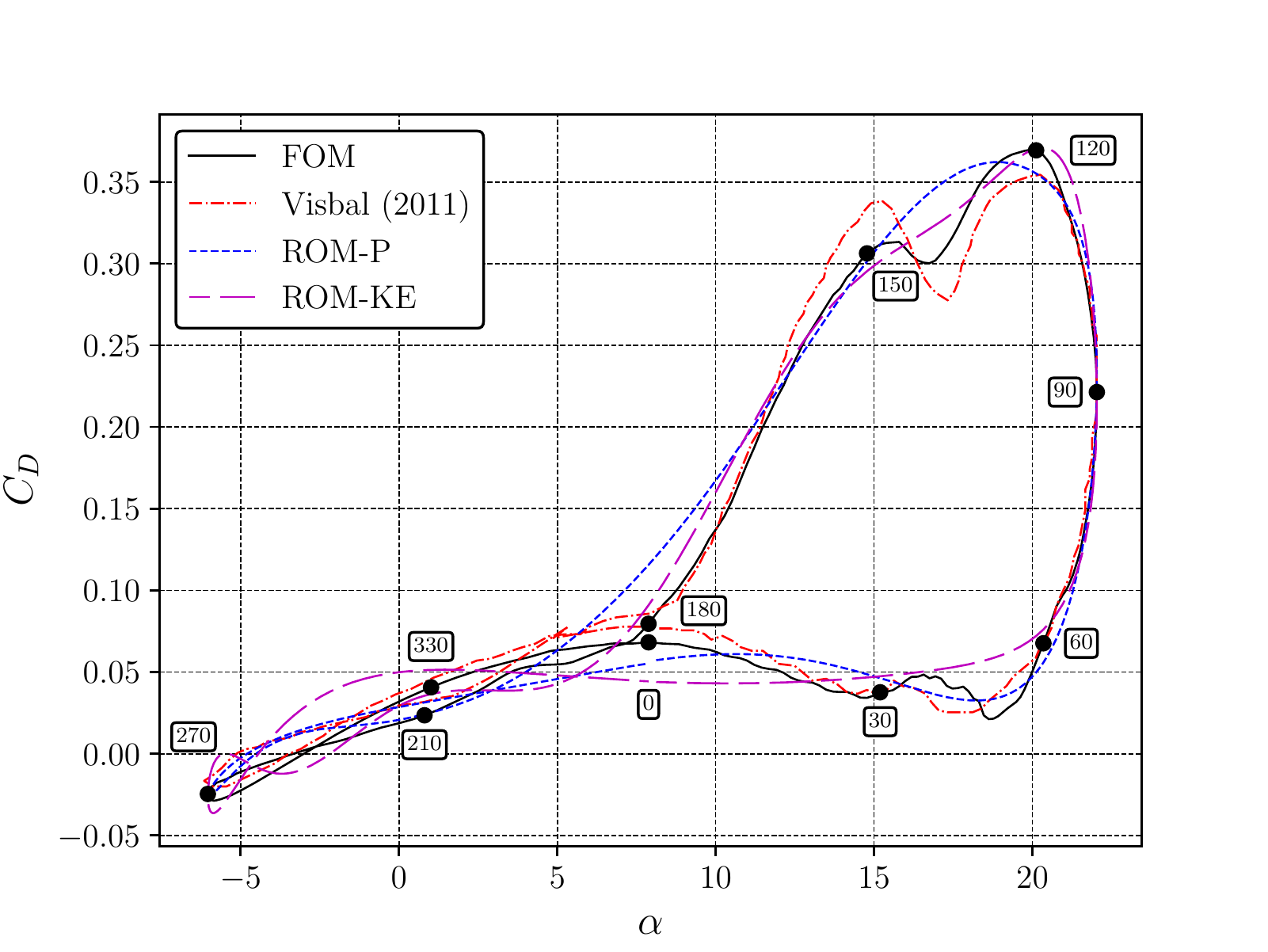}}
	\subfigure[Drag coefficient $C_D$.]{\includegraphics[trim = 1mm 1mm 1mm 1mm, clip,width=.49\linewidth]{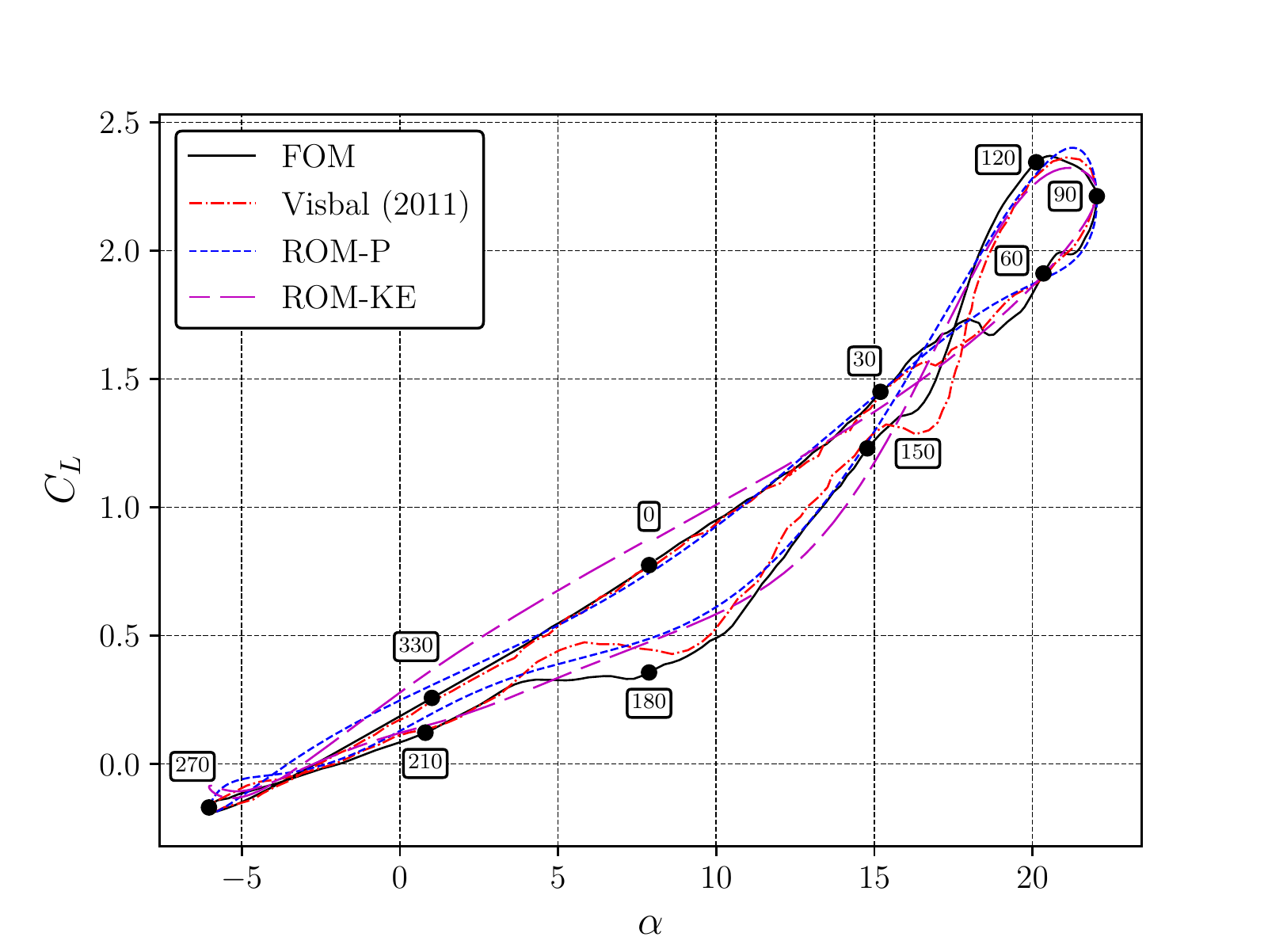}}
	\caption{Phase-averaged aerodynamic coefficients computed by \citet{Visbal:2011}, the current LES (FOM), and the DNN approach with POD norms based on pressure (P) and kinetic energy (KE).}
	\label{fig:16}
\end{figure}

Finally, we compute the phase-averaged aerodynamic coefficients for the plunging SD7003 airfoil. Lift and drag coefficients are shown in figure \ref{fig:16} for the FOM and ROM using the six cycles computed in the training and test sets. Both the pressure and friction forces are accounted for in the calculation of the aerodynamic coefficients. However, for this case, the pressure component dominates the aerodynamic loads. The effective incidence angles are depicted in the figures and the aerodynamic coefficients are calculated using the 10th order compact scheme for the derivatives of the temporal modes and the first 10 SPOD modes. As a validation of the current solutions, the results from \citet{Visbal:2011} are also shown. Resuls are presented for the reduced order models computed using both the pressure and kinetic energy norms. The DNNs could find stable and accurate models for both POD norms and the aerodynamic coefficients obtained for the best models are compared in the figure. As one can see, the present ROM solutions show good comparisons to the current LES and \citet{Visbal:2011}.
The total simulation cost of the FOM for this case was 100,000 hours. On the other hand, the full cost of the ROM, including the optimization of the hyperparameters, plus the training and evaluation of 700 models, required approximately 7 hours. Hence, the computational cost for the training procedure is around 100 models per hour in a single GPU. Once a reduced order model is chosen, the cost of the simulation is reduced to a few seconds.

\section{Conclusions}

We present a methodology for constructing reduced order models combining flow modal decomposition and regression analysis via deep feedforward neural networks (DNNs). The framework is implemented in a context similar to that of the sparse identification of non-linear dynamics (SINDy) algorithm recently proposed in the literature. 
The details of the methodology are described including algorithm charts which facilitate the understanding and implementation of the proposed framework. The source code can be downloaded from \url{<http://cces.unicamp.br/software/>}.

The method is tested for different problems involving nonlinear dynamical systems: the canonical nonlinear damped oscillator, the flow past a circular cylinder at a low Reynolds number and the turbulent flow past a plunging SD7003 airfoil under deep dynamic stall. For the previous two cases, the compressible Navier Stokes equations are solved in full contravariant form and additional non-inertial terms are added to the equations to simulate the airfoil plunging motion. Numerical simulations are performed using a high order compact finite difference flow solver. Then, the high fidelity results from the simulations are used as the input data for the construction of the reduced order models. 

To create stable and accurate ROMs of more complex flows, the application of the spectral proper orthogonal decomposition was found necessary. Hence, in order to filter high frequency content present in the temporal modes of the classical snapshot POD, we apply a filter function to the correlation matrix (SPOD approach). The proposed numerical framework allows the prediction of the flowfield beyond the training window using larger time increments than those employed by the full order model, which demonstrates the robustness of the current ROMs constructed via deep feedforward neural networks. The resolution of the numerical schemes used for computation of the POD temporal mode derivatives is shown to have an important role when larger time increments are employed in the construction of the reduced order models. In this case, the higher POD modes are composed by a broad range of frequencies and the accurate representation of the temporal derivatives is crucial for obtaining ROMs via regression analysis.

A discussion on the optimization of hyperparameters for obtaining the best ROMs via deep neural networks is provided together with a description of the effects of individual hyperparameters on model performance. Here, we test the random search and the Bayesian optimization, which are the most widely used procedures for hyperparameter optimization. To report the performance of each model from a set of parameters, we compute the mean absolute error (MAE) over the training data and choose the candidate models with lower MAE values to provide the best parameters. Following this approach, the random search produced the best models at lower computational costs for all cases investigated in this work. It is worth mentioning that, in order to reduce the set of parameters for optimization, we first selected the hyperparameters related to the optimization step, such as the learning rate and the number of iterations, by manual search. We also choose the exponential linear unit activation function over the hyperbolic tangent function for the current problems since it provided results with fewer iterations.

Using 10 POD modes, 99\% of the flow energy is recovered in the cylinder flow study. The ROM obtained for this case shows an excellent agreement with the FOM. Even when 2 POD modes are employed in the reconstruction, the ROM is still stable beyond the training set and captures most of the dynamics of the vortex shedding and sound wave propagation. Both the DNN and the original SINDy approaches are tested for the reconstruction of the transient regime of an incompressible flow past a cylinder. In this case, regression is performed for the 2 most energetic POD modes of the flow, obtained from \cite{Brunton3932}. The DNN model is able to accurately recover both the transient and limit cycle solutions of the test data for both POD modes. On the other hand, sparse regression reconstructs the transient portion of the dynamics but not the long term prediction of the limit cycle.

In the dynamic stall configuration, the flow is turbulent along the airfoil suction side, mostly during the downstroke motion. The complex flow dynamics of this case exhibit unpaired POD modes with high frequency noise. We show how the SPOD approach modifies the temporal modes for this study. Reduced order models are constructed both for the full 3D and the spanwise-averaged flow solutions. In both cases, the ROMs are able to capture  the dynamics of the leading edge stall vortex, including its formation, transport and ejection, and of the trailing edge vortex. 
The total simulation cost of the FOM for this case was 100,000 hours. On the other hand, the full cost of the ROM was approximately 7 hours in a single GPU. This cost already includes the optimization of hyperparameters, plus the training and evaluation of 700 models. When the best model is chosen, the cost of the simulation is reduced to a few seconds.
The data obtained by the DNN-ROMs were used to compute aerodynamic coefficients of the dynamic stall and good agreement was found compared to the LES used as full order model.
Again, we compare models obtained by DNNs and sparse regression and show that, despite the higher computational cost, the DNN approach presents good long-time predictive capabilities with stable and accurate solutions beyond the training window. On the other hand, the best model obtained from sparse regression could learn the dynamics during the training window and also for a few cycles in the test set but its solution was not stable for long-time predictions. We expect that, in a future work, the current methodology can be further improved for applications in flow control and extrapolation of flow configurations other than those used for training. 

\section*{Acknowledgements}

The authors acknowledge the financial support received from Funda\c{c}\~ao de Amparo \`a Pesquisa do Estado de S\~ao Paulo, FAPESP, under Grants No.\ 2013/08293-7 and No.\ 2013/07375-0, and from Conselho Nacional de Desenvolvimento Cient\'ifico e Tecnol\'ogico, CNPq, under Grants No.\ 304335/2018-5 and 407842/2018-7. The support provided by CNPq through a scholarship for the first author is likewise gratefully appreciated. We also thank CEPID-CeMEAI for providing the computational resources used in this work. We gratefully acknowledge Brener D'Lelis and Tulio Ricciardi for providing the high fidelity dynamic stall and cylinder data employed in the current work. 

\bibliographystyle{jfm}
\bibliography{JFM_Lui_Wolf}

\begin{thebibliography}{63}
\expandafter\ifx\csname natexlab\endcsname\relax\def\natexlab#1{#1}\fi
\def\au#1{#1} \def\ed#1{#1} \def\yr#1{#1}\def\at#1{#1}\def\jt#1{\textit{#1}}
  \def\bt#1{#1}\def\bvol#1{\textbf{#1}} \def\vol#1{#1} \def\pg#1{#1}
  \def\publ#1{#1}\def\arxiv#1{#1}\def\org#1{#1}\def\st#1{\textit{#1}}

\bibitem[Abadi \& et~al.(2015)]{tensorflow2015-whitepaper}
{\sc \au{Abadi, M.} \& \au{et~al.}} \yr{2015} {TensorFlow}: Large-scale machine
  learning on heterogeneous systems. Software available from tensorflow.org.

\bibitem[Akaike(1973)]{akaike}
{\sc \au{Akaike, H.}} \yr{1973} {\em Information Theory and an Extension of the
  Maximum Likelihood Principle\/},  \pg{pp. 199--213}.  \publ{New York, NY:
  Springer New York}.

\bibitem[Aris(1989)]{Aris:1989}
{\sc \au{Aris, R.}} \yr{1989} {\em Vectors, Tensors, and the Basic Equations of
  Fluid Mechanics\/}.  \publ{Dover Publications}.

\bibitem[Baik {\em et~al.\/}(2009)Baik, Rausch, Bernal \& Ol]{Baik:2009}
{\sc \au{Baik, Y.~S.}, \au{Rausch, J.}, \au{Bernal, L.} \& \au{Ol, M.~V.}}
  \yr{2009} Experimental investigation of pitching and plunging airfoils at
  reynolds number between 1 x 10\textsuperscript{4} and 6 x
  10\textsuperscript{4}.  \bt{In {\em AIAA Paper\/}}.

\bibitem[Balajewicz {\em et~al.\/}(2016)Balajewicz, Tezaur \&
  Dowell]{Balajewicz2016}
{\sc \au{Balajewicz, M.}, \au{Tezaur, I.} \& \au{Dowell, E.}} \yr{2016}
  \at{Minimal subspace rotation on the {S}tiefel manifold for stabilization and
  enhancement of projection-based reduced order models for the compressible
  {N}avier-{S}tokes equations}.  \jt{Journal of Computational Physics}
  \bvol{321},  \pg{224--241}.

\bibitem[Beam \& Warming(1978)]{Beam1978}
{\sc \au{Beam, R.~M} \& \au{Warming, R.~F}} \yr{1978}  \at{An implicit factored
  scheme for the compressible navier-stokes equations}.  \jt{AIAA Journal}
  \bvol{16}~(4),  \pg{393--402}.

\bibitem[Bengio(2009)]{Bengio}
{\sc \au{Bengio, Y.}} \yr{2009}  \at{Learning deep architectures for ai}.
  \jt{Found. Trends Mach. Learn.}  \bvol{2}~(1),  \pg{1--127}.

\bibitem[Bergstra \& Bengio(2012)]{Bergstra}
{\sc \au{Bergstra, J.} \& \au{Bengio, Y.}} \yr{2012}  \at{Random search for
  hyper-parameter optimization}.  \jt{J. Mach. Learn. Res.}  \bvol{13},
  \pg{281--305}.

\bibitem[Brochu {\em et~al.\/}(2010)Brochu, Cora \& Freitas]{bayesian}
{\sc \au{Brochu, E.}, \au{Cora, V.~M.} \& \au{Freitas, N.}} \yr{2010}  \at{A
  tutorial on bayesian optimization of expensive cost functions, with
  application to active user modeling and hierarchical reinforcement learning}.
   \jt{CoRR}  \bvol{abs/1012.2599},  \arxiv{arXiv: 1012.2599}.

\bibitem[Brunton \& Noack(2015)]{BruntonControl}
{\sc \au{Brunton, S.~L.} \& \au{Noack, B.~R.}} \yr{2015}  \at{Closed-loop
  turbulence control: Progress and challenges}.  \jt{Applied Mechanics Reviews}
   \bvol{67},  \pg{050801}.

\bibitem[Brunton {\em et~al.\/}(2016)Brunton, Proctor \& Kutz]{Brunton3932}
{\sc \au{Brunton, S.~L.}, \au{Proctor, J.~L.} \& \au{Kutz, N.~J.}} \yr{2016}
  \at{Discovering governing equations from data by sparse identification of
  nonlinear dynamical systems}.  \jt{Proceedings of the National Academy of
  Sciences}  \bvol{113}~(15),  \pg{3932--3937},  \arxiv{arXiv:
  http://www.pnas.org/content/113/15/3932.full.pdf}.

\bibitem[Carlberg {\em et~al.\/}(2011)Carlberg, Bou-Mosleh \&
  Farhat]{Carlberg2011}
{\sc \au{Carlberg, K.}, \au{Bou-Mosleh, C.} \& \au{Farhat, C.}} \yr{2011}
  \at{Efficient non-linear model reduction via a least-squares
  petrov–galerkin projection and compressive tensor approximations}.
  \jt{International Journal for Numerical Methods in Engineering}
  \bvol{86}~(2),  \pg{155--181},  \arxiv{arXiv:
  https://onlinelibrary.wiley.com/doi/pdf/10.1002/nme.3050}.

\bibitem[Cazemier {\em et~al.\/}(1998)Cazemier, Verstappen \&
  Veldman]{Cazemier1998}
{\sc \au{Cazemier, W.}, \au{Verstappen, R. W. C.~P.} \& \au{Veldman, A. E.~P.}}
  \yr{1998}  \at{Proper orthogonal decomposition and low-dimensional models for
  driven cavity flows}.  \jt{Physics of Fluids}  \bvol{10}~(7),  \pg{1685}.

\bibitem[Chaturantabut \& Sorensen(2010)]{Saifon:2010}
{\sc \au{Chaturantabut, S.} \& \au{Sorensen, D.}} \yr{2010}  \at{Nonlinear
  model reduction via discrete empirical interpolation}.  \jt{SIAM Journal of
  Scientific Computing}  \bvol{32},  \pg{2737--2764}.

\bibitem[Clainche \& Vega(2017)]{HODMD}
{\sc \au{Clainche, S.} \& \au{Vega, J.~M.}} \yr{2017}  \at{Higher order dynamic
  mode decomposition to identify and extrapolate flow patterns}.  \jt{Physics
  of Fluids}  \bvol{29},  \pg{084102}.

\bibitem[Clevert {\em et~al.\/}(2015)Clevert, Unterthiner \&
  Hochreiter]{Clevert}
{\sc \au{Clevert, D.}, \au{Unterthiner, T.} \& \au{Hochreiter, S.}} \yr{2015}
  \at{Fast and accurate deep network learning by exponential linear units
  (elus)}.  \jt{CoRR}  \bvol{abs/1511.07289},  \arxiv{arXiv: 1511.07289}.

\bibitem[D'Lelis {\em et~al.\/}(2019)D'Lelis, Wolf, Yeh \& Taira]{Brener:2019}
{\sc \au{D'Lelis, B. O.~R.}, \au{Wolf, W.~R.}, \au{Yeh, C.} \& \au{Taira, K.}}
  \yr{2019} Modal analysis of deep dynamic stall for a plunging airfoil.
  \bt{In {\em AIAA Scitech 2019\/}}.  \publ{AIAA}.

\bibitem[Goodfellow {\em et~al.\/}(2016)Goodfellow, Bengio \&
  Courville]{Goodfellow-et-al}
{\sc \au{Goodfellow, I.}, \au{Bengio, Y.} \& \au{Courville, A.}} \yr{2016} {\em
  Deep Learning\/}.  \publ{The MIT Press}.

\bibitem[Green {\em et~al.\/}(2007)Green, Rowley \& Haller]{GreenLCS}
{\sc \au{Green, M.~A.}, \au{Rowley, C.~W.} \& \au{Haller, G.}} \yr{2007}
  \at{Detecting vortex formation and shedding in cylinder wakes using
  lagrangian coherent structures}.  \jt{Journal of Fluid Mechanics}
  \bvol{572},  \pg{111--120}.

\bibitem[Haller(2015)]{Haller2015}
{\sc \au{Haller, G.}} \yr{2015}  \at{Lagrangian coherent structures}.
  \jt{Annual Review of Fluid Mechanics}  \bvol{47},  \pg{137--162}.

\bibitem[Juang \& Pappa(1985)]{ERA1985}
{\sc \au{Juang, J.~N.} \& \au{Pappa, R.~S.}} \yr{1985}  \at{An eigensystem
  realization algorithm for modal parameter identification and model
  reduction}.  \jt{Journal of Guidance, Control, and Dynamics}  \bvol{8},
  \pg{620--627}.

\bibitem[Kennedy {\em et~al.\/}(1999)Kennedy, Carpenter \& Lewis]{Kennedy1999}
{\sc \au{Kennedy, C.~A.}, \au{Carpenter, M.~H.} \& \au{Lewis, R.~M.}} \yr{1999}
  Low-storage, explicit {R}unge-{K}utta schemes for the compressible
  {N}avier-{S}tokes equations. ICASE Report No. 99-22.

\bibitem[Kingma \& Ba(2014)]{Adam}
{\sc \au{Kingma, D.~P.} \& \au{Ba, J.}} \yr{2014}  \at{Adam: {A} method for
  stochastic optimization}.  \jt{Proceedings of the 3rd International
  Conference on Learning Representations (ICLR)} ,  \arxiv{arXiv: 1412.6980}.

\bibitem[Krizhevsky(2009)]{Krizhevsky}
{\sc \au{Krizhevsky, A.}} \yr{2009}  \bt{Learning multiple layers of features
  from tiny images}.  \org{{\em Tech. Rep.\/}}.

\bibitem[Kutz(2017)]{Kutz2017}
{\sc \au{Kutz, J.}} \yr{2017}  \at{Deep learning in fluid dynamics}.
  \jt{Journal of Fluid Mechanics}  \bvol{814},  \pg{1–4}.

\bibitem[Lele(1992)]{Lele1992}
{\sc \au{Lele, S.~K.}} \yr{1992}  \at{Compact finite difference schemes with
  spectral-like resolution}.  \jt{Journal of Computational Physics}
  \bvol{103}~(1),  \pg{16--42}.

\bibitem[Ling {\em et~al.\/}(2016)Ling, Kurzawski \& Templeton]{Ling:2016}
{\sc \au{Ling, J.}, \au{Kurzawski, A.} \& \au{Templeton, J.}} \yr{2016}
  \at{Reynolds averaged turbulence modelling using deep neural networks with
  embedded invariance}.  \jt{Journal of Fluid Mechanics}  \bvol{807},
  \pg{155--166}.

\bibitem[Ling \& Templeton(2015)]{Ling:2015}
{\sc \au{Ling, J.} \& \au{Templeton, J.}} \yr{2015}  \at{Evaluation of machine
  learning algorithms for prediction of regions of high {R}eynolds averaged
  {N}avier {S}tokes uncertainty}.  \jt{Physics of Fluids}  \bvol{27},
  \pg{085103}.

\bibitem[Luhar {\em et~al.\/}(2014)Luhar, Sharma \& McKeon]{Mckeon2014}
{\sc \au{Luhar, M.}, \au{Sharma, A.~S.} \& \au{McKeon, B.~J.}} \yr{2014}
  \at{On the structure and origin of pressure fluctuations in wall turbulence:
  predictions based on the resolvent analysis}.  \jt{Journal of Fluid
  Mechanics}  \bvol{751},  \pg{38--70}.

\bibitem[Lumley(1967)]{Lumley1967}
{\sc \au{Lumley, J.~L.}} \yr{1967}  \at{The structure of inhomogeneous
  turbulence}.  \bt{In {\em Atmospheric Turbulence and Wave Propagation\/}},
  \pg{pp. 166--178}.  \publ{Moscow: Nauka}.

\bibitem[Nagarajan {\em et~al.\/}(2003)Nagarajan, Lele \&
  Ferziger]{Nagarajan2003}
{\sc \au{Nagarajan, S.}, \au{Lele, S.~K.} \& \au{Ferziger, J.~H.}} \yr{2003}
  \at{A robust high-order compact method for large eddy simulation}.
  \jt{Journal of Computational Physics}  \bvol{191}~(2),  \pg{392--419}.

\bibitem[Nair \& Hinton(2010)]{RELU}
{\sc \au{Nair, V.} \& \au{Hinton, G.~E.}} \yr{2010} Rectified linear units
  improve restricted boltzmann machines.  \bt{In {\em Proceedings of the 27th
  International Conference on International Conference on Machine Learning\/}},
   \pg{pp. 807--814}.  \publ{USA: Omnipress}.

\bibitem[Noack {\em et~al.\/}(2003)Noack, Afanasiev, Morzynski, Tadmor \&
  Thiele]{Noack:2003}
{\sc \au{Noack, B.~R.}, \au{Afanasiev, K.}, \au{Morzynski, M.}, \au{Tadmor, G.}
  \& \au{Thiele, F.}} \yr{2003}  \at{A hierarchy of low-dimensional models for
  the transient and post-transient cylinder wake}.  \jt{Journal of Fluid
  Mechanics}  \bvol{497},  \pg{335--363}.

\bibitem[Ol {\em et~al.\/}(2009)Ol, Bernal, Kang \& Shyy]{Ol:2009}
{\sc \au{Ol, M.~V.}, \au{Bernal, L.}, \au{Kang, C.} \& \au{Shyy, W.}} \yr{2009}
   \at{Shallow and deep dynamic stall for flapping low reynolds number
  airfoils}.  \jt{Experiments in Fluids}  \bvol{46}~(5),  \pg{883--901}.

\bibitem[Olson(2012)]{Britton}
{\sc \au{Olson, B.~J.}} \yr{2012}  \at{Large-eddy simulation of multi-material
  mixing and over-expanded nozzle flow}. PhD thesis, Stanford University.

\bibitem[{\" O}sth {\em et~al.\/}(2014){\" O}sth, Noack, Krajnović, Barros \&
  Bor{\' e}e]{Osth2014}
{\sc \au{{\" O}sth, J.}, \au{Noack, B.~R.}, \au{Krajnović, S.}, \au{Barros,
  D.} \& \au{Bor{\' e}e, J.}} \yr{2014}  \at{On the need for a nonlinear
  subscale turbulence term in {POD} models as exemplified for a
  high-{R}eynolds-number flow over an {A}hmed body}.  \jt{Journal of Fluid
  Mechanics}  \bvol{747},  \pg{518--544}.

\bibitem[{Pan} \& {Duraisamy}(2018)]{Karthik2018}
{\sc \au{{Pan}, S.} \& \au{{Duraisamy}, K.}} \yr{2018}  \at{{Long-time
  predictive modeling of nonlinear dynamical systems using neural networks}}.
  \jt{ArXiv e-prints} ,  \arxiv{arXiv: 1805.12547}.

\bibitem[Protas {\em et~al.\/}(2015)Protas, Noack \& {\" O}sth]{Protas2015}
{\sc \au{Protas, B.}, \au{Noack, B.~R.} \& \au{{\" O}sth, J.}} \yr{2015}
  \at{Optimal nonlinear eddy viscosity in {G}alerkin models of turbulent
  flows}.  \jt{Journal of Fluid Mechanics}  \bvol{766},  \pg{337--367}.

\bibitem[Ribeiro \& Wolf(2017)]{Jean2017}
{\sc \au{Ribeiro, J. H.~M.} \& \au{Wolf, W.~R.}} \yr{2017}  \at{Identification
  of coherent structures in the flow past a naca0012 airfoil via proper
  orthogonal decomposition}.  \jt{Physics of Fluids}  \bvol{29}~(8),
  \pg{085104}.

\bibitem[Rockwood {\em et~al.\/}(2016)Rockwood, Taira \& Green]{TairaLCS}
{\sc \au{Rockwood, M.~P.}, \au{Taira, K.} \& \au{Green, M.~A.}} \yr{2016}
  \at{Detecting vortex formation and shedding in cylinder wakes using
  lagrangian coherent structures}.  \jt{AIAA Journal}  \bvol{55},  \pg{15--23}.

\bibitem[Rowley {\em et~al.\/}(2004)Rowley, Colonius \& Murray]{Rowley2004}
{\sc \au{Rowley, C.~W.}, \au{Colonius, T.} \& \au{Murray, R.~M.}} \yr{2004}
  \at{Model reduction for compressible flows using {POD} and galerkin
  projection}.  \jt{Physica D: Nonlinear Phenomena}  \bvol{189}~(1–2),
  \pg{115 -- 129}.

\bibitem[Rudy {\em et~al.\/}(2017)Rudy, Brunton, Proctor \& Kutz]{Rudy2017}
{\sc \au{Rudy, S.~H.}, \au{Brunton, S.~L.}, \au{Proctor, J.~L.} \& \au{Kutz,
  N.~J.}} \yr{2017}  \at{Data-driven discovery of partial differential
  equations}.  \jt{Science Advances}  \bvol{3}~(4),  \arxiv{arXiv:
  http://advances.sciencemag.org/content/3/4/e1602614.full.pdf}.

\bibitem[{Rudy} {\em et~al.\/}(2018){Rudy}, {Kutz} \& {Brunton}]{Rudy2018}
{\sc \au{{Rudy}, S.~H.}, \au{{Kutz}, J.~N.} \& \au{{Brunton}, S.~L.}} \yr{2018}
   \at{{Deep learning of dynamics and signal-noise decomposition with
  time-stepping constraints}}.  \jt{ArXiv e-prints} ,  \arxiv{arXiv:
  1808.02578}.

\bibitem[San \& Maulik(2018)]{San2018}
{\sc \au{San, O.} \& \au{Maulik, R.}} \yr{2018}  \at{Extreme learning machine
  for reduced order modeling of turbulent geophysical flows}.  \jt{Phys. Rev.
  E}  \bvol{97},  \pg{042322}.

\bibitem[Schmid(2010)]{Schmid2010}
{\sc \au{Schmid, P.~J.}} \yr{2010}  \at{Dynamic mode decomposition of numerical
  and experimental data}.  \jt{Journal of Fluid Mechanics}  \bvol{656},
  \pg{5–28}.

\bibitem[Schwarz(1978)]{schwarz}
{\sc \au{Schwarz, G.}} \yr{1978}  \at{Estimating the dimension of a model}.
  \jt{The Annals of Statistics}  \bvol{6},  \pg{461--464}.

\bibitem[Sharma {\em et~al.\/}(2016)Sharma, Moarref, McKeon, Park, Graham \&
  Willis]{Mckeon2016}
{\sc \au{Sharma, A.~S.}, \au{Moarref, R.}, \au{McKeon, B.~J.}, \au{Park,
  J.~S.}, \au{Graham, M.~D.} \& \au{Willis, A.~P.}} \yr{2016}
  \at{Low-dimensional representations of exact coherent states of the
  navier-stokes equations from the resolvent model of wall turbulence}.
  \jt{Physical Review E}  \bvol{93},  \pg{021102}.

\bibitem[Sieber {\em et~al.\/}(2016)Sieber, Paschereit \&
  Oberleithner]{sieber2016}
{\sc \au{Sieber, M.}, \au{Paschereit, C.~O.} \& \au{Oberleithner, K.}}
  \yr{2016}  \at{Spectral proper orthogonal decomposition}.  \jt{Journal of
  Fluid Mechanics}  \bvol{792},  \pg{798–828}.

\bibitem[Sirovich(1986)]{Sirovich1986}
{\sc \au{Sirovich, L.}} \yr{1986}  \at{Turbulence and the dynamics of coherent
  structures. {P}art {I}: Coherent structures}.  \jt{Quarterly of Applied
  Mathematics}  \bvol{45},  \pg{561--571}.

\bibitem[Snoek {\em et~al.\/}(2012)Snoek, Larochelle \& Adams]{Snoek}
{\sc \au{Snoek, J.}, \au{Larochelle, H.} \& \au{Adams, R.~P.}} \yr{2012}
  \at{Practical bayesian optimization of machine learning algorithms}.  \bt{In
  {\em Advances in Neural Information Processing Systems 25\/}},  \pg{pp.
  2951--2959}.  \publ{Curran Associates, Inc.}

\bibitem[Str\"{o}fer {\em et~al.\/}(2019)Str\"{o}fer, Wu, Xiao \&
  Paterson]{Strofer:2019}
{\sc \au{Str\"{o}fer, C.~M.}, \au{Wu, J.~L.}, \au{Xiao, H.} \& \au{Paterson,
  E.}} \yr{2019}  \at{Data-driven, physics based feature extraction from fluid
  flow fields using convolutional neural networks}.  \jt{Communications in
  Computational Physics}  \bvol{25},  \pg{625--650}.

\bibitem[Taira {\em et~al.\/}(2017)Taira, Brunton, Dawson, Rowley, Colonius,
  McKeon, Schmidt, Gordeyev, Theofilis \& Ukeiley]{Taira2017}
{\sc \au{Taira, K.}, \au{Brunton, S.~L.}, \au{Dawson, S. T.~M.}, \au{Rowley,
  C.~W.}, \au{Colonius, T.}, \au{McKeon, B.~J.}, \au{Schmidt, O.~T.},
  \au{Gordeyev, S.}, \au{Theofilis, V.} \& \au{Ukeiley, L.~S.}} \yr{2017}
  \at{Modal analysis of fluid flows: An overview}.  \jt{AIAA Journal}
  \bvol{47},  \pg{1--28}.

\bibitem[Towne {\em et~al.\/}(2018)Towne, Schmidt \& Colonius]{towne2018}
{\sc \au{Towne, A.}, \au{Schmidt, O.~T.} \& \au{Colonius, T.}} \yr{2018}
  \at{Spectral proper orthogonal decomposition and its relationship to dynamic
  mode decomposition and resolvent analysis}.  \jt{Journal of Fluid Mechanics}
  \bvol{847},  \pg{821–867}.

\bibitem[Tu {\em et~al.\/}(2014)Tu, Rowley, Luchtenburg, Brunton \&
  Kutz]{bruntonDMD}
{\sc \au{Tu, J.~H.}, \au{Rowley, C.~W.}, \au{Luchtenburg, D.~M.}, \au{Brunton,
  S.~L.} \& \au{Kutz, N.~J.}} \yr{2014}  \at{On dynamic mode decomposition:
  Theory and applications}.  \jt{Journal of Computational Dynamics}  \bvol{1},
  \pg{391--421}.

\bibitem[Visbal(2011)]{Visbal:2011}
{\sc \au{Visbal, M.~R.}} \yr{2011}  \at{Numerical investigation of deep dynamic
  stall of a plunging airfoil}.  \jt{AIAA Journal}  \bvol{49},
  \pg{2152--2170}.

\bibitem[Vlachas {\em et~al.\/}(2018)Vlachas, Byeon, Wan, Sapsis \&
  Koumoutsakos]{Vlachas2018}
{\sc \au{Vlachas, P.~R.}, \au{Byeon, W.}, \au{Wan, Z.~Y.}, \au{Sapsis, T.~P.}
  \& \au{Koumoutsakos, P.}} \yr{2018}  \at{Data-driven forecasting of
  high-dimensional chaotic systems with long short-term memory networks}.
  \jt{Proceedings of the Royal Society of London A: Mathematical, Physical and
  Engineering Sciences}  \bvol{474}~(2213),  \arxiv{arXiv:
  http://rspa.royalsocietypublishing.org/content/474/2213/20170844.full.pdf}.

\bibitem[Wan {\em et~al.\/}(2018)Wan, Vlachas, Koumoutsakos \& Sapsis]{Wan2018}
{\sc \au{Wan, Z.~Y.}, \au{Vlachas, P.}, \au{Koumoutsakos, P.} \& \au{Sapsis,
  T.}} \yr{2018}  \at{Data-assisted reduced-order modeling of extreme events in
  complex dynamical systems}.  \jt{PLOS ONE}  \bvol{13}~(5),  \pg{1--22}.

\bibitem[Wang {\em et~al.\/}(2017)Wang, Wu \& Xiao]{Wang:2017}
{\sc \au{Wang, J.~X.}, \au{Wu, J.~L.} \& \au{Xiao, H.}} \yr{2017}
  \at{Physics-informed machine learning approach for reconstructing reynolds
  stress modeling discrepancies based on dns data}.  \jt{Physical Review
  Fluids}  \bvol{2},  \pg{034603}.

\bibitem[Wolf(2011)]{Wolf:2011}
{\sc \au{Wolf, W.~R.}} \yr{2011}  \at{Airfoil aeroacoustics: Les and acoustic
  analogy predictions}. PhD thesis, Stanford University.

\bibitem[Wolf {\em et~al.\/}(2012{\natexlab{{\em a\/}}})Wolf, Azevedo \&
  Lele]{Wolf2012}
{\sc \au{Wolf, W.~R.}, \au{Azevedo, J. L.~F.} \& \au{Lele, S.~K.}}
  \yr{2012{\natexlab{{\em a\/}}}}  \at{Convective effects and the role of
  quadrupole sources for aerofoil aeroacoustics}.  \jt{Journal of Fluid
  Mechanics}  \bvol{708},  \pg{502--538}.

\bibitem[Wolf {\em et~al.\/}(2012{\natexlab{{\em b\/}}})Wolf, Lele,
  Jothiprasard \& Cheung]{Wolf:DU96}
{\sc \au{Wolf, W.~R.}, \au{Lele, S.~K.}, \au{Jothiprasard, G.} \& \au{Cheung,
  L.}} \yr{2012{\natexlab{{\em b\/}}}} Investigation of noise generated by a
  du96 airfoil.  \bt{In {\em 18th AIAA/CEAS Aeroacoustics Conference (33th AIAA
  Aeroacoustics Conference)\/}}.

\bibitem[Xavier \& Yoshua(2010)]{Xavier}
{\sc \au{Xavier, G.} \& \au{Yoshua, B.}} \yr{2010} Understanding the difficulty
  of training deep feedforward neural networks.  \bt{In {\em Proceedings of the
  Thirteenth International Conference on Artificial Intelligence and
  Statistics\/} (ed. \ed{Yee~Whye Teh \& Mike Titterington})},  \st{Proceedings
  of Machine Learning Research},  \vol{vol.~9},  \pg{pp. 249--256}.  \publ{Chia
  Laguna Resort, Sardinia, Italy: PMLR}.

\bibitem[Zimmermann \& Willcox(2016)]{Zimmermann:2016}
{\sc \au{Zimmermann, R.} \& \au{Willcox, K.}} \yr{2016}  \at{An accelerated
  greedy missing point estimation procedure}.  \jt{SIAM Journal of Scientific
  Computing}  \bvol{38},  \pg{2827--2850}.

\end{thebibliography}

\end{document}